\documentclass[5pt,a4paper]{article}%
\usepackage{bm}
\usepackage{eurosym}
\usepackage{amsmath}
\usepackage{amsfonts}
\usepackage{amssymb}
\usepackage{graphicx}
\usepackage{fullpage}
\usepackage{url}
\usepackage{amsthm}
\usepackage{caption}
\usepackage{subcaption}
\usepackage{comment}%
\usepackage{lineno}
\usepackage[title]{appendix}
\usepackage{cite}
\usepackage{hyperref}

\providecommand{\U}[1]{\protect\rule{.1in}{.1in}}
\newtheorem*{theorem}{Principal Result}
\newtheorem{remark}{Remark}

\newtheorem{lemma}{Lemma}
\newtheorem{proposition}{Proposition}
\providecommand{\keywords}[1]
{
  \small	
  \textbf{\textit{Keywords---}} #1
}

\numberwithin{equation}{section}

\begin{document}

\title{Complex oscillatory motion of multiple spikes in a three-component Schnakenberg system}
\author{Shuangquan Xie\footnote{WPI-AIMR center, Tohoku university, Japan.}, Theodore Kolokolnikov \footnote{Department of Mathematics and Statistics, Dalhousie University, Canada.} and Yasumasa Nishiura \footnote{Research Institute for Electronic Science, Hokkaido University,
WPI-AIMR center, Tohoku University, MathAM-OIL,
Tohoku University and AIST, Japan.}}
\date{\today}
\maketitle

\begin{abstract}
In this paper, we introduce a three-component Schnakenberg model, whose key feature is that it has a solution consisting of $N$ spikes that undergoes Hopf bifurcations with respect to $N$ distinct modes nearly simultaneously. This results in complex oscillatory dynamics of the spikes, not seen in
typical two-component models. For parameter values beyond the Hopf bifurcations, we derive reduced equations of motion which consist of coupled ordinary differential equations (ODEs) of dimension $2N$ for spike positions and their velocities. These ODEs fully describe the slow-time evolution of the spikes near the Hopf bifurcations. We then apply the method of multiple scales to the resulting ODEs to derive the long-time dynamics. For a single spike, we find that its long-time motion consists of oscillations near the steady state whose amplitude can be computed explicitly. For two spikes, the long-time behavior can be either in-phase or out-of-phase oscillations. Both in-phase and out-of-phase oscillations are stable, coexist for the same parameter values, and the fate of motion depends solely on the initial conditions. Further away from the Hopf bifurcation points, we offer numerical experiments indicating the existence of highly complex  oscillations.
\end{abstract}
\section{Introduction}
Nonlinear reaction-diffusion (RD) systems are commonly used to model self-organized phenomena in nature, such as vegetation patterns \cite{klausmeier1999regular}, species invasion phenomena \cite{mcgillen2014general}, chemical reactions \cite{schnakenberg1979simple,
pearson1993complex}, animal skin patterns \cite{kondo1995reaction,
barrio1999two, maini2012turing, shaw1990analysis}, and morphogenesis
\cite{benson1993diffusion}. One of the simplest RD models is the Schnakenberg system \cite{iron2004stability, wei2008stationary, kolokolnikov2009spot,
kolokolnikov2018pattern}. It is a two-component model of a simplified
activator-substrate reaction, and is often used as one of the simplest models for studying spike dynamics in reaction-diffusion systems; it is also a limiting case of both the Gray-Scott model \cite{pearson1993complex}
and the Klausmeyer model for vegetation \cite{klausmeier1999regular}. It
describes the space-time dependence of the concentrations of the intermediate
products $u$ (the activator) and $v$ (the substrate) in a sequence of
reactions, and has the form%
\begin{equation}
\left\{
\begin{array}
[c]{c}%
u_{t}=D_{u}u_{xx}-u+u^{2}v,\\
v_{t}=D_{v}v_{xx}+A-u^{2}v.
\end{array}
\right.  
\end{equation}
Here, $u$ and $v$ represent the concentrations of the activator $P$ and substrate $Q$ respectively, and $D_{\left\{  u,v\right\}  }$ denotes their respective diffusion
coefficients. The activator decays exponentially whereas the substrate $v$ is fed into the system with a constant rate $A$. The nonlinear term corresponds
to the reaction $P+2Q\rightarrow3Q.$

In this paper, motivated by the three-component model of gas discharge phenomena \cite{schenk1997interacting, bode2002interaction, or1998spot,
teramoto2004phase}, we introduce an \textquotedblleft
extended\textquotedblright\ Schnakenberg model having the following form:%
\begin{equation}
\left\{
\begin{array}{rl}
u_{t}=&D_{u}u_{xx}-k_{1}u-k_{2}w+u^{2}v,\\
\theta v_{t}=&D_{v}v_{xx}+A-u^{2}v,\\
\tau w_{t}=&D_{w}w_{xx}+u-k_{3}w.
\end{array}
\right.   \label{eq0p}%
\end{equation}
The system (\ref{eq0p}) is a three-component activator-substrate-inhibitor system. It has an additional reactant $w$ acting as an inhibitor to $u$, which interacts with $v$ indirectly through the intake of activator $u$. The reactant $w$ can be treated as an out-product that is being removed continuously during the reaction process, see \cite{mahara2004three}. The \textquotedblleft classical\textquotedblright~Schnakenberg model corresponds to choosing $k_2=0,$ so that $w$ plays no role in the reaction for $u$ and $v.$

\begin{figure}[th]
\includegraphics[width=1\textwidth]{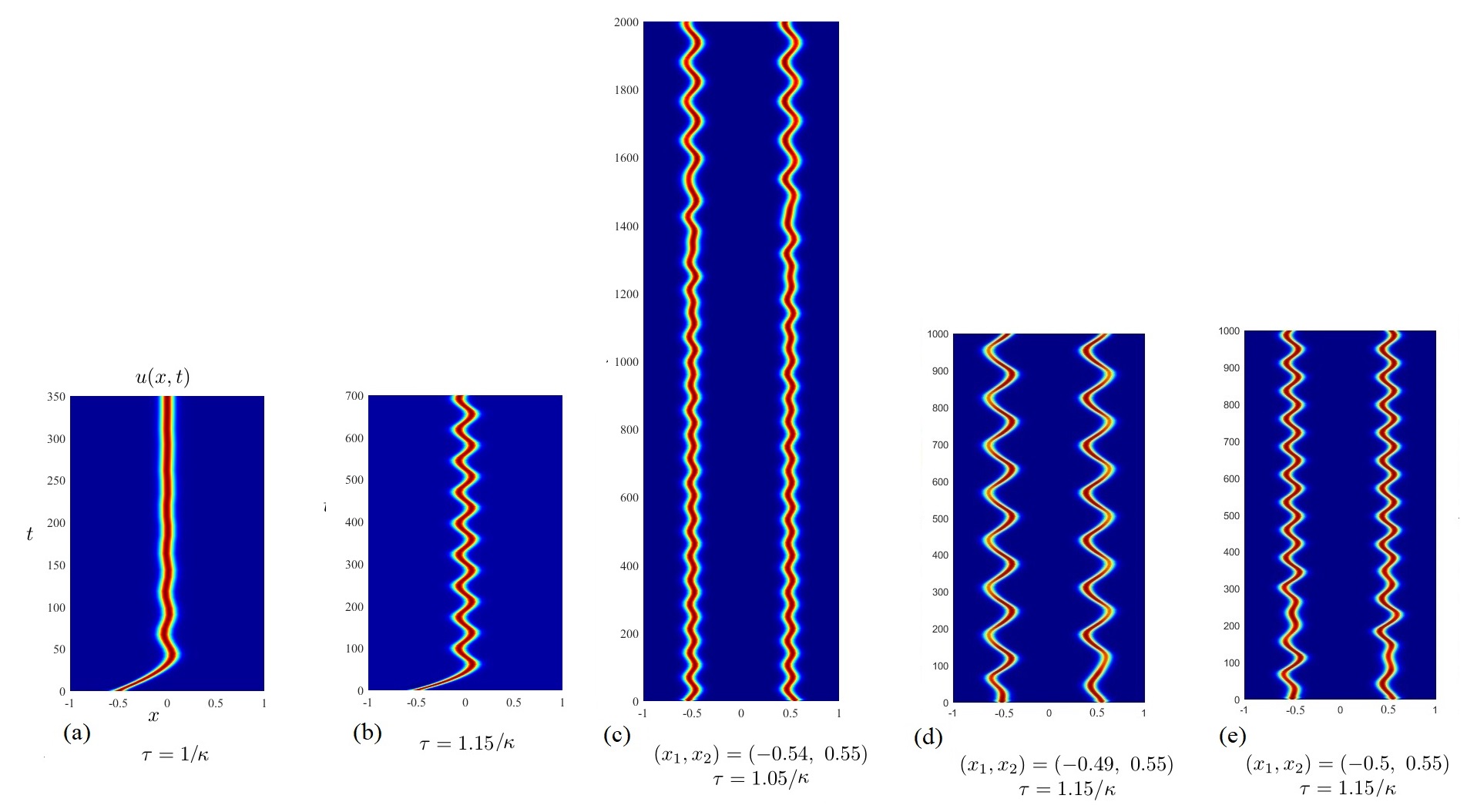}\caption{(Color online) Simulation of
the PDE system~(\ref{eq0})\ with $\varepsilon=0.025,\ \ D=0.00625,$ $\kappa=0.5$ with one or
two spikes. Initial conditions are $u(x,0)=0.1\exp(-100(x-x_{1})^{2}%
)+0.1\exp(-100(x-x_{2})^{2}),\ \ v(x,0)=1,\ \ w(x,0)=u(x,0).$ In (a-b),
$x_{1}=x_{2}=-0.5$ and $\tau$ is as indicated. There is a Hopf bifurcation at $\tau\sim1/\kappa+O(\varepsilon^{2}).$ Decaying oscillations in spike position are observed in (a) with sustained oscillations in (b). (c):\ with $\tau=1.05/\kappa,$ even oscillations are unstable with solution eventually settling into odd oscillation (in-phase oscillation) after (possibly long)\ initial transient. (d-e): For $\tau=1.15/\kappa,$ the system converges to either odd or even oscillation depending only on initial conditions. Both even (out-of-phase) and odd (in-phase) oscillations are stable in this case.}%
\label{fig:intro}%
\end{figure}
We are interested in the new dynamics that the extra variable $w$ introduces, as compared to the previous studies of dynamics in RD systems. We will use $\tau$ as the bifurcation parameter, which controls the reaction speed ratio of $u$ and $w$. It has been shown in many RD systems \cite{nishiura1989layer,xie2016oscillations,mckay2012interface,ikeda1994pattern} that oscillatory behaviors of the fronts and localized patterns are observed when $\tau$ varies. Oscillation of either the pattern tip or position occurs.  We expect similar behavior in this three-component system.  Numerical evidence shows that varying $\tau$ will lead to oscillation of the spike center, accompanied by  oscillation of the spike tip. Therefore, unless otherwise stated, the spike oscillation in this paper refers to the oscillation of the spike centers instead of the spike tips. To simplify the analysis, we will also assume that $\theta$ and $D_{w}$ are sufficiently small and can be set to zero. We remark that this assumption leads to a singular reduction of the system since it alters the order of differential equation.  It is not clear whether the conclusion we make in this paper can be extended to the case when $\theta$ and $D_w$ are small but not zero.  In addition, we will write $\varepsilon^{2}=D_{u}$ and assume that $\varepsilon\ll 1$. Hereafter we use $D$ instead of $D_v$ for simplicity. By further rescaling, $\hat{u}=\frac{1}{\varepsilon} u,~\hat{v}=\varepsilon v, \text{and}~\hat{w}= \frac{1}{\varepsilon}w$; normalizing the coefficients $k_1, k_2, k_3,$ and $A$ in a standard way \cite{murray2007mathematical}; and dropping the hat, we consider the following system as our starting point without loss of generality:%
\begin{equation}\label{eq0}%
\left\{
\begin{array}
{l}%
\begin{array}
{rl}%
u_{t}&=\varepsilon^{2}u_{xx}-(1-\kappa)u-\kappa w+u^{2}v,\\
0&=D v_{xx}+\frac{1}{2}-u^{2}v/\varepsilon,\\
\tau w_{t}&=u-w,
\end{array}
\ \ x\in\left(  -1,1\right)  ,\ \ t\geq0.\\
\text{Neumann boundary conditions at }x=\pm1.
\end{array}
\right. 
\end{equation}
This scaling simplifies the calculation; in particular, the Schnakenberg model is a special case corresponding to the singular limit $\tau \rightarrow 0$. In this case, the solution is well known to consist of $N$ spikes whose stability and dynamics have been extensively studied
\cite{kolokolnikov2005existence,iron2004stability,kolokolnikov2009spot,ward2002existence,doelman2000slowly,xie2017moving,kolokolnikov2018pattern,bellsky2013adiabatic}. The basic steady state consisting of $N$ stationary  spikes persists even when $\tau>0.$ On the other hand, we will show using a simple argument in \S \ref{sec2} that this basic state undergoes a Hopf bifurcation when $\tau$ is increased to slightly more than $1/\kappa,$ leading to oscillatory behavior in the spike positions. Moreover, $N$ small eigenvalues (controlling the motion of $N$ spikes) undergo Hopf bifurcations nearly simultaneously. Consequently, a complex interaction between the different modes can be observed, leading to the coexistence of multiple possible stable oscillating patterns. The main goal of this paper is to shed light on this complex behavior using finite-dimensional reduction and multiple scales techniques.

Fig.~\ref{fig:intro} illustrates the aforementioned phenomenon. For a single spike, there is a single small eigenvalue that undergoes Hopf bifurcation for values of $\tau$ slightly greater than $1/\kappa$, which causes the spike center to oscillate periodically.  In \S \ref{sec3}, we compute the amplitude modulation of this position oscillation as a function of $\tau.$ For two spikes, the long-time dynamics are even more interesting. We show in \S \ref{sec3} that for $\tau$ slightly above the Hopf bifurcation, the dynamics settle into one of two possible patterns, corresponding to either odd (in-phase) or even (out-of-phase) oscillations in spike positions (see Figs.~\ref{fig:intro}(d)\ and (e), respectively). Which pattern is chosen depends on both $\tau$ and the initial conditions. When $\tau=\frac{1.05}{\kappa},$ only even oscillations are stable (Fig.~\ref{fig:intro}(c)). On the other hand, when $\tau=\frac{1.15}{\kappa}$ both even and odd oscillations coexist for the same parameter values and the pattern selection mechanism depends only on the initial conditions.

The main result of this paper is as follows.
\begin{theorem}
\label{thm:main}
Let
\[
\tau=\frac{1}{\kappa}+\varepsilon^{2}\hat{\tau},%
\]
and assume that $\hat{\tau}=O(1)$ as $\varepsilon\rightarrow0.$ Then there exists a solution consisting of $N$ spikes nearly-uniformly spaced, but whose
centers evolve near the symmetric configurations on a slow time-scale
according to the following. Let $\hat{x}_{k}$ be the center of the $k$-th spike.
Then $\hat{x}_{k}\sim-1+\frac{2k-1}{N}+\varepsilon p_{k}$ where%
\begin{equation}\label{postion}
p_{k}=\sum_{j=1}^{N}Q_{kj}\mathcal{B}_{j}(\varepsilon^{2}%
t)\cos\left(  \varepsilon\omega_{j}t+\theta_{j}(\varepsilon^2 t)\right).
\end{equation}
In Eq.~(\ref{postion}), $Q_{kj}$ is the element of the matrix $Q$ defined by Eq.~(\ref{defQ}), the amplitude modulations $\{ B_{j}(s),~j=1,\ldots, N \}$ and phase modulations $\{ \theta_{j}(s),~j=1,\ldots, N \}$ solve the ODE system Eq.~(\ref{NspikesB}), and the frequencies $ \{ \omega_{j},~j=1,\ldots, N \} $ satisfy $\omega_{j}=\sqrt
{\frac{-\kappa\lambda_{j,0}}{3N}}$, where $\lambda_{j,0}$ is defined by
Eq.~(\ref{lambda}).
\end{theorem}

The remainder of this article is structured as follows. In \S \ref{sec2}, we describe the instabilities of multiple-spike patterns triggered by increasing $\tau$. It turns out that in the regime $\tau\sim\frac{1}{\kappa}%
+\hat{\tau}\varepsilon^{2}$, only eigenvalues that are asymptotically small as $\varepsilon\rightarrow 0$ can become unstable. In the case at hand, all small eigenvalues correspond to translational modes to leading order, and their instability induces a slow (possibly periodic) motion of the spikes.  In section \S \ref{sec3}, we present the derivation of a reduced ODE system that describes the spike motion for $\tau$ near $\frac{1}{\kappa
}.$ The reduced system contains $2N$ variables corresponding to both the positions and velocities of $N$ spikes. In \S \ref{sec4}, we apply the method of multiple
scales to the resulting reduced system, which yields the Principal Result. We conclude with some open questions in \S \ref{sec:discuss}. Finally, in the Appendix, we present the detailed evaluations of several integrals used in \S \ref{sec3}.

\section{Stability analysis for N-spike solution}
\label{sec2}
In this section, we describe the stability of N-spike profiles for the system (\ref{eq0}).  We begin by formulating the linear stability problem of the steady state. As the equilibrium solution is the same as the two-component system, we follow the conclusion in \cite{iron2004stability}.

\begin{lemma}
As $\varepsilon\rightarrow0$, the system (\ref{eq0}) admits a N-spike solution $(u_s,v_s,w_s)$, whose leading order is given by%
\begin{equation}
w_{0}=u_{0}=\frac{1}{\xi_{0}}\sum_{j=1}^{N}\rho\left(  \frac{x-x_{j}%
}{\varepsilon}\right),
\end{equation}
where $x_{j}=-1+\frac{2j-1}{N},~j=1,\ldots, N$, $\rho(y)=\frac{3}{2}%
\operatorname{sech}{}^{2}(y/2)$ is the unique positive solution to%
\begin{equation}
\rho_{yy}-\rho+\rho^{2}=0,\ \ \rho^{\prime}(0)=0,\ \rho\rightarrow
0~\text{as}~y\rightarrow\pm\infty; \label{rho}%
\end{equation}%
\begin{equation}
\xi_{0}=N\int_{-\infty}^{\infty}\rho^2(y)dy;
\end{equation}
and $v_{0}$ satisfies%
\begin{equation}
\left\{
\begin{array}
[c]{l}%
D v_{0xx}+\frac{1}{2}-\frac{1}{N}\sum_{j=1}^{N}\delta(x-x_{j})=0,\\
v_{0}(x_{j})=\xi_{0},\\
v_{0}^{\prime}(-1)=v_{0}^{\prime}(1)=0.
\end{array}
\right. 
\end{equation}
\end{lemma}
We are interested in the stability of the N-spike equilibrium solution.  To analyze the stability of the equilibrium solution, we introduce small perturbations 
\begin{equation}
u=u_{s}+e^{\lambda t}\phi(x),~~v=v_{s}+e^{\lambda t}\eta(x),~~w=w_{s}%
+e^{\lambda t}\psi(x). \label{perb}%
\end{equation}
Substituting Eq.~(\ref{perb}) into the system (\ref{eq0}) gives the following eigenvalue problem for an eigenfunction $[\phi ,\eta ,\psi]$
\begin{subequations}
\label{eigen_0}%
\begin{align}
\lambda\phi &  =\varepsilon^{2}\phi_{xx}-(1-{\kappa})\phi+u_{s}^{2}\eta
+2u_{s}v_{s}\phi-{\kappa}\psi,\label{phi}\\
0  &  =D\eta_{xx}-\varepsilon^{-1}\left(  u_{s}^{2}\eta+2u_{s}v_{s}\phi\right),
\label{eta}\\
\tau\lambda\psi &  =\phi-\psi\label{psi},%
\end{align}
with Neumann boundary conditions. From (\ref{psi}), we obtain
\end{subequations}
\begin{equation}
\label{temp1}\psi=\frac{\phi}{1+\tau\lambda}.
\end{equation}
Using (\ref{temp1}) to remove $\psi$ in (\ref{phi}) yields
\begin{subequations}
\label{eigenpro}%
\begin{align}
\lambda(1-\frac{\tau{\kappa}}{1+\tau\lambda})\phi &  =\varepsilon^{2}\phi
_{xx}-\phi+u_{s}^{2}\eta+2u_{s}v_{s}\phi,\label{phi2}\\
0  &  =D\eta_{xx}-\varepsilon^{-1}\left(  u_{s}^{2}\eta+2u_{s}v_{s}%
\phi\right).  \label{eta2}%
\end{align}
\end{subequations}
 As the terms in the right-hand sides of Eqs~(\ref{eigenpro}) are the same for $\tau=0$ and $\tau \neq 0$, the eigenfunction are also the same. Let $\gamma$ and $[\phi^{0},\eta^{0}]$ be the eigenvalue and eigenfunction of the eigenvalue problem Eqs.~(\ref{eigenpro}) at $\tau=0$. By comparing Eqs.~(\ref{eigenpro}) for $\tau=0$ and $\tau \neq 0$, it can readily be seen that the roots of
\begin{equation}\label{quar}
\lambda(1-\frac{\tau{\kappa}}{1+\tau\lambda})=\gamma,%
\end{equation}
are the eigenvalues of Eqs.~(\ref{eigenpro}). Solving Eq.~(\ref{quar}) for $\lambda$ yields
\begin{equation}\label{lambdaMain}
    \lambda=\frac{\tau\kappa+\tau \gamma-1}{2\tau}\pm \sqrt{\frac{\gamma}{\tau} +\left(\frac{\tau\kappa+\tau \gamma-1}{2\tau}\right)^2 }.
\end{equation}
In addition, the eigenfunction of Eqs~(\ref{eigen_0}) is
\begin{equation}
    [\phi,\eta,\psi]=[\phi^0,\eta^0, \frac{\phi^{0}}{1+\tau\lambda}].
\end{equation}
As we are interested in the role of parameter $\tau$ on the stability of the N-spike solution, we require the N-spike solution to be stable with respect to $D$ when $\tau=0$.
 Let us recall the following lemma on the stability of N-spike solution corresponding to $\tau=0$ from \cite{iron2004stability},
\begin{lemma} \label{lemmaDv}
For $N\geq2$, let
\begin{equation}
D_{N}:=\frac{1}{2\int\rho^{2}dy}\frac{1}{N^{3}},%
\end{equation}
and suppose that $  \varepsilon^{2}  \ll D .$ Then for
$\varepsilon\ll1,$
\begin{itemize}
\item the one-spike solution is stable.
\item for $D<D_{N},$ the N-spike solution is stable, whereas for $D>D_{N},$ the N-spike solution is unstable.
\end{itemize}
\end{lemma}
Lemma \ref{lemmaDv} implies that the real part of $\gamma$ is negative when $D<D_{N}$. In addition, it was shown in \cite{iron2004stability} that, the first $N$ eigenvalues with the largest real parts are real and of order $ \varepsilon^{2}$, with eigenfunctions that are the translational modes to leading order. We denote $\gamma_j$ the sorted eigenvalues of Eqs.~(\ref{eigen_0}) at $\tau=0$ in order of decreasing real part, then it follows from Proposition 3.3 in \cite{iron2004stability} that $\gamma_{j}=\frac{\lambda_{j-1,0}\varepsilon^{2}}{3N}$ for $j=2,\ldots, N$ and $\gamma_N=\lambda_{1,0}$ to leading order, where $\lambda_{1,0}:=-\frac{1}{2D}$ and 
\begin{equation}
\lambda_{j,0}:=-\frac{1}{2D}-\frac{1}{24N^{3}D^{2}}\frac{1}{\tan^{2}{\frac
{\pi(j-1)}{2N}}}\left(  1-\frac{1}{12DN^{3}\sin^{2}{\frac{\pi(j-1)}{2N}}%
}\right)  ^{-1}\quad \text{for} \quad j=2,\ldots,N. \label{lambda}%
\end{equation}
We remark that the expression we use here is different from the one in Proposition 3.3 of \cite{iron2004stability}.  As the expression in the reference contains some typo, we correct it here.  With these facts in mind, we can conclude that $\lambda$ in Eq.~(\ref{lambdaMain}) has negative real part only when
\begin{equation}
0<\tau<|\frac{1}{{\kappa}+\gamma_{j}}|\leq\frac{1}{{\kappa}+\gamma_{1}}.
\end{equation}
Thus, we arrive at the following proposition,
\begin{proposition}
\label{pro1} For $N \ge 2$, the N-spike solution of the system (\ref{eq0}) is stable when $D<D_{N}$ and  $\tau<\frac{1}{{\kappa}+\gamma_{1}}$.
\end{proposition}

 When $\tau=\tau_{1}:=\frac{1}{{\kappa}+\gamma_{1}}$, the first two eigenvalues with the largest real parts, $\lambda_{1,2}=\pm i\sqrt{\frac{-\gamma_1}{\tau}}$, are purely imaginary numbers, indicating that the system undergoes a Hopf bifurcation as $\tau$ passes through $\tau_{1}$.  It is worth to note that there exist another $N-1$ threshold values $\tau_k:=\frac{1}{{\kappa}+\gamma_{k}}\sim \frac{1}{\kappa}-\frac{ \lambda_{k-1,0} \varepsilon^2 }{3N \kappa^2}+\cdots$ for $k=2 \ldots N$ in a small neighbourhood of $\tau_1$ such that $k$ eigenvalues have positive real part when $\tau>\tau_k$, indicating that $k$ translational modes become unstable almost simultaneously. It is natural to further investigate the dynamics of the N-spike solution when all of the translational modes become unstable.  We conduct the multiple-scale analysis and the method of matched asymptotic expansions to study the dynamics of the spikes beyond the Hopf bifurcations, as described in the next section.
\begin{remark}
Note that all the $\tau_k$ have distances of $\mathcal{O} (\varepsilon^2)$ to the value $\tau_c:=\frac{1}{\kappa}$. If we confine our analysis close to $\tau_c$, the large eigenvalues will remain strictly negative for all small $\varepsilon$, enabling us to focus only on how different translational modes interact.
\end{remark}

\section{Dynamics of multiple spikes near the threshold} \label{sec3}
In this section, we study the dynamical solutions that appear through the Hopf bifurcation near $\tau_{c}:=\frac{1}{{\kappa}}$. For convenience, we rewrite $\tau$ as $\tau=\tau_{c}+\varepsilon^{2}\hat{\tau}$, where $\hat{\tau}\sim \mathcal{O}(1)$. With this notation, the first $2N$ eigenvalues of the problem (\ref{eigen_0}) with the largest real parts are
\begin{equation}
    \lambda_k \sim \frac{1}{2}\varepsilon^2\left(\hat{\tau}\kappa^2+\frac{\lambda_{k,0}}{3N} \right)\pm i\varepsilon \sqrt{\frac{-\kappa \lambda_{k,0}}{3N} },~\text{for}~k=1,\ldots, N,
\end{equation} to leading order, whereas the real parts of the remaining eigenvalues are strictly negative. Note that the real and imaginary parts of the first $2N$ eigenvalues are small and of different orders with respect to $\varepsilon$, which suggests that the problem could be handled by using a multiple-time-scale method.  Thus, following the multiple-time-scale perturbation approach described in the framework of front bifurcations \cite{bode2002interaction,gurevich2013moving}, and the matched asymptotic method used in studying the two-component RD model \cite{kolokolnikov2005existence},  we derive a reduced dynamical system describing the locations and velocities of each spike near the bifurcation points.

We first rescale the time as $\hat{t}=\varepsilon t$. After dropping the hat, the system (\ref{eq0}) becomes
\begin{equation}
\left\{
\begin{array}
[c]{c}%
\varepsilon u_{t}=\varepsilon^{2}u_{xx}-(1-{\kappa})u+u^{2}v-{\kappa}w,\\
0=D v_{xx}+\frac{1}{2}-u^{2}v/\varepsilon,\\
\tau\varepsilon w_{t}=u-w, \\
\text{Neumann boundary conditions at }x=\pm1.
\end{array}
\right.  \label{eq1.0}%
\end{equation}
The spikes will oscillate around their equilibrium positions $\{x_k=-1+\frac{2k}{N},~k=1 \ldots N\} $; thus we assume the $k$-th spike to be located at $\hat{x}_k=x_k+\varepsilon p_{k}$.  Then, we calculate the solution in the inner region near the $k$-th spike where $|x-\hat{x}_k|\sim \mathcal{O}(\varepsilon)$, and in the outer region away from the $k$-th spike where $|x-\hat{x}_k| \gg \mathcal{O}(\varepsilon)$. The equations for the position and velocity of each spike are determined by matching the outer and inner solutions.

\textbf{Inner region}: Near the $k$-th spike, we introduce variable $y=\frac{x-x_{k}-\varepsilon p_{k}
(t)}{\varepsilon}$, and rewrite $u,v$ and $w$ as
\begin{equation}
u(x,t)=U(y,t),~~~v(x,t)=V(y,t),~~~w(x,t)=W(y,t).
\end{equation}
Then, the system (\ref{eq1.0}) becomes
\begin{subequations}\label{eq2}
\begin{align}%
-U_{y}\varepsilon \dot{p}_{k}+\varepsilon \frac{\partial U}{\partial t}&=U_{yy}-(1-{\kappa})U+U^{2}V-{\kappa}W,\\
0&=DV_{yy}+\frac{1}{2}\varepsilon^{2}-\varepsilon U^{2}V,\\
-W_{y}\varepsilon \dot{p}_{k}+\varepsilon \frac{\partial W}{\partial t}&=\frac{1}{\tau}(U-W).  
\end{align}
\end{subequations}
The far-field conditions as $y \rightarrow \infty$ are that $U$ and $W$ tend to zero, whereas the conditions for $V$ can be obtained by matching with the outer solution.

We introduce slow time scales
\begin{equation*}
T_{1}=\varepsilon t,\quad T_{2}=\varepsilon^{2}t,
\end{equation*}
and the following ansatz to facilitate the analysis
\begin{equation}
\begin{bmatrix}
U\\
V\\
W
\end{bmatrix}
=%
\begin{bmatrix}
U_{0}\\
V_{0}\\
W_{0}%
\end{bmatrix}
+\varepsilon\left(
\begin{bmatrix}
U_{1}\\
V_{1}\\
W_{1}%
\end{bmatrix}
+\alpha_{k}%
\begin{bmatrix}
0\\
0\\
U_{0y}%
\end{bmatrix}
\right)  +\varepsilon^{2}%
\begin{bmatrix}
U_2\\
V_2\\
W_2%
\end{bmatrix}
+\varepsilon^{3}%
\begin{bmatrix}
U_3\\
V_3\\
W_3%
\end{bmatrix}
+h.o.t\quad . \label{expansion}%
\end{equation}
For the expansion of different orders, one must be aware of the following dependence:
\begin{equation}
\begin{split}
    &p_k=p_k(t,T_1,T_2),\quad \alpha_k=\alpha_k(t,T_1,T_2),\\
        &\begin{bmatrix}
U_{k}\\
V_{k}\\
W_{k}%
\end{bmatrix}=        \begin{bmatrix}
U_{k}(y,t)\\
V_{k}(y,t)\\
W_{k}(y,t)%
\end{bmatrix},\quad \text{for} \quad k\geq 1.
\end{split}
\end{equation}
To make the expansion of $U$ and $W$ unique, we also require that
\begin{equation}\label{orth}
\begin{split}
 \int_{-\infty}^{\infty} U_k U_{0y} dy=0, \\
\int_{-\infty}^{\infty} W_k U_{0y} dy=0,
\end{split}
\quad \quad \text{for} \quad k\geq 1.
\end{equation}
We remark that the ansatz (\ref{expansion}) includes an extra term $[0,0,\alpha_{k}
U_{0y}]^{T}$ in addition to the usual expansion, which is attributed to an important feature of the linearized operator (\ref{L}) around the steady state. We will explain this extra term later in the analysis of the $\mathcal{O}{(\varepsilon)}$ terms. 

We expand the system (\ref{eq2}) in the power of $\varepsilon$ and collect terms with equal powers of $\varepsilon$.

$\bullet$ In the leading order, we obtain
\begin{subequations}\label{order1}
\begin{align}
0&=U_{0yy}-(1-{\kappa})U_{0}+U_{0}^{2}V_{0}-{\kappa}U_{0},\\
0&=V_{0yy},\\
0&=U_{0}-W_{0}.%
\end{align}
\end{subequations}
It follows that
\begin{equation} \label{order0}
V_{0}=C_{0,k},\quad U_{0}=\rho(y)/C_{0,k}, \quad W_0=\rho(y)/C_{0,k} .
\end{equation}
where $\rho(y)=\frac{3}{2}\text{sech}^2{(\frac{y}{2})}$ and $C_{0,k}$ is a constant to be determined by matching with the outer solution near $x_{k}$. 

$\bullet$ In the order of $\varepsilon,$ we obtain%
\begin{subequations}\label{order2}
\begin{align}
-\frac{\partial p_{k}}{\partial t}U_{0y}+\frac{\partial U_0}{\partial t} & =U_{1yy}-(1-{\kappa})U_{1}%
+U_{0}^{2}V_{1}+2U_{0}V_{0}U_{1}-{\kappa}(\alpha_{k}U_{0y}+W_{1})\label{order2_1}, \\
0 & =DV_{1yy}-U_{0}^{2}V_{0} \label{order2_2},\\ 
-\frac{\partial p_{k}}{\partial t}W_{0y}+\frac{\partial W_0}{\partial t} & =\kappa \left(
U_{1}-\alpha_{k}U_{0y}-W_{1}\right). \label{order2_3}
\end{align}
\end{subequations}
A key observation is that $V_1$ only relies on $U_0$ and $V_0$ from Eq.~(\ref{order2_2}). One can solve for $V_1$ first and reduce the system (\ref{order2}) to a two-component system. Integrating both sides of Eq.~(\ref{order2_2}) yields
\begin{equation} \label{V1y}
    \frac{\partial V_1}{\partial y} =\frac{1}{D}  \int_0^y U_0^2V_0 dz+B_{1,k}.
\end{equation}
As $\int_0^y U_0^2V_0 dz$ is odd, the constant $B_{1,k}$ can be determined by the far-field behavior as follows
\begin{equation}
    B_{1,k}=\frac{1}{2}\left( \frac{\partial V_1}{\partial y}(+\infty)+ \frac{\partial V_1}{\partial y}(-\infty) \right) .
\end{equation}
Integrating Eq.~(\ref{V1y}), we obtain
\begin{equation}\label{V1}
    V_1=\frac{1}{D} \int_0^y \int_0^{\hat{y}} U_0^2V_0 dz d\hat{y} +B_{1,k}y+C_{1,k},
\end{equation}
where $C_{1,k}$ is determined by matching with the outer solution near $x_k$.  We remark that the far field behavior of $V_1$ is linear in $y$ as,
\begin{equation}\label{V1far}
    V_1(y)  \sim y \left(\frac{1}{D}  \int_0^{\pm \infty} U^2_0 V_0 dz  + B_{1,k} \right)+\left[ C_{1,k}+\frac{1}{D}\int_0^{\pm\infty}  \left( \int_0^{\hat{y}} U^2_0 V_0 dz-\int_0^{\pm \infty} U^2_0 V_0 dz  \right) d\hat{y}\right],~\text{as}~y \rightarrow \pm \infty, 
\end{equation} 
where the term in square brackets is the constant term that will be used in the later matching procedure. By taking $V_1$ as a known function in the system (\ref{order2}) and noting that $\frac{\partial W_0}{\partial t}=\frac{\partial U_0}{\partial t}=0$, we obtain
\begin{equation}\label{tc1}
\begin{bmatrix}
\left(  \frac{-\partial p_{k}}{\partial t}+{\kappa}\alpha_{k}\right)  U_{0y}-U_0^2V_1\\
\left(  \frac{-\partial p_{k}}{\partial t}+\kappa\alpha_{k}\right)
W_{0y}%
\end{bmatrix}
=\mathcal{L}%
\begin{bmatrix}
U_{1}\\
W_{1}%
\end{bmatrix},
\end{equation}
where
\begin{equation}
\mathcal{L}=%
\begin{pmatrix}
\frac{\partial^{2}}{\partial y^{2}}-(1-{\kappa})+2U_{0}V_{0} &
-{\kappa}\\
\kappa & -\kappa%
\end{pmatrix}.
\label{L}%
\end{equation}
Inspection of the linear operator $\mathcal{L}$ reveals that it has an eigenfunction with eigenvalue $0$, 
\begin{equation}
    G:=[U_{0y},W_{0y}]=[U_{0y},U_{0y}],
\end{equation}
such that
\begin{equation}
    \mathcal{L}G=0,
\end{equation}
which is referred to as ``goldstone mode" in \cite{gurevich2013moving,bode1997front}. 
However, the eigenvalue zero is not simple but has an algebraic multiplicity of two. Indeed, we can find a generalized eigenfunction
\begin{equation}\label{P}
    P:=[0,-\frac{1}{\kappa}U_{0y}],
\end{equation}
such that
\begin{equation}
    \mathcal{L}P=G,
\end{equation}
which is referred as ``propagator mode" in \cite{gurevich2013moving,bode1997front}.
The set of eigenfunctions is incomplete and must be supplemented with a generalized eigenfunction $P$ with the eigenvalue zero. Note that $P$ is not unique since any addition $cG$ to it is also a generalized eigenfunction. Here we have chosen P as Eq (\ref{P}) for the convenience of computation. Then the solution of the inhomogeneous equation (\ref{tc1}) can be represented as a linear combination of eigenfunctions and the generalized eigenfunction. Thus we include the term $[0,0,\alpha_{k} U_{0y}]^{T}$ in our ansatz (\ref{expansion}). The homogeneous solution $[U_{0y},0,U_{0y}]^{T}$ of the system (\ref{order2}) has been also implicitly incorporated into the ansatz.   

As the operator $\mathcal{L}$ is not self-adjoint, its eigenfunctions do not have an orthogonality relation. The equations for $p_k$ and $\alpha_k$ have to be calculated by projection onto the eigenfunction $G^{\dagger}$ and generalized function $P^{\dagger}$ of the adjoint operator $\mathcal{L}^{\dagger}$ to the eigenvalue zero.

For clarity, we use the following definitions for the inner product and adjoint operator.  Set $Z=L^2(\mathbb{R}) \times L^2(\mathbb{R})$, we define the inner product of two function pair, $H_j=(\phi_j,\psi_j) \in Z ,j= 1,2$, as
\begin{equation}
\langle H_1, H_2 \rangle_{Z}= \int_{\mathbb{R}} \phi_1 \bar{\psi_1} + \phi_2 \bar{\psi_2} dx,
\end{equation}
where the overbar denotes the complex conjugate.
The adjoint operator of $\mathcal{L}$ is defined by the linear operator $\mathcal{L}^{\dagger}$ fulfilling
\begin{equation}
\langle \mathcal{L} H_1, H_2 \rangle_{Z}= \langle  H_1, \mathcal{L}^{\dagger} H_2 \rangle_{Z}.
\end{equation}
With these definitions, the adjoint operator of $\mathcal{L}$ is
\begin{equation}
\mathcal{L}^{\dagger}=%
\begin{pmatrix}
\frac{\partial^{2}}{\partial y^{2}}-(1-{\kappa})+2U_{0}V_{0} &\kappa
\\
-{\kappa} & -\kappa%
\end{pmatrix}.
\label{L*}%
\end{equation}
By inspection,  
\begin{equation}
G^{\dagger}:=[U_{0y},-U_{0y}]^T \quad \text{and} \quad  P^{\dagger}:=[\frac{1}{\kappa} U_{0y},0]^T
\end{equation}
are the eigenfunction and generalized eigenfunction of  $\mathcal{L}^{\dagger}$ to the eigenvalue $0$ such that
\begin{equation}
    \mathcal{L}^\dagger G^{\dagger}=0, \quad \mathcal{L}^\dagger P^{\dagger}=G.
\end{equation}
To make the expansion of $U$ and $W$ unique, we also demand the orthogonality relations $\int_{-\infty}^{\infty} [U_k,W_k] \cdot P^{\dagger} dy=0$ and $\int_{-\infty}^{\infty} [U_k,W_k] \cdot G^{\dagger}dy=0$, resulting in the condition (\ref{orth}). 

By projecting Eqs.~(\ref{tc1}) onto $G^{\dagger}=[U_{0y},-U_{0y}]^T
$, we obtain the first solvability condition
\begin{equation}
\label{solv1}
0=\int_{-\infty}^{\infty} U^{2}_{0} V_1 U_{0y} dy.
\end{equation}
We now show that Eq.~(\ref{solv11}) is an identity and yields no information about the dynamics. Integrating by parts and substituting Eq.~(\ref{V1y}) into Eq.~(\ref{solv1}) yields%
\begin{equation}
\label{solv11}
0=\int_{-\infty}^{\infty} U^{2}_{0} V_1 U_{0y} dy=-\frac{1}{3}\int_{-\infty}^{\infty} U_0^3V_{1y} dy=-\frac{1}{3} \int_{-\infty}^{\infty} U_{0}^{3} \int_{0}^{y} U_{0}^{2}V_{0}dz dy-\frac{B_{1,k}}{3}\int_{-\infty}^\infty U^{3}_{0}(y) dy.
\end{equation}
From Eqs.~(\ref{order0}), $U_0=\rho (y)/C_{0,k}$ and $V_0=C_{0,k}$ are even functions; then, $\int_{0}^{y} U_{0}^{2}V_{0} dz$  and $U_{0}^{3} \int_{0}^{y} U_{0}^{2}V_{0} dz$ are odd, it can be concluded that
 \begin{equation}\label{int1s}
 \int_{-\infty}^{\infty} U_{0}^{3} \int_{0}^{y} U_{0}^{2}V_{0} dy=0.
 \end{equation}
We will see later that $B_{1,k}=0$ from Eq.~(\ref{B1k}) by matching with the outer solution; thus, by Eq.~(\ref{B1k}) and Eq.~(\ref{int1s}), Eq.~(\ref{solv11}) becomes an identity.

Projection of  Eqs.~(\ref{tc1}) onto $P^{\dagger
}=[\frac{1}{\kappa} U_{0y},0]^T$ yields the second solvability condition
\begin{equation}\label{TT11}
\left(\frac{\partial p_{k}}{\partial t}-{\kappa} \alpha_{k}\right) \int_{-\infty}^\infty U_{0y}^2~ dy=\int_{-\infty}^{\infty} U^{2}_{0} V_1 U_{0y} dy.%
\end{equation}
Since $\int_{-\infty}^{\infty} U^{2}_{0} V_1 U_{0y} dy=0$ from Eq.~(\ref{solv1}), Eq.~(\ref{TT11}) becomes
 \begin{equation}\label{T11}
    \frac{\partial p_{k}}{\partial t}={\kappa} \alpha_{k}.
\end{equation}
For later use, we solve for $U_1$ and $W_1$ explicitly as
\begin{equation}\label{U_1}
    W_1=U_1=-\frac{ C_{1k} \rho}{C_{0,k}^2} -\frac{ \rho \int_0^y \int_0^z  \rho^2 d\hat{y} dz-f }{DC_{0,k}^3} ,
\end{equation}
where $f$ is defined as
\begin{equation}\label{Deff}
    f=-\frac{7}{4}\rho^2+\frac{5}{4}\rho-3y\rho'.
\end{equation}
Further details are given in the Appendix (\ref{Af}).

$\bullet$ In the order of $\varepsilon^{2}$, we obtain%
\begin{subequations}\label{order3}
\begin{equation}\label{order3-1}
-\frac{\partial p_{k}}{\partial T_{1}}U_{0y}-\frac{\partial p_{k}}{\partial t}U_{1y}+\frac{\partial U_1}{\partial t}=\frac{\partial^{2}U_2}{\partial
y^{2}}-(1-{\kappa})U_2+2U_{0}V_{0}U_2+U_{0}^{2}V_2-{\kappa}W_2
+U_{1}^{2}V_{0}+2U_{0}U_{1}V_{1},
\end{equation}
\begin{equation}\label{order3-2}
0=D\frac{\partial^{2}V_2}{\partial y^{2}}-2U_{0}V_{0}U_{1}-U_{0}^{2}V_{1} ,
\end{equation}
\begin{equation}\label{order3-3}
\left(  -\frac{\partial p_{k}}{\partial T_{1}}+\frac{\partial\alpha_{k}%
}{\partial t}\right)  U_{0y}-\alpha_{k}\frac{\partial p_{k}}{\partial t}
U_{0yy}-\frac{\partial p_{k}}{\partial t}W_{1y}+\frac{\partial W_1}{\partial t}=\kappa(W_2-U_2).
\end{equation}
\end{subequations}
 From (\ref{order3-2}), we obtain
\begin{equation}
\frac{\partial V_2}{\partial y}=\frac{1}{D}\int_{0}^{y}(2U_{0}V_{0}%
U_{1}+U_{0}^{2}V_{1})dz+B_{2,k}.%
\end{equation}
Since $\int_{0}^{y}(2U_{0}V_{0}%
U_{1}+U_{0}^{2}V_{1})dz $ is odd, the constant $B_{2,k}$ can be determined as follows,
\begin{equation}
B_{2,k}=\frac{\frac{\partial V_2}{\partial y}(+\infty)+\frac{\partial V_2%
}{\partial y}(-\infty)}{2} .\label{C2}%
\end{equation}
Noting that $\frac{\partial W_1}{\partial t}=\frac{\partial U_1}{\partial t}=0$, the following system can be obtained for $U_2$ and $W_2$:
\begin{equation}\label{tc2}
\begin{bmatrix}
-\frac{\partial p_{k}}{\partial T_{1}}U_{0y}-U_0^2V_2-(U_{1}^{2}V_{0}+2U_{0}U_{1}V_{1})-\frac{\partial p_{k}}{\partial t}U_{1y}\\
\left(  -\frac{\partial p_{k}}{\partial T_{1}}+\frac{\partial\alpha_{k}%
}{\partial t}\right)  U_{0y}-\alpha_{k}\frac{\partial p_{k}}{\partial t}%
U_{0yy}-\frac{\partial p_{k}}{\partial t}W_{1y}
\end{bmatrix}
=\mathcal{L}%
\begin{bmatrix}
U_2\\
W_2%
\end{bmatrix}.
\end{equation}

Projection of Eqs.~(\ref{tc2}) onto $G^{\dagger}=(U_{0y},-U_{0y})$ yields
\begin{equation}\label{alphat}
    \frac{\partial\alpha_{k}}{\partial t}\int_{-\infty}^\infty U_{0y}^2 dy=\frac{1}{3}\int_{-\infty
}^{\infty} U^{3}_{0}(y) \frac{\partial V_2} {\partial y} dy-\int_{-\infty}^{\infty}(U_{1}^{2}V_{0}+2U_{0}U_{1}V_{1})U_{0y}dy+\int_{-\infty}^{\infty}\alpha_{k}\frac{\partial p_{k}}{\partial t} %
U_{0yy}U_{0y}dy.
\end{equation}
Noting that $U_{0yy}U_{0y}$ is an odd function, we have
\begin{equation}
\int_{-\infty}^{\infty}\alpha_{k}\frac{\partial p_{k}}{\partial t} U_{0yy}U_{0y}dy=0.
\end{equation}
From Eqs.~(\ref{A8},\ref{A9}) in the Appendix, we have
\begin{equation}
    \int_{-\infty}^{\infty}(U_{1}^{2}V_{0}+2U_{0}U_{1}V_{1})U_{0y}dy=0,\quad \int_{-\infty}^{\infty} U^{3}_{0}(y) \int_{0}^y (2U_{0}V_{0}%
U_{1}+U_{0}^{2}V_{1})d\hat{y} dy=0.
\end{equation}
Thus, Eq.~(\ref{alphat}) becomes
\begin{equation}\label{aT1}
\frac{\partial\alpha_{k}}{\partial t}=\frac{\int_{-\infty
}^{\infty} U^{3}_{0}(y) \frac{\partial V_2} {\partial y} dy}{3 \int_{-\infty}^{\infty} U^{2}_{0y} dy}=\frac{B_{2,k} \int_{-\infty}^{\infty} U^{3}_{0}(y) dy}%
{3 \int_{-\infty}^{\infty} U^{2}_{0y} dy}.
\end{equation}

Projection of  Eqs.~(\ref{tc2}) onto
$P^{\dagger}=(\frac{1}{{\kappa}}U_{0y},0)$ yields
\begin{equation}
\frac{\partial p_{k}}{\partial T_{1}}\int_{-\infty}^{\infty} U^{2}_{0y} dy=\frac{1}{3}\int_{-\infty
}^{\infty} U^{3}_{0}(y) \frac{\partial V_2} {\partial y} dy-\int_{-\infty}^{\infty}(U_{1}^{2}V_{0}+2U_{0}U_{1}V_{1})U_{0y}dy-\int_{-\infty}^{\infty}\frac{\partial p_k}{\partial t} U_{1y}U_{0y}dy.
\end{equation}
Thus,
\begin{equation}
\label{pT1}\frac{\partial p_{k}}{\partial T_{1}}=\frac{B_{2,k}\int_{-\infty}^{\infty} U^{3}_{0}(y) dy
}{3\int_{-\infty}^{\infty} U^{2}_{0y} dy }-\frac{\partial p_k}{\partial t} \frac{\int_{-\infty}^{\infty} U_{1y}U_{0y}dy}{\int_{-\infty}^{\infty} U^{2}_{0y} dy}.
\end{equation}

$\bullet$ In the order of $\varepsilon^{3}$, we obtain
\begin{subequations}\label{order4}
\begin{equation}\label{order4-1}
-\frac{\partial p_{k}}{\partial T_{2}}U_{0y}-\frac{\partial p_{k}}{\partial T_{1}}U_{1y}-\frac{\partial p_{k}}{\partial t} U_{2y}+\frac{\partial U_2}{\partial
t}=\frac{\partial^{2}U_3}{\partial y^{2}}-(1-{\kappa})U_3+2U_{0}V_{0}%
U_3+U_{0}^{2}V_3-{\kappa}W_3+U_{1}^{2}V_{1}+2U_{0}V_{1}U_2+2U_{1}%
V_{0}U_2+2U_{0}U_{1}V_2,
\end{equation}
\begin{equation}\label{order4-2}
0=D\frac{\partial^{2}V_3}{\partial y^{2}}-U_{0}^{2}V_2-2U_{0}V_{0}%
U_2-U_{1}^{2}V_{0}-2U_{0}U_{1}V_{1},
\end{equation}
\begin{equation}\label{order4-3}
-\kappa\hat{\tau}\frac{\partial p_{k}}{\partial t}{U_{0y}}-\left(
\frac{\partial p_{k}}{\partial T_{2}}-\frac{\partial\alpha_{k}}{\partial
T_{1}}\right)  U_{0y}-\alpha_{k}\frac{\partial p_{k}}{\partial T_{1}}%
U_{0yy}-\frac{\partial p_{k}}{\partial T_{1}}W_{1y}-\frac{\partial p_{k}}{\partial t} W_{2y}+\frac{\partial W_2}{\partial
t}=\kappa(U_3-W_3).
\end{equation}
\end{subequations}
From Eq~(\ref{order4-2}), we obtain
\begin{equation}
\frac{\partial V_3}{\partial y}=\frac{1}{D}\int_{0}^{y}(U_{0}^{2}%
V_2+2U_{0}V_{0}U_2+U_{1}^{2}V_{0}+2U_{0}U_{1}V_{1})dy+B_{3,k}.
\end{equation}
The constant $B_{3,k}$ can be determined by using the following equation
\begin{equation}
2B_{3,k}+\frac{1}{D}\left( \int_{0}^{ \infty}+\int_{0}^{ -\infty} \right)(U_{0}^{2} %
V_2+2U_{0}V_{0}U_2+U_{1}^{2}V_{0}+2U_{0}U_{1}V_{1})dy=\frac{\partial V_3}{\partial y}(-\infty
)+\frac{\partial V_3}{\partial y}(+\infty
),  \label{C3}%
\end{equation}
where evaluations of the integral $\left( \int_{0}^{ \infty}+\int_{0}^{ -\infty} \right)(U_{0}^{2} 
V_2+2U_{0}V_{0}U_2+U_{1}^{2}V_{0}+2U_{0}U_{1}V_{1})dy$ is shown by Eqs.~(\ref{A20}) in the Appendix.

Then, $U_3$ and $W_3$ satisfy
\begin{equation}\label{tc3}
\begin{bmatrix}
-\frac{\partial p_{k}}{\partial T_{2}}U_{0y}+\frac{\partial U_2}{d
t}-\frac{\partial p_k}{\partial t}U_{2y}-U_0^2V_3-(U_{1}^{2}V_{1}+2U_{0}V_{1}U_2+2U_{1}V_{0}%
U_2+2U_{0}U_{1}V_2)\\
-\kappa\hat{\tau}\frac{\partial p_{k}}{\partial t}{U_{0y}}-\left(
\frac{\partial p_{k}}{\partial T_{2}}-\frac{\partial\alpha_{k}}{\partial
T_{1}}\right)  U_{0y}-\alpha_{k}\frac{\partial p_{k}}{\partial T_{1}}%
U_{0yy}+\frac{\partial W_2}{\partial
t}-\frac{\partial p_k}{\partial t}W_{2y}
\end{bmatrix}
=\mathcal{L}%
\begin{bmatrix}
U_3\\
W_3%
\end{bmatrix}.
\end{equation}
Projection of Eq.~(\ref{tc3}) onto $G^{\dagger}=(U_{0y},-U_{0y})$ yields
\begin{equation}
\int_{-\infty}^\infty  \left[ \hat{\tau}\kappa \frac{\partial p_{k}}{\partial
t}-\frac{\partial\alpha_{k}}{\partial T_{1}} \right] U_{0y}^2+\frac{\partial p_{k}%
}{\partial T_{2}}U_{0yy}U_{0y} +\left[\frac
{\partial U_2}{\partial t}-\frac{\partial W_2}{\partial t}-\frac{\partial p_k}{\partial t}U_{2y}+\frac{\partial p_k}{\partial t}W_{2y}\right]U_{0y} 
~dy=I_1+I_2+I_3,
\label{pj3}%
\end{equation}
where
\begin{subequations}\label{order4ID}
\begin{equation}
I_1=\int_{-\infty}^{\infty}U_{0y}(U_{1}^{2}V_{1}+2U_{0}V_{1}U_2+2U_{1}V_{0}%
U_2+2U_{0}U_{1}V_2)dy,
\end{equation}
\begin{equation}
   I_2= \int_{-\infty}^\infty  \alpha_{k}\frac{\partial p_{k}}{\partial T_{1}} %
U_{0yy}  U_{0y} dy,
\end{equation}
\begin{equation}
   I_3= \int_{-\infty}^{\infty} U_0^2V_3 U_{0y} dy.
\end{equation}
\end{subequations}
Since $U_{0yy}  U_{0y}$ is odd, we have
\begin{equation}
    I_2=0.
\end{equation}
As shown by Eqs.~(\ref{I1A}, \ref{I2A}) in the Appendix, $I_1$ and $I_3$ are evaluated as
\begin{equation}
    I_1=-(\frac{144}{5}\ln{2}-\frac{3438}{175})\frac{B_{2,k}}{DC_{0,k}^5}+\frac{24B_{2,k}C_{1k}}{5C_{0,k}^4},\quad I_3=(\frac{72}{5}\ln{2}-\frac{4143}{350})\frac{B_{2,k}}{DC_{0,k}^5}-\frac{12B_{3,k}}{5C_{0,k}^3}.
\end{equation}
We use Eq.~(\ref{orth}) to find
\begin{equation}
0=\frac{d}{d t}\int_{-\infty}^\infty(U_2-W_2)U_{0y}dy=\int_{-\infty}^\infty\frac
{d(U_2-W_2)}{d t}U_{0y}-\frac{\partial p_{k}}{\partial
t}(U_2-W_2)U_{0yy}dy.
\end{equation}
Thus,
\begin{equation}\label{eff0}
    \int_{-\infty}^\infty\frac
{d(U_2-W_2)}{d t}U_{0y}dy= \int_{-\infty}^\infty \frac{\partial p_{k}}{\partial
t}(U_2-W_2)U_{0yy}dy.
\end{equation}
Finally, we utilize Eqs.~(\ref{eff0}), (\ref{order3-3}), and (\ref{T11}) to obtain
\begin{equation}\label{eff}%
\begin{split}
&\int_{-\infty}^{\infty}\left[\frac
{\partial U_2}{\partial t}-\frac{\partial W_2}{\partial t}-\frac{\partial p_k}{\partial t}U_{2y}+\frac{\partial p_k}{\partial t}W_{2y}\right]U_{0y} dy\\
&= \int_{-\infty}^\infty \frac{d(U_2-W_2)}{d t}U_{0y}dy\\
&= \int_{-\infty}^\infty \frac{\partial p_{k}}{\partial
t}(U_2-W_2)U_{0yy}dy\\
&=\frac{1}{\kappa}\int_{-\infty}^\infty \frac{\partial p_{k}}{\partial
t}\left( \left[-\frac{\partial p_{k}}{\partial T_{1}}+\frac{\partial\alpha_{k}%
}{\partial t}\right]  U_{0y}-\alpha_{k}\frac{\partial p_{k}}{\partial t}
U_{0yy}-\frac{\partial p_{k}}{\partial t}W_{1y}+\frac{\partial W_1}{\partial t} \right)U_{0yy}  \\
&=-{\kappa}\alpha_{k}^{3}\int_{-\infty}^\infty U_{0yy}^{2}dy.
\end{split}
\end{equation}
Employing Eq.~(\ref{eff}) to eliminate $U_2$ and $W_2$ in Eq.~(\ref{pj3}) gives
\begin{equation}
\frac{\partial\alpha_{k}}{\partial T_{1}}=\hat{\tau}{\kappa^2
}\alpha_{k}-\frac{\left(I_1+I_2+I_3\right)}{\int_{-\infty}^\infty U_{0y}^{2}dy}-{\kappa}%
\alpha_{k}^{3}\frac{\int_{-\infty}^\infty U_{0yy}^{2}dy}{\int_{-\infty}^\infty U_{0y}^{2}dy}. \label{alphak}%
\end{equation}%

Projecting Eqs.~(\ref{tc3}) onto
$P^{\dagger}=(\frac{1}{{\kappa}}U_{0y},0)$ yields
\begin{equation}
-\frac{\partial p_{k}}{\partial T_{2}}\int_{-\infty}^\infty U_{0y}^{2}dy-\int_{-\infty}^\infty \frac{\partial p_{k}}{\partial T_{1}}U_{1y}U_{0y}dy+\int\frac{d
U_2}{d t}U_{0y}dy=I_1+I_2+I_3, \label{pT3}%
\end{equation}
Similarly, using orthogonal condition (\ref{orth}), it can be concluded that
\begin{equation}
\int_{-\infty}^\infty \frac{d U_2}{d t}U_{0y}dy=\int_{-\infty}^\infty U_2\frac{\partial p_{k}%
}{\partial t}U_{0yy}dy.
\end{equation}
Thus,
\begin{equation}
\frac{\partial p_{k}}{\partial T_{2}}=-\frac{1}{\int_{-\infty}^\infty U_{0y}^{2}dy}\left( I_1+I_2+I_3 +\frac{\partial p_{k}}{\partial T_{1}}\int_{-\infty}^\infty U_{1y}U_{0y}dy-\frac{\partial p_{k}}{\partial t}\int_{-\infty}^\infty U_2U_{0yy}dy\right).
\end{equation}

Hence, from Eqs.~(\ref{T11}), (\ref{aT1}), (\ref{pT1}) and (\ref{alphak}), we obtain our main result for the first two leading-order dynamics of the $k$-th spike.
\begin{equation}
\label{innerdynamic}\left\{
\begin{array}
[c]{l}%
\frac{\partial p_{k}}{\partial t}={\kappa}\alpha_{k},\\
\frac{\partial\alpha_{k}} {\partial t} =\frac{B_{2,k}}{3}\frac{\int_{-\infty}^{\infty}
U^{3}_{0}(y) dy}{\int_{-\infty}^{\infty} U_{0y}^2 dy};
\end{array}
\right.
\end{equation}

\begin{equation}
\label{innerdynamic2}\left\{
\begin{array}
[c]{l}%
\frac{\partial p_{k}}{\partial T_{1}}=\frac{B_{2,k}}{3} \frac{\int_{-\infty}^{\infty}
U^{3}_{0}(y) dy}{\int_{-\infty}^{\infty} U_{0y}^2 dy}-\kappa \alpha_k \frac{\int_{-\infty}^{\infty} U_{1y}U_{0y}dy}{\int_{-\infty}^{\infty} U^{2}_{0y} dy},\\
\frac{\partial\alpha_{k}} {\partial T_{1}} =\hat{\tau}
{\kappa}^2 \alpha_{k}-\frac{\left(I_1+I_2+I_3\right)}{\int_{-\infty}^\infty U_{0y}^{2}dy}-  {\kappa
}\alpha_{k}^{3} \frac{\int_{-\infty}^\infty U^{2}_{0yy}dy }{\int_{-\infty}^\infty U^{2}_{0y}dy }.%
\end{array}
\right.
\end{equation}
Summing the different orders and evaluating all the integrals yields
\begin{subequations}
\begin{equation}
        \frac{d p_k}{dt}=\frac{\partial p_{k}}{\partial t}+\varepsilon \frac{\partial p_{k}}{\partial T_{1}}+\cdots =\kappa{\alpha_k}+ \varepsilon \left[ \frac{2B_{2,k}}{C_{0,k}} +\kappa\alpha_k\left(\frac{C_{1,k}}{C_{0,k}}+(\frac{873}{140}-6\ln{2} )\frac{1}{DC_{0,k}^2}\right) \right]+\mathcal{O}(\varepsilon^2),
\end{equation}
\begin{equation}
        \frac{d \alpha_k}{dt}=\frac{\partial \alpha_{k}}{\partial t}+\varepsilon \frac{\partial \alpha_{k}}{\partial T_{1}} +\cdots
        =\frac{2B_{2,k}}{C_{0,k}} +\varepsilon \left[ \hat{\tau}
{\kappa}^2 \alpha_{k}-\left((-24\ln{2}+\frac{1453}{70})\frac{B_{2,k}}{DC_{0,k}^3}+\frac{4B_{2,k}C_{1k}}{C_{0,k}^2}\right)-\frac{5}{7} \kappa
\alpha_{k}^{3}  \right]+\mathcal{O}(\varepsilon^2). 
\end{equation}
\end{subequations}

\textbf{Outer region}:  Away from the
spike centers, $u(x)$ is assumed to be exponentially small so that
$Dv_{xx}+\frac{1}{2}=0$ for $x\neq x_{k}$. Near $x_{k}+\varepsilon%
p_{k}$, the term $\frac{u^{2}v}{\varepsilon}$ in (\ref{eq1.0}) acts like a Dirac delta function, producing
\begin{equation}
Dv_{xx}+\frac{1}{2}=\sum_{j=1}^{N}s_{j}\delta(x-x_{j}-\varepsilon p_{j}).
\label{outeq}%
\end{equation}
Here, the weights $s_{j}$ are defined as
\begin{align}
s_{j}  &  =\int_{-\infty}^{\infty}U^{2}Vdy\nonumber\label{s_j}\\
&  =\int_{-\infty}^\infty U_{0}^{2}V_{0}dy+\varepsilon\int_{-\infty}^\infty(U_{0}^{2}V_{1}+2U_{0}V_{0}%
U_{1})dy+\cdots\nonumber\\
&  =\frac{6}{C_{0,j}}-\varepsilon\left(  \frac{6C_{1,j}}%
{C_{0,j}^{2}}+\frac{E}{DC_{0,j}^3}\right)  +\cdots,
\end{align}
where $U$ and $V$ are the inner solution near the $k$-th spike, and $E$ is a constant defined by
\begin{equation}
E:=\int_{-\infty}^{\infty} \rho \left(\rho \int_0^y \int_0^z  \rho^2 d\hat{y} dz-2f\right)  dy=24-36\ln{2}.
\end{equation}
Integrating Eq.~(\ref{outeq}) produces
\begin{equation}
\int_{-1}^{1}\frac{1}{2}dx=\sum_{j=1}^{N}s_{j}. \label{eqc}%
\end{equation}
The solution of Eq.~(\ref{outeq}) is then given by%
\begin{equation}
v(x)=\sum_{j=1}^{N}s_{j}G(x,x_{j}+\varepsilon p_{j})+\bar{v} \label{veq},%
\end{equation}
where $\bar{v}$ is a constant to be determined by Eq.~(\ref{eqc}) and $G$ is the Green's function satisfying
\begin{align}
&  DG_{xx}+\frac{1}{2}=\delta(x-z)\label{Greenf},\\
&  G_{x}(-1)=G_{x}(1)=0,~~~~\int_{-1}^{1}Gdx=0.
\end{align}
We can decompose $G(x,z)$ as follows
\begin{equation}
G(x,z)=\frac{|x-z|}{2D}+H(x,z),
\end{equation}
where \begin{equation}
    H=\frac{1}{2D}\left[-\frac{1}{3}-\frac{x^2}{2}-\frac{z^2}{2} \right],
\end{equation}
is the regular part of $G$. 

Following the notations in \cite{iron2004stability}, we define matrix $\mathcal{G}$ as
\begin{equation}
\mathcal{G}=(G(x_{k},x_{j})).
\end{equation}
Let us denote $\frac{\partial}{\partial x_{j}}$as $\nabla_{x_{k}}$. When
$k\neq j$, we can define $\nabla_{x_{k}}G(x_{k},x_{j})$ and $\nabla_{x_{j}}G(x_{k},x_{j})$ in the classical way. When $k=j$, we define
\begin{equation}
\nabla_{x_{k}}G(x_{k},x_{k}):=\frac{\partial}{\partial x}\big\vert_{x=x_{k}%
}H(x,x_{k}).
\end{equation}
We also define the derivative of matrix $\mathcal{G}$ as follows,
\begin{equation}
\nabla\mathcal{G}:=(\nabla_{x_{k}}G(x_{k},x_{j})).
\end{equation}
From \cite{iron2004stability}, we have the following identities related to $G$
\begin{equation}
    \sum_{j=1}^{N} G(x_k,x_j) =-\frac{1}{6DN},
\end{equation}

\begin{equation}
\label{Gkj}\sum_{j=1}^{N} \nabla_{x_{k}} G(x_{k},x_{j})=0,~~~~~\sum_{k=1}^{N}
\nabla_{x_{k}} G(x_{k},x_{j})=0,~~~~\nabla_{x_{k}} G(x_{k},x_{j}%
)=\nabla_{x_{k}} G(x_{j},x_{k}).
\end{equation}
Then, near the $k$-{th} spike $x=x_{k}+\varepsilon p_{k}+\varepsilon y$, we have
\begin{equation}\label{vG}
\begin{split}
v(x) &= \sum_{j=1}^{N} s_{j} G(x_{k}+\varepsilon p_{k}+\varepsilon y,
x_{j}+\varepsilon p_{j})+\bar{v}\\
    &=v_{k,0}+\varepsilon v_{k,1}(y)+\varepsilon^2 v_{k,2}(y)+\varepsilon^3 v_{k,3}(y) +\cdots,
\end{split}
\end{equation}
where 
\begin{subequations}
\begin{equation}
    v_{k,0}=\sum_{j=1}^{N} \frac{6}{C_{0,j}} G(x_{k},x_{j})+\bar{v}_0,
\end{equation}
\begin{equation}
\begin{split}
    v_{k,1}(y)=&\left[ \frac{3}{ C_{0,k}D}\text{sign}(y)+\sum_{j=1}^{k}%
\frac{6}{C_{0,j}} \nabla_{x_{k}}G(x_{k},x_{j})\right]y-\sum_{j=1}^{N}  \left[  \frac{6C_{1,j}}{C_{0,j}^{2}} +\frac{E}{DC_{0,j}^3} \right]
G(x_{k},x_{j}) \\
&+\sum_{j=1}^{N}\frac{6}{C_{0,j}} \left[ \nabla_{x_{k}}
G(x_{k},x_{j}) p_{k}+\nabla_{x_{j}} G(x_{k},x_{j}) p_{j} \right]  +\bar{v}_{1}.
\end{split}
\end{equation}
Since only the derivatives of $v_{k,2}$ and $v_{k,3}$ are needed in the later matching procedure, we compute only $\frac{\partial v_{k,2}}{\partial y}$ and $\frac{\partial v_{k,3}}{\partial y}$ as follows,
\begin{equation}
    \frac{\partial v_{k,2}}{\partial y} =\sum_{j=1}^{N}\left[  \frac{6}{C_{0,j}}\left(  \nabla_{x_{k}}\nabla
_{x_{k}}G(x_{k},x_{j})p_{k}+\nabla_{x_{j}}\nabla_{x_{k}}G(x_{k},x_{j}%
)p_{j}\right)  \right]  -\left[ \frac{6C_{1,j}}{C_{0,j}^{2}} +\frac{E}{DC_{0,j}^3} \right]\nabla_{x_{k}}G(x_{k}%
,x_{j}),
\end{equation}
\begin{equation}
    \frac{\partial v_{k,3}}{\partial y}=0.
\end{equation}
\end{subequations}

\textbf{Matching:} We match Eq.~(\ref{vG}) with the far field behavior of the inner solution  and determine the constants $B_{j,k}$ and $C_{j,k}$.

$\bullet$ To leading order, we obtain%
\begin{equation}
\label{C0k1}\sum_{j=1}^{N} \frac{6}{C_{0,j}} G(x_{k},x_{j})+\bar{v}_{0}
=C_{0,k},\quad k=1,\ldots,N.%
\end{equation}
On the other hand, the leading order of Eq.~(\ref{eqc}) implies
\begin{equation}
\label{C0k2}\sum_{j=1}^{N} \frac{6}{C_{0,j}}=1.
\end{equation}
Eqs.~(\ref{C0k1}) and Eq.~(\ref{C0k2}) can be combined to obtain
\begin{equation}\label{C0k}
C_{0,j}=6N,\quad j=1,\ldots, N.
\end{equation}

$\bullet$ Matching the constant terms in the order $\varepsilon$ yields
\begin{equation}
\label{C1k1}
\begin{split}
\sum_{j=1}^{N} -\left(  \frac{6C_{1,j}}{C_{0,j}^{2}} + \frac{E}{DC_{0,j}^3} \right)
G(x_{k},x_{j})+\sum_{j=1}^{N} \frac{6}{C_{0,j}} \left(  \nabla_{x_{k}}
G(x_{k},x_{j}) p_{k}+\nabla_{x_{j}} G(x_{k},x_{j}) p_{j} \right)  +\bar{v}_{1}\\
=C_{1,k}+\frac{1}{DC_{0,k}}  \int_0^{+\infty}\left( \int_{0}^y \rho^2 dz-\int_{0}^\infty \rho^2 dz \right) dy,\quad k=1,\ldots, N.
\end{split}
\end{equation}
The order $\varepsilon$ of Eq.~(\ref{eqc}) reads,
\begin{equation}
\label{C1k2}\sum_{j=1}^{N} -\left(  \frac{6C_{1,j}}{C_{0,j}^{2}} +\frac{E}{DC_{0,j}^3}  \right)
=0.
\end{equation}
From Eqs.~(\ref{C1k2}) and (\ref{C0k}), one can obtain
\begin{equation}\label{EE}
    E=-36D \sum_{j=1}^N C_{1,j}.
\end{equation}
Summing Eq.~(\ref{C1k1}) with respect to $k$ and using Eqs.~(\ref{C0k}),(\ref{C1k2}) and (\ref{Gkj}) produces
\begin{equation}\label{v1bar}
\bar{v}_{1}=\frac{1}{6DN}\int_0^{+\infty}\int_{+\infty}^y \rho^2 dz dy+ \frac{1}{N} \sum_{k=1}^{N} C_{1,k} =\frac{1}{6DN}\int_0^{+\infty}\int_{+\infty}^y \rho^2 dz dy-\frac{E}{36DN}.
\end{equation}
Substituting Eqs.~(\ref{v1bar}) and (\ref{EE}) back into Eq.~(\ref{C1k1}) and solving for $C_{1,k}$, we obtain
\begin{equation}\label{C1}
C_1=%
\begin{pmatrix}
C_{1,1}\\
\cdots\\
C_{1,N}%
\end{pmatrix}
= \mathcal{M}_1
\begin{pmatrix}
p_{1}\\
\cdots\\
p_{N}%
\end{pmatrix}-\frac{E}{36DN},
\end{equation}
where $\mathcal{M}_1$ is defined as
\begin{equation}\label{definitionM1}
    \mathcal{M}_1=\frac{1}{N}(I+\frac{1}{6N^{2}}\mathcal{G})^{-1} ( \nabla\mathcal{G})^{T}.
\end{equation}

Other important constants $B_{1,k},B_{2,k},$ and $B_{3,k}$ depend on the derivative of $\left\{ v_{k,j}(x),~j=1,\ldots,3 \right\}$ by matching as follows
\begin{align}
\begin{split}\label{B1k}
B_{1,k}&=\frac{1}{2}\left(  \frac{\partial v_{k,1}(0^{+})}{\partial y}%
+\frac{\partial v_{k,1}(0^{-})}{\partial y}\right)  =\sum_{j=1}^{k}%
C_{0,j}\nabla_{x_{k}}G(x_{k},x_{j})=0 ,
\end{split}\\
\begin{split}\label{B2k}
 B_{2,k} &  =\frac{1}{2}\left(  \frac{\partial v_{k,2}(0^{+})}{\partial
y}+\frac{\partial v_{k,2}(0^{-})}{\partial y}\right) \\
&  =\sum_{j=1}^{N}\left[  \frac{6}{C_{0,k}}\left(  \nabla_{x_{k}}\nabla
_{x_{k}}G(x_{k},x_{j})p_{k}+\nabla_{x_{j}}\nabla_{x_{k}}G(x_{k},x_{j}%
)p_{j}\right)  \right]  -\left(  \frac{6C_{1,j}}{C_{0,j}^{2}} + \frac{E}{DC_{0,j}^3} \right)\nabla_{x_{k}}G(x_{k}%
,x_{j})\\
&  =-\frac{1}{2D}p_{k}-\sum_{j=1}^{N}\frac{1}{6N^{2}}\left(C_{1,j}+\frac{E}{36DN}\right)\nabla_{x_{k}%
}G(x_{k},x_{j}),
\end{split}\\
\begin{split}\label{B3k}%
B_{3,k}&=\frac{1}{2} \left( \frac{\partial v_{k,3}(0^{+})}{\partial y}+\frac{\partial v_{k,3}(0^{-})}{\partial y}\right)-\frac{1}{2D}\left( \int_{0}^\infty+\int_{0}^{-\infty}\right) \left( U_0^2V_2+2U_0V_0U_2\right)dy =(6\ln{2}-\frac{57}{8})\frac{B_{2,k}}{D C_{0,k}^2 }.
\end{split}
\end{align}
Substituting Eq.~(\ref{C1}) into Eq.~(\ref{B2k}), we obtain
\begin{equation} \label{C2def}
B_{2}=%
\begin{pmatrix}
B_{2,1}\\
\cdots\\
B_{2,N}%
\end{pmatrix}
=\mathcal{M}_2
\begin{pmatrix}
p_{1}\\
\cdots\\
p_{N}%
\end{pmatrix},
\end{equation}
where $\mathcal{M}_2$ is defined as
\begin{equation}
\mathcal{M}_2=-\frac{1}{2D}I-\frac{1}{6N^{3}}\nabla\mathcal{G}(I+\frac{1}%
{6N^{2}}\mathcal{G})^{-1}(\nabla\mathcal{G})^{T} \label{definitionM2}.%
\end{equation}
Substituting all the constants back into Eqs.~(\ref{innerdynamic}) and (\ref{innerdynamic2}),
one can obtain the reduced dynamic system for the spike locations, as formally stated in the following proposition.
\begin{proposition}
 Assume that $\varepsilon \ll 1$ and $\tau=\tau_c+\varepsilon^2 \hat{\tau}$. Then, the equations for $p_k$ and $\alpha_k$  are approximately governed by
 \begin{subequations}\label{2NODE}
\label{dynamicnn}%
\begin{align}
&  \frac{d p_{k}}{d t}=\frac{\partial p_{k}}{\partial t}+\varepsilon
\frac{\partial p_{k}}{\partial T_{1}}+\cdots={\kappa} \alpha_{k}
+\varepsilon\left( \frac{B_{2,k} }{3N }+\kappa\alpha_k\left[\frac{C_{1,k}}{6N}+(\frac{873}{140}-6\ln{2})\frac{1}{6^2DN^2}\right] \right)+\cdots,\\
&  \frac{d \alpha_{k}}{d t}=\frac{\partial\alpha_{k}} {\partial t}%
+\varepsilon\frac{\partial\alpha_{k}}{\partial T_{1}}+\cdots=\frac{B_{2,k}
}{3N } + \varepsilon\left(  \hat{\tau} {\kappa}^{2}\alpha_{k}- \frac{5 }{
7}{\kappa}\alpha_{k}^{3}-\left[(\frac{1453}{70}-24\ln{2})\frac{B_{2,k}}{6^3DN^3}+\frac{4B_{2,k}C_{1,k}}{6^2N^2}\right] \right)+\cdots,
\end{align}
\end{subequations}
where $B_{2,k}$ and $C_{1,k}$ are the k-${th}$ element of $B_2$ defined by Eq~(\ref{C2def}) and $C_1$ defined by Eq~(\ref{C1}), respectively.
\end{proposition}
The 2N-dimensional system of ODEs (\ref{2NODE}) describes the motion of the N-spike solution observed in the PDEs (\ref{eq1.0}) when the
spikes are sufficiently close to the equilibrium and move slowly, with $x_k+\varepsilon p_k$ being the location of the $k$-{th} spike.

\begin{remark}
The terms $B_{2,k}$ (defined in Eq.~(\ref{C2def})), which are related to the Green's function, serve as the weak interactions between the spikes to leading order even though they are far away from each other. 
\end{remark}

\section{Analysis of the ODE system} \label{sec4}
The reduced ODE system (\ref{2NODE}) is a linear system to leading order with a weakly nonlinear term. The existence of small order nonlinear terms makes a further approximation possible.  In this section,  we apply the method of multiple-time-scale analysis to obtain a leading order approximation of the reduced system (\ref{2NODE}). We remark that other perturbation methods, such as the averaging method, normal forms theory, renormalization group method also produce the same results though procedures are different, see \cite{chiba2009extension}. After the approximate solution is obtained, the numerical comparisons between the simulation of ODEs and PDEs are provided to validate our results for the case of $N=1,2,3$. Within this section, the PDE simulations are conducted by using Flexpde $7$ \cite{flexpde} with an accuracy control setting of $10^{-5}$ .  The ODE simulation is conducted by using the MATLAB \cite{matlab} function ode45 , with the default settings that have a relative error tolerance of $10^{-5}$.  In addition, the codes for the PDE and ODE simulations are provided on GitHub at \url{https://github.com/KaleonXie/Complex-motion-of-spikes}.

\subsection{One-spike dynamics}
In the case of one spike, after discarding the high order term in the system (\ref{2NODE}), one can obtain 
\begin{equation} 
\left\{
\begin{array}
[c]{l}%
\frac{dp}{dt}={\kappa}\alpha+\varepsilon\left(-\frac{p}{6D}+\frac{313}{140}\frac{\kappa\alpha}{36D}\right),\\
\frac{d\alpha}{dt}=-\frac{p}{6D}+\varepsilon\left(  \hat{\tau}{\kappa}%
^{2}\alpha-\frac{5}{7}{\kappa}\alpha^{3}+\frac{333}{140}\frac{p}{6^3D^2} \right).
\end{array}
\right. \label{dynamic1}%
\end{equation}

\begin{figure}[th]
\centering
\includegraphics[width=0.8\textwidth]{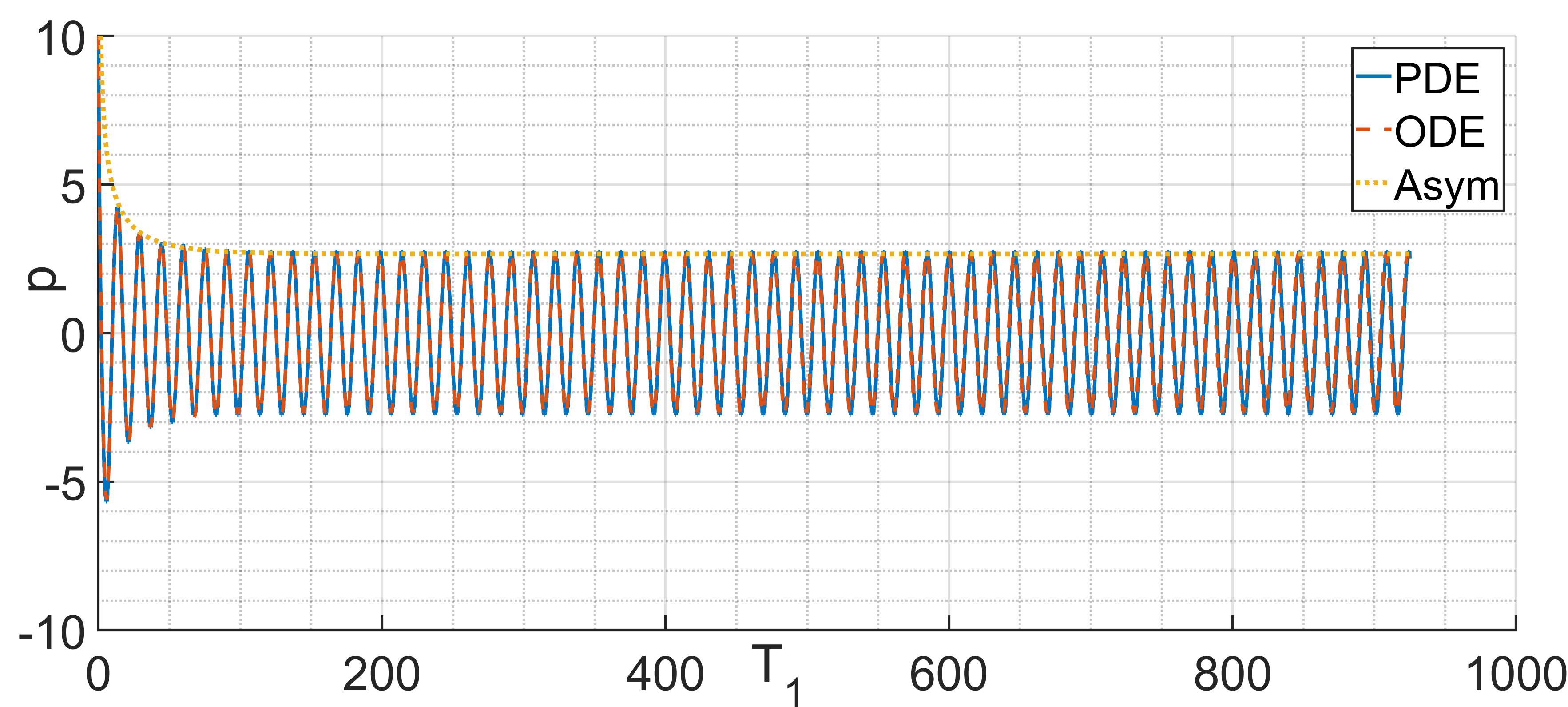}\caption{Direct
comparison of the spike locations between the simulations of the PDE system (\ref{eq1.0}), ODE system (\ref{dynamic1}) and the amplitude modulation equation~(\ref{mts1}). The parameters are $\hat{\tau}=100,~\varepsilon=0.01,~D=0.2,~{\kappa}=0.2$. The center of the spike oscillates as time progresses. }\label{onespike}
\end{figure}

The ODE~(\ref{dynamic1}) and original PDE simulation results agrees well as shown in Fig.~\ref{onespike}. For convenience, we further approximate the ODEs~(\ref{dynamic1}) using the following single equation of $p$ up to $\mathcal{O}(\varepsilon)$,
\begin{equation}
\ddot{p}+\frac{{\kappa}p}{6D}-\varepsilon\left[  (\hat{\tau}{\kappa}^{2}%
-\frac{1}{6D})\dot{p}-\frac{5}{7{\kappa}}\dot{p}^{3}+\nu p \right]  =0.
\label{secondorder}%
\end{equation}
where $\nu$ is a constant defined by 
\begin{equation}
    \nu=\frac{333-313\kappa}{140} \frac{1}{6^3D^2}.
\end{equation}

Eq.~(\ref{secondorder}) resembles a linear oscillator with weakly nonlinear damping. We proceed to use multiple-time-scale analysis to construct uniformly valid approximations to the solution of Eq.~(\ref{secondorder}) when $t< \frac{C}{\varepsilon}$ for some constant $C$. We define slow time scale as
\begin{equation}
T_{1}=\varepsilon t,
\end{equation}
and seek a solution of the form:
\begin{equation}
p=q_{0}(t,T_{1})+\varepsilon q_{1}(t,T_{1})+\cdots .\label{qexp}%
\end{equation}
Substituting expansion (\ref{qexp}) into Eq.~(\ref{secondorder}) and separating at
each order in $\varepsilon$ yields the problems of different orders:
\begin{align}
\mathcal{O}(1)  &  ~~~~~~~\frac{\partial^{2}q_{0}}{\partial t^{2}}+\frac{{\kappa}%
q_{0}}{6D}=0,\\
\mathcal{O}(\varepsilon)  &  ~~~~~~~\frac{\partial^{2}q_{1}}{\partial t^{2}}+\frac{{\kappa
}q_{1}}{6D}=-2\frac{\partial^{2}}{\partial t\partial T_{1}}q_{0}+\left[
(\hat{\tau}{\kappa}^{2}-\frac{1}{6D})\dot{q}_{0}-\frac{5}{7{\kappa}}\dot
{q}_{0}^{3}+\nu q_0\right]. \label{q1eq}%
\end{align}
The $\mathcal{O}(1)$ solution is
\begin{equation}
q_{0}=\mathcal{A}(T_{1})e^{i\omega t}+\mathcal{A}^{\ast}(T_{1})e^{-i\omega t}
\label{q0}%
\end{equation}
where $\omega^{2}=\frac{{\kappa}}{6D}$. Substituting Eq.~(\ref{q0}) into
Eq.~(\ref{q1eq}) produces
\begin{equation}
\frac{\partial^{2}q_{1}}{\partial t^{2}}+\frac{{\kappa}q_{1}}{6D}=\left[
-2i\omega\frac{\partial\mathcal{A}}{\partial T_{1}}e^{i\omega t}+i\tilde{\tau
}{\kappa}^{2}\omega\mathcal{A}e^{i\omega t}-\frac{5}{7{\kappa}}\left(
-i\omega^{3}\mathcal{A}^{3}e^{3i\omega t}+3\mathcal{A}^{2}A^{\ast}i\omega
^{3}e^{i\omega t}\right)+\nu\mathcal{A}e^{i\omega t}  \right]  +c.c
\end{equation}
where $c.c$ means the complex conjugate of the term inside the square brackets.
To remove the secular terms at $\mathcal{O}(\varepsilon)$, the condition%

\begin{equation}
\frac{\partial\mathcal{A}}{\partial T_{1}}=\frac{1}{2}(\hat{\tau}{\kappa}%
^{2}-\frac{1}{6D})\mathcal{A}-\frac{5}{28D}\mathcal{A}^{\ast}\mathcal{A}^{2}+\frac{i\nu}{2\omega} \mathcal{A},
\label{mts}%
\end{equation}
must be satisfied. Solving explicitly for $\mathcal{A}$ by setting
$\mathcal{A}=\frac{\mathcal{B}e^{i\theta}}{2}$ yields the amplitude and phase modulation equations on the $T_1$ scale,
\begin{subequations} 
\begin{align}
&  \frac{\partial\mathcal{B}}{\partial T_{1}}=\frac{1}{2}(\hat{\tau}{\kappa
}^{2}-\frac{1}{6D})\mathcal{B}-\frac{5}{112D}\mathcal{B}^{3},\label{mts1}\\
&  \frac{\partial\theta}{\partial T_{1}}=\frac{\nu}{2\omega}.
\end{align}
\end{subequations}

\begin{figure}[th]
\includegraphics[width=0.45\textwidth]{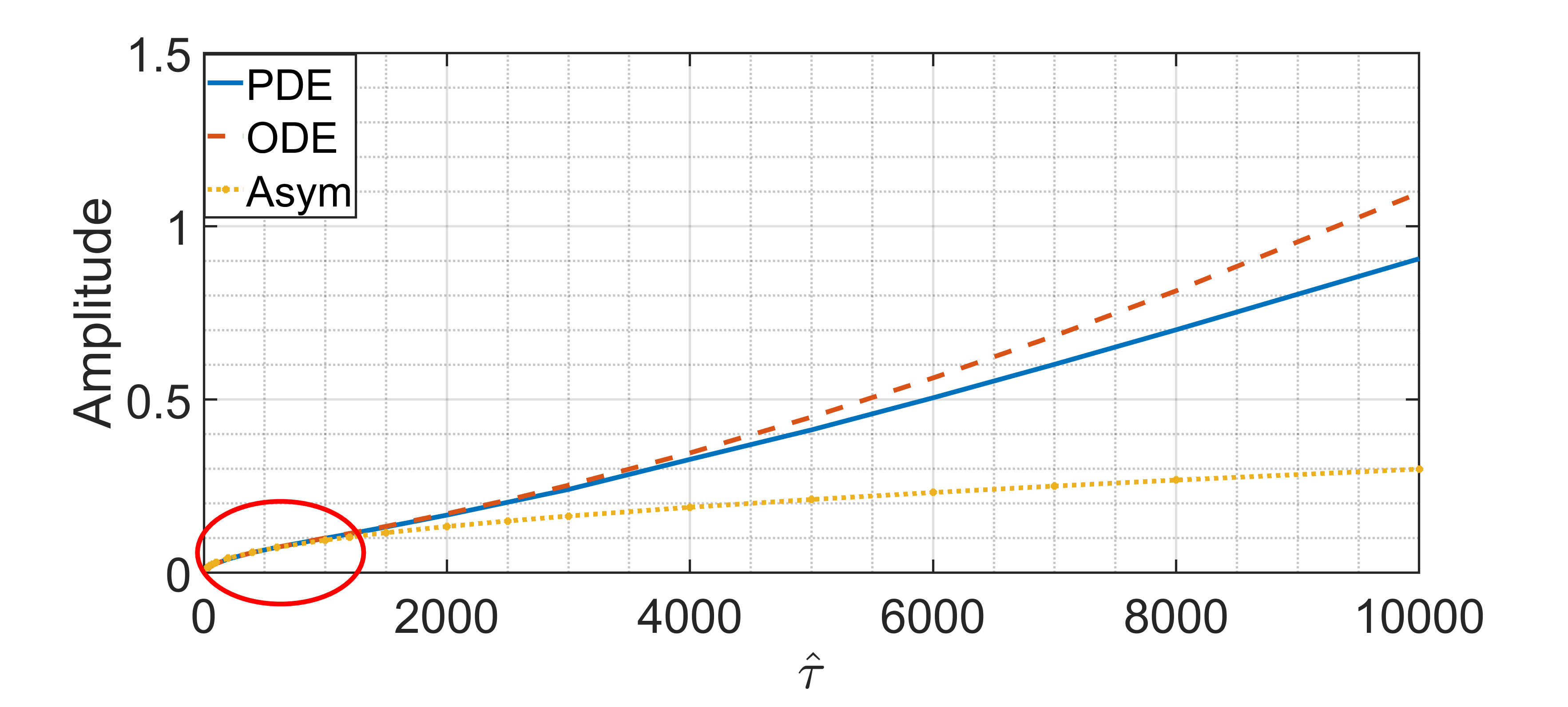}
\includegraphics[width=0.45\textwidth]{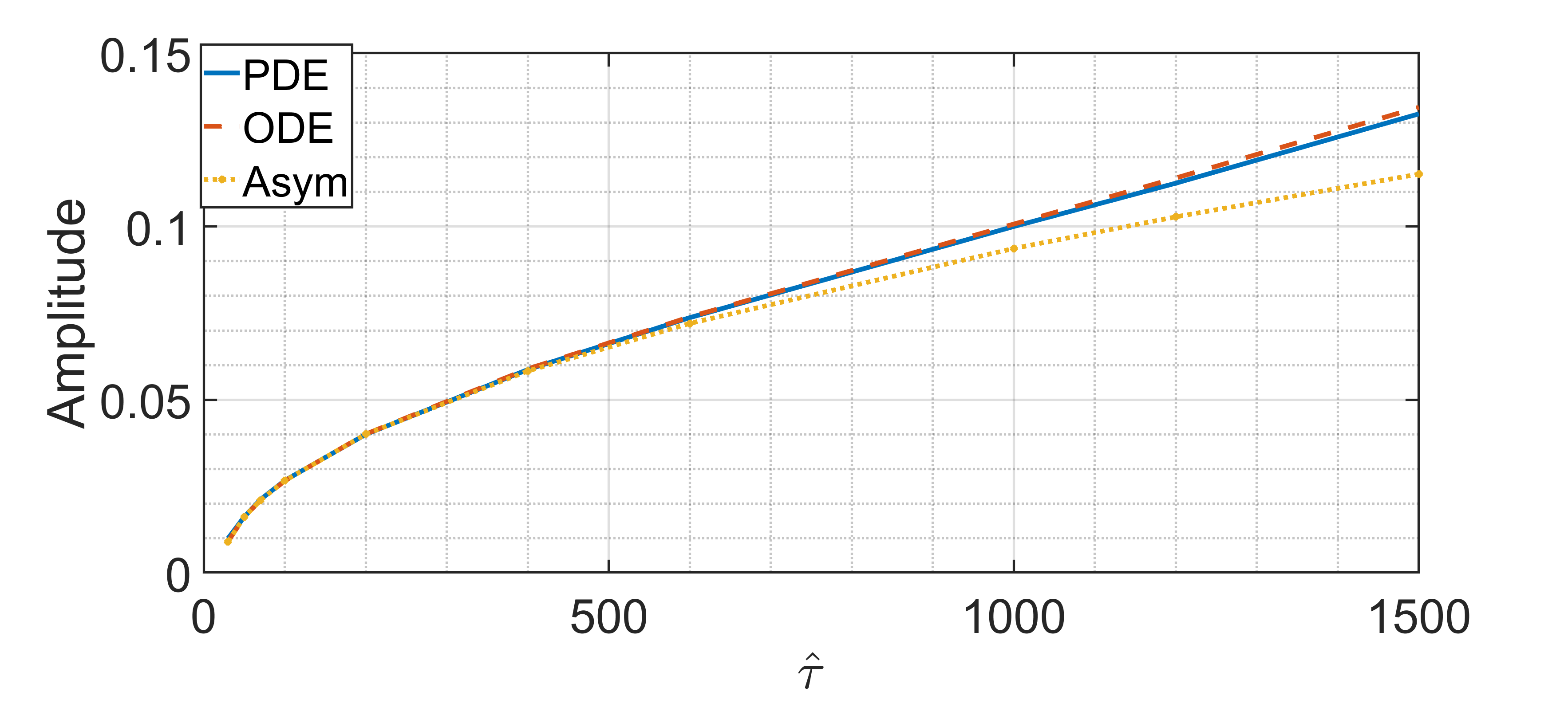}\caption{(Color online) The limit of the amplitude modulation vs. $\hat{\tau}
$ between the simulation of the PDE~(\ref{eq1.0}), ODE~(\ref{dynamic1}) and prediction ($\varepsilon\lim_{T_1\rightarrow\infty} \mathcal{B}$) from Eq.~(\ref{asym}) for the one-spike case in the original coordinate.  Parameters are $\varepsilon=0.01,~D=0.2,~{\kappa}=0.2$.
The left figure shows the correspondence  between the limit of the amplitude modulation and parameter $\hat{\tau}$, and the part inside the red circle is zoomed in in the right figure. The agreement is reasonably good even when $\hat{\tau}$ is big.} \label{com1}%
\end{figure}

It is easy to see that the phase modulation $\theta=\theta_0+\frac{\nu}{2\omega} T_1$  and the amplitude modulation will converge to $0$ or $\sqrt{\frac{56D}{5}\left( \hat{\tau}{\kappa}^{2}-\frac{1}{6D}\right)}$ as time approaches infinity, that is, 
\begin{equation}\label{asym}
\lim_{T_1 \rightarrow \infty} \mathcal{B}=\sqrt{\max \left( \frac{56D}{5} \left( \hat{\tau}{\kappa}^{2}-\frac{1}{6D}\right), 0\right)}.
\end{equation}
Fig.~\ref{com1} and Fig.~\ref{onespike} present direct comparisons between the PDE simulation, ODE simulation and asymptotic prediction (\ref{asym}) from the amplitude modulation equation  to validate our results.

\subsection{ N-spike dynamics}
In this subsection, we use the same method as what we have applied to the one-spike dynamics to construct an approximate solution to the N-spike dynamics. In general, the dynamic system of N-spike motion up to $\mathcal{O}(\varepsilon)$ can be written as
\begin{equation} \label{rawN}%
\begin{pmatrix}
\frac{\partial ^{2} p_{1} }{\partial t^{2}}\\
\vdots\\
\frac{\partial^{2} p_{N}}{\partial t^{2}}%
\end{pmatrix}
= \frac{{\kappa}\mathcal{M}_2}{3N}%
\begin{pmatrix}
p_{1}\\
\vdots\\
p_{N}%
\end{pmatrix}
+\varepsilon\left[  (\hat{\tau}{\kappa}^{2}+\frac{\mathcal{M}_2}{3N})
\begin{pmatrix}
\frac{\partial p_{1} }{\partial t}\\
\vdots\\
\frac{\partial p_{N} }{\partial t}%
\end{pmatrix}
- \frac{5}{7{\kappa}}
\begin{pmatrix}
\frac{\partial p_{1} }{\partial t}\\
\vdots\\
\frac{\partial p_{N} }{\partial t}%
\end{pmatrix}
^{\circ3} - \nu_1 \mathcal{M}_2 \begin{pmatrix}
p_{1}\\
\vdots\\
p_{N}%
\end{pmatrix}-F(\mathbf{p},\frac{\partial \mathbf{p}}{\partial t})  \right]  =0,
\end{equation}
where 
\begin{equation}
   F(\mathbf{p},\frac{\partial \mathbf{p}}{\partial t}) =-\frac{1}{6N} \mathcal{M}_1
\begin{pmatrix}
\frac{\partial p_{1} }{\partial t}\\
\vdots\\
\frac{\partial p_{N} }{\partial t}%
\end{pmatrix} \odot 
\begin{pmatrix}
\frac{\partial p_{1} }{\partial t}\\
\vdots\\
\frac{\partial p_{N} }{\partial t}%
\end{pmatrix}+\left(\frac{4}{6^2N^2}-\frac{\kappa}{18N^2}\right) \mathcal{M}_1
\begin{pmatrix}
p_{1}\\
\vdots\\
p_{N}%
\end{pmatrix}
 \odot \mathcal{M}_2
\begin{pmatrix}
p_{1}\\
\vdots\\
p_{N}%
\end{pmatrix},
\end{equation}
$\sim^{\circ}$ and $\odot$ are the Hadamard power and product symbol respectively; $\mathcal{M}_1$ is
defined by Eq.~(\ref{definitionM1}); $\mathcal{M}_2$ is
defined by Eq.~(\ref{definitionM2}); and $\nu_1$ is a constant defined as follows:
\begin{equation}
\nu_1=\left( \frac{333}{70} - \frac{313\kappa}{70} \right) \frac{1}{6^3DN^3}. 
\end{equation}

The properties of  $\mathcal{M}_2$ are well studied in Appendix C of \cite{iron2004stability}. We summarize them in the following lemma,
\begin{lemma}
Define
\begin{equation}
\label{defQ}Q=(\mathbf{q}_{1},\cdots,\mathbf{q}_{N}),
\end{equation}
where
\begin{align}
\mathbf{q}_{1}=\sqrt{\frac{1}{N}}(1,-1,1,\cdots,(-1)^{N+1})^{\prime};\\
\mathbf{q}_{k}=(q_{k,1},\cdots,q_{k,N})^{\prime},~~~k=2,\cdots,N;\\
q_{k,j}=\sqrt{\frac{2}{N}}\sin\left(  \frac{\pi(j-1)}{N}(k-\frac{1}{2})
\right).
\end{align}
Then 
\begin{equation}
Q^{T}Q=I,~~~~~~~Q^{T} \mathcal{M}_2Q=\Lambda:=%
\begin{pmatrix}
\lambda_{1,0} & ~~ & ~~\\
~~ & \ddots & ~~\\
~~ & ~~ & \lambda_{N,0}%
\end{pmatrix},
\end{equation}
where $\lambda_{j,0}$ is defined by Eq. (\ref{lambda}).
\end{lemma}

Define $\bm{\xi}=Q^{T}\mathbf{p}$. Then, $Q^{T}\mathcal{M}_2 \mathbf{p}=\Lambda Q^{T} \mathbf{p}=\Lambda \bm{\xi}$.
Replacing $\mathbf{p}$ in Eq.~(\ref{rawN}) with $\bm{\xi}$ yields
\begin{equation}
\ddot{\bm{\xi}}=\frac{{\kappa}}{3N}\Lambda\bm{\xi}+\varepsilon\left[  (\hat{\tau}%
{\kappa}^{2}+\frac{\Lambda}{3N} ) \dot{\bm{\xi}}-\frac{5}{7{\kappa}}Q^{T}(Q\dot
{\bm{\xi}})^{\circ3}-\nu_1\Lambda \bm{\xi}- Q^T F( Q\bm{\xi},Q\dot{\bm{\xi}})  \right].
\end{equation}
Noting that the function $F( Q\bm{\xi},Q\dot{\bm{\xi}} )$ is composed of quadratic terms that do not contribute to the secular term, by following the same procedure as in the one-spike case, one can find the general equations for the amplitude and phase modulations of $\xi_k$
\begin{subequations}
\begin{equation}
\label{NspikesB}\frac{\partial\mathcal{B}_{k}}{\partial T_1}=\mathcal{B}_{k}
\left[  \frac{1}{2} (\hat{\tau}{\kappa}^{2}+\frac{\lambda_{k,0}}{3N})+\frac
{5}{56}\sum_{j=1}^{N} b_{k j}\mathcal{B}_{j}^{2} \right],
\end{equation}
\begin{equation}
    \frac{\partial \theta_{k}}{\partial T_1}=-\frac{\nu_1\lambda_{k,0}}{2\kappa\omega_k},
\end{equation}
\end{subequations}
where
\begin{equation}
b_{kj}=\left\{
\begin{array}
[c]{ll}%
\frac{1}{N}\left(  \sum_{l=1}^{N} Q_{lj}^{4} \right)  \lambda_{j,0} & ~~~j= k\\
\frac{1}{N}\left(  2\sum_{l=1}^{N} Q_{lj}^{2}Q_{lk}^{2} \right)  \lambda_{j,0} &
~~~j\neq k
\end{array}
\right.,\quad \text{and} \quad \omega_k=\sqrt{-\frac{\kappa \lambda_{k,0}}{3N}}  .
\end{equation}
Constant solutions of Eq.~(\ref{NspikesB}) can be determined
by setting $\frac{\partial\mathcal{B}_{k}}{\partial T_1}=0$. These solutions correspond to periodic or quasi-periodic motions of the reduced system in Eq.~(\ref{2NODE}). Moreover, Eq.~(\ref{NspikesB}) makes it possible to detect the stability of periodic and quasi-periodic motions of the spike center by analyzing the stability of the equilibrium points.

We remark that all the eigenvalues $\lambda_{j,0}$ are negative when $D<D_N$, which results in $b_{kj}<0$.  Thus, the sign of $(\hat{\tau}{\kappa}^{2}+\frac{\lambda_{j,0}}{3N})$ will determine whether the system (\ref{NspikesB}) admits non-zero equilibrium points.  We will further address whether non-zero equilibrium points are stable in the next subsection for $N=2$ and $3$.
\begin{remark}
When $\hat{\tau}>-\frac{\lambda_{j,0}}{3N \kappa^2}$, which corresponds to $\tau>\tau_j$ in the original variable, the zero equilibrium points of the system (\ref{NspikesB}) become unstable and the system (\ref{NspikesB}) admits at least $j$ different non-zero equilibrium points. In this way, we recover the instability result in Proposition \ref{pro1}.
\end{remark}

\begin{figure}[t!]
\centering
\includegraphics[width=0.6\textwidth]{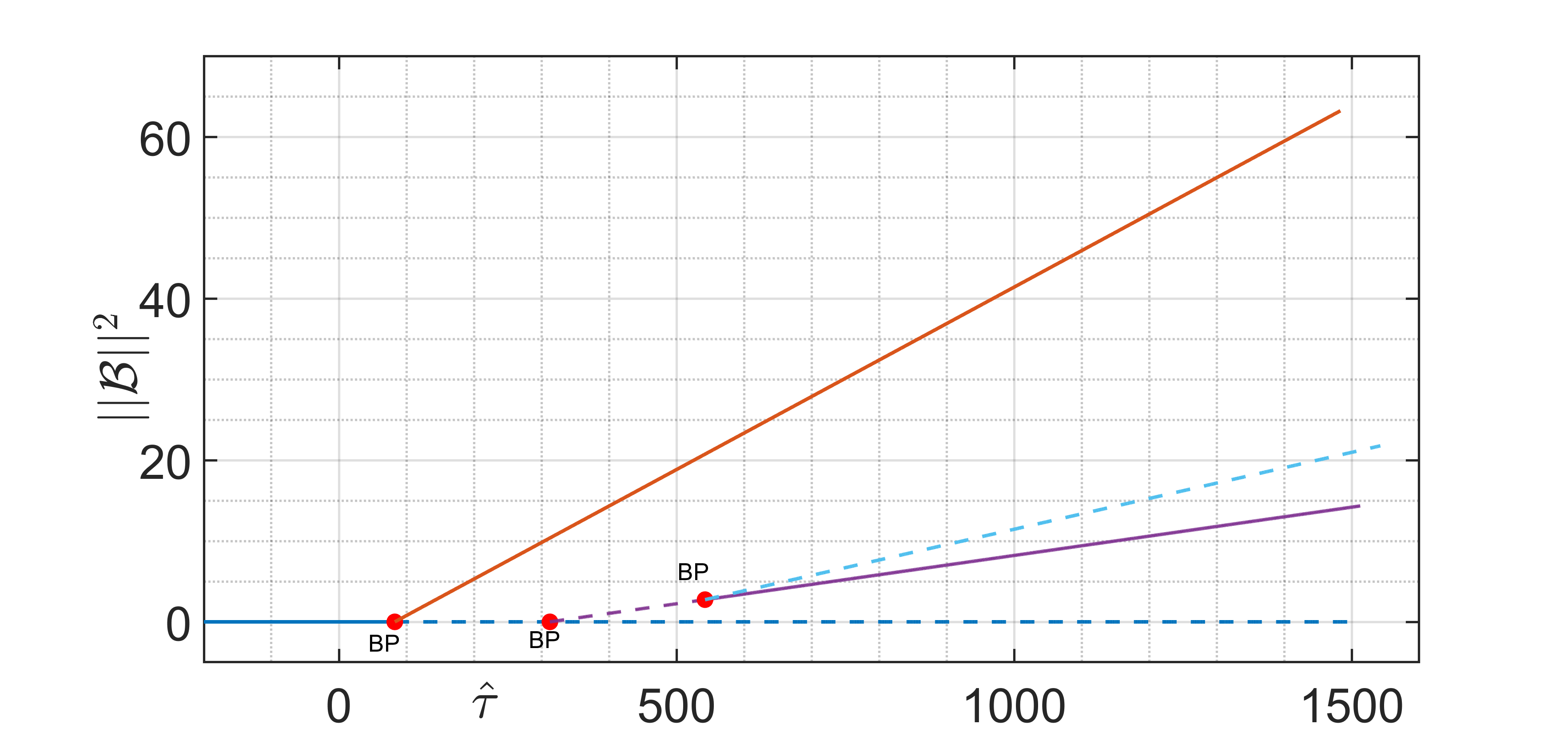}\caption{(Color online) Bifurcation
diagram for parameter $\hat{\tau}$ of Eq.~(\ref{2spikesB}). The horizontal axis is
$\hat{\tau}$, and the vertical axis is $||\mathcal{B}||^{2}=\mathcal{B}_{1}%
^{2}+\mathcal{B}_{2}^{2}$. The solid lines are the stable parts and dash lines are the unstable parts. BP indicates the bifurcation branch point. The parameters are $D=\frac {1}{150},~{\kappa}=0.2$.}%
\label{bd}%
\end{figure}
\subsection{Detailed analysis of amplitude modulation equations for two spikes}
In this subsection, we explicitly state and classify the equilibrium points of the amplitude modulation equations in the case of $N=2$.  

When $N=2$, the constants in Eqs.~(\ref{NspikesB}) are as follows:
\begin{align}
& q_{1}=\frac{\sqrt{2}}{2}(1,-1)^{\prime},~~q_{2}=\frac{\sqrt{2}}{2}%
(1,1)^{\prime},\\%
&  \lambda_{1,0}=-\frac{1}{2D},~~  \lambda_{2,0}=-\frac{1}{2D}\left(  1+\frac{1}{2D(48D-1)}\right),\\
& \mathbf{b}=%
\begin{pmatrix}
\frac{1}{4}\lambda_{1,0} & \frac{1}{2}\lambda_{2,0}\\
\frac{1}{2}\lambda_{1,0} & \frac{1}{4}\lambda_{2,0}%
\end{pmatrix}.
\end{align}
Eq.~(\ref{NspikesB}) becomes:
\begin{subequations}\label{2spikesB}
\begin{align}
\frac{\partial\mathcal{B}_{1}}{\partial t}=\mathcal{B}_{1}
\left[  \frac{1}{2} (\hat{\tau}{\kappa}^{2}+\frac{\lambda_{1,0}}{3N})+\frac{5}{224}(2\lambda_{1,0} \mathcal{B}_{1}^2+\lambda_{2,0}\mathcal{B}_{2}^2\right], \\
\frac{\partial\mathcal{B}_{2}}{\partial t}=\mathcal{B}_{2} \left[  \frac{1}{2}
(\hat{\tau}{\kappa}^{2}+\frac{\lambda_{2,0}}{3N})+\frac{5}{224}(2\lambda_{1,0} \mathcal{B}_{1}^2+\lambda_{2,0}\mathcal{B}_{2}^2  \right],
\end{align}
\end{subequations}
with
\[
\xi_{k} \sim\mathcal{B}_{k} \cos(\varepsilon\omega_{k} t+ \theta_{k}),
\]
and
\[
p_{1} = \frac{1}{\sqrt{2}}(\xi_{1}+\xi_{2}),~~ p_{2} = \frac{1}{\sqrt{2}}%
(\xi_{2}-\xi_{1}).
\]
The equilibrium points satisfy
\begin{subequations}\label{Equil}
\begin{align}
\mathcal{B}_{1} \left[  \frac{1}{2} (\hat{\tau}{\kappa}^{2}%
+\frac{\lambda_{1,0}}{6})+\frac{5}{224}(\lambda_{1,0} \mathcal{B}_{1}^2+2\lambda_{2,0}\mathcal{B}_{2}^2  )
\right]  =0,\\
\mathcal{B}_{2} \left[  \frac{1}{2} (\hat{\tau}{\kappa}^{2}+\frac{\lambda_{2,0}%
}{6})+\frac{5}{224}(2\lambda_{1,0} \mathcal{B}_{1}^2+\lambda_{2,0}\mathcal{B}_{2}^2  ) \right]  =0.
\end{align}
\end{subequations}
The system~(\ref{Equil}) decouples when one of $\mathcal{B}_{1}$ or $\mathcal{B}_{2}$ is
zero. When $\mathcal{B}_{1}\ne0$ and $\mathcal{B}_{2}=0$, the corresponding solution exhibits ``even'' (out-of-phase) oscillations, as shown in Fig.~\ref{fig:intro}(e). When $\mathcal{B}_{2}\ne0$ and $\mathcal{B}_{1}=0$, the corresponding solution
exhibits ``odd'' (in-phase) oscillations, as shown in Fig.~\ref{fig:intro}(d).

\begin{figure}[th!]
\begin{subfigure}[b]{0.23\textwidth}
\includegraphics[width=\textwidth]{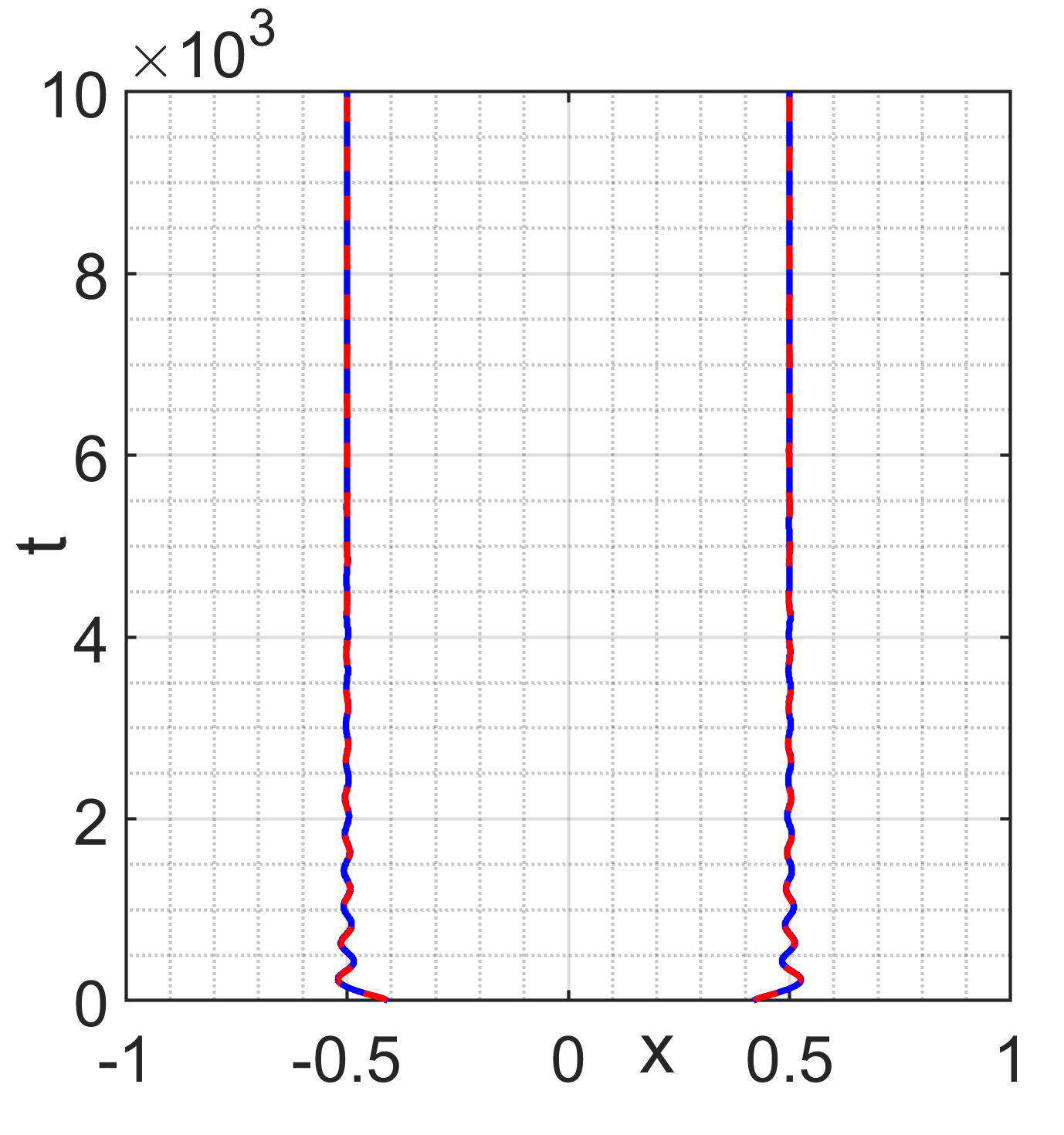}
\end{subfigure}
\hfill\begin{subfigure}[b]{0.23\textwidth}
\includegraphics[width=\textwidth]{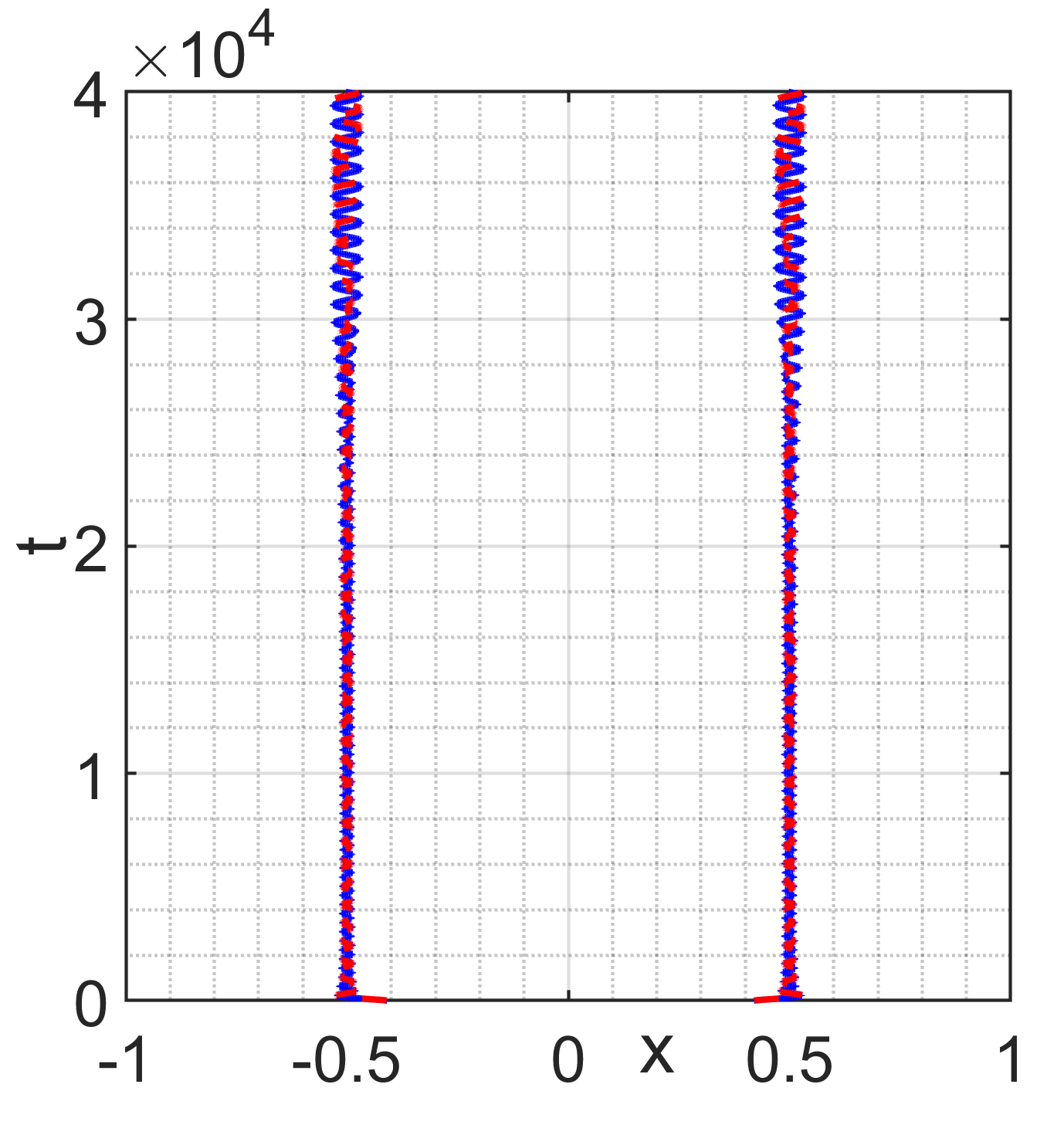}
\end{subfigure}
\hfill\begin{subfigure}[b]{0.22\textwidth}
\includegraphics[width=\textwidth]{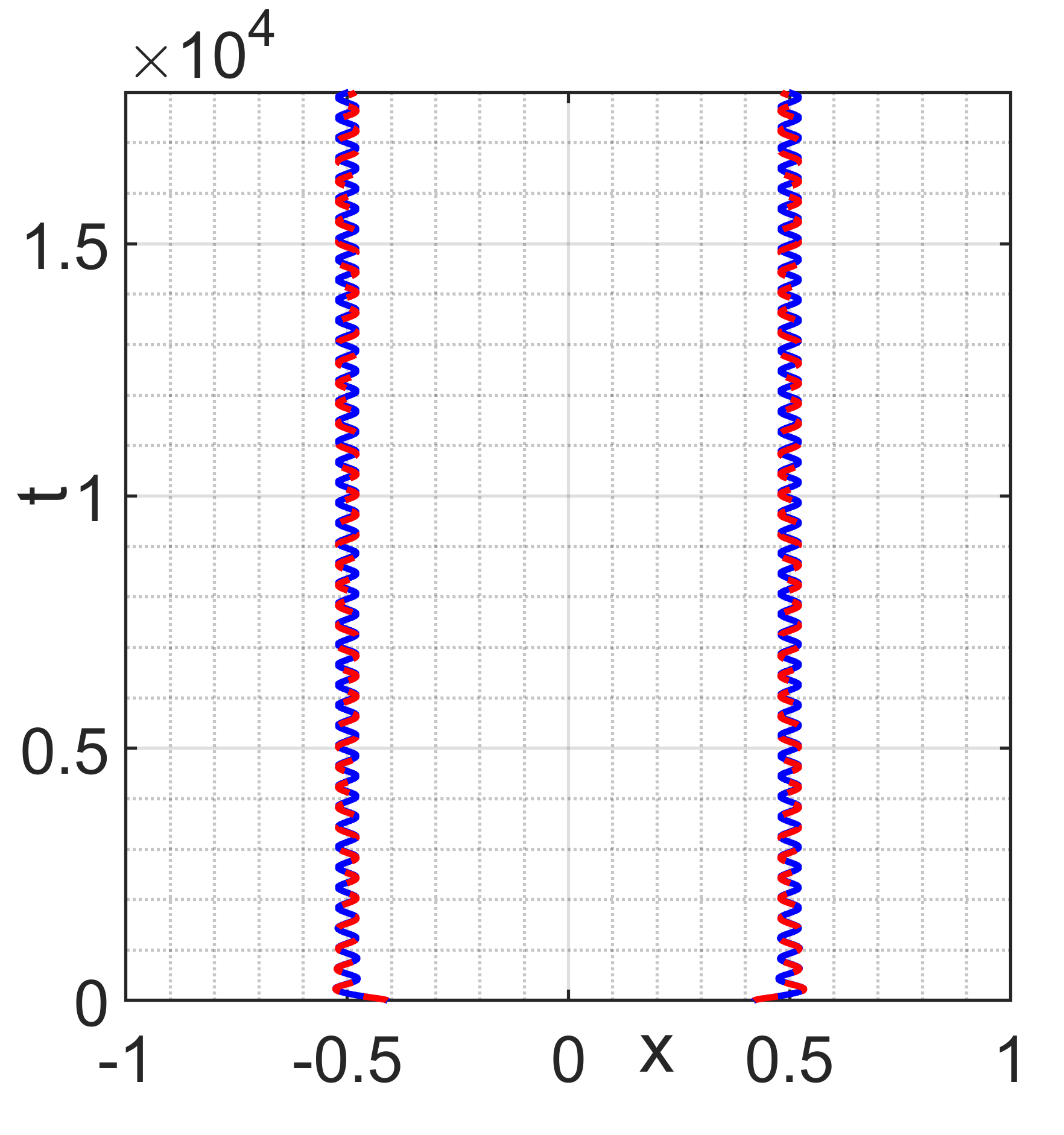}
\end{subfigure}
\hfill\begin{subfigure}[b]{0.23\textwidth}
\includegraphics[width=\textwidth]{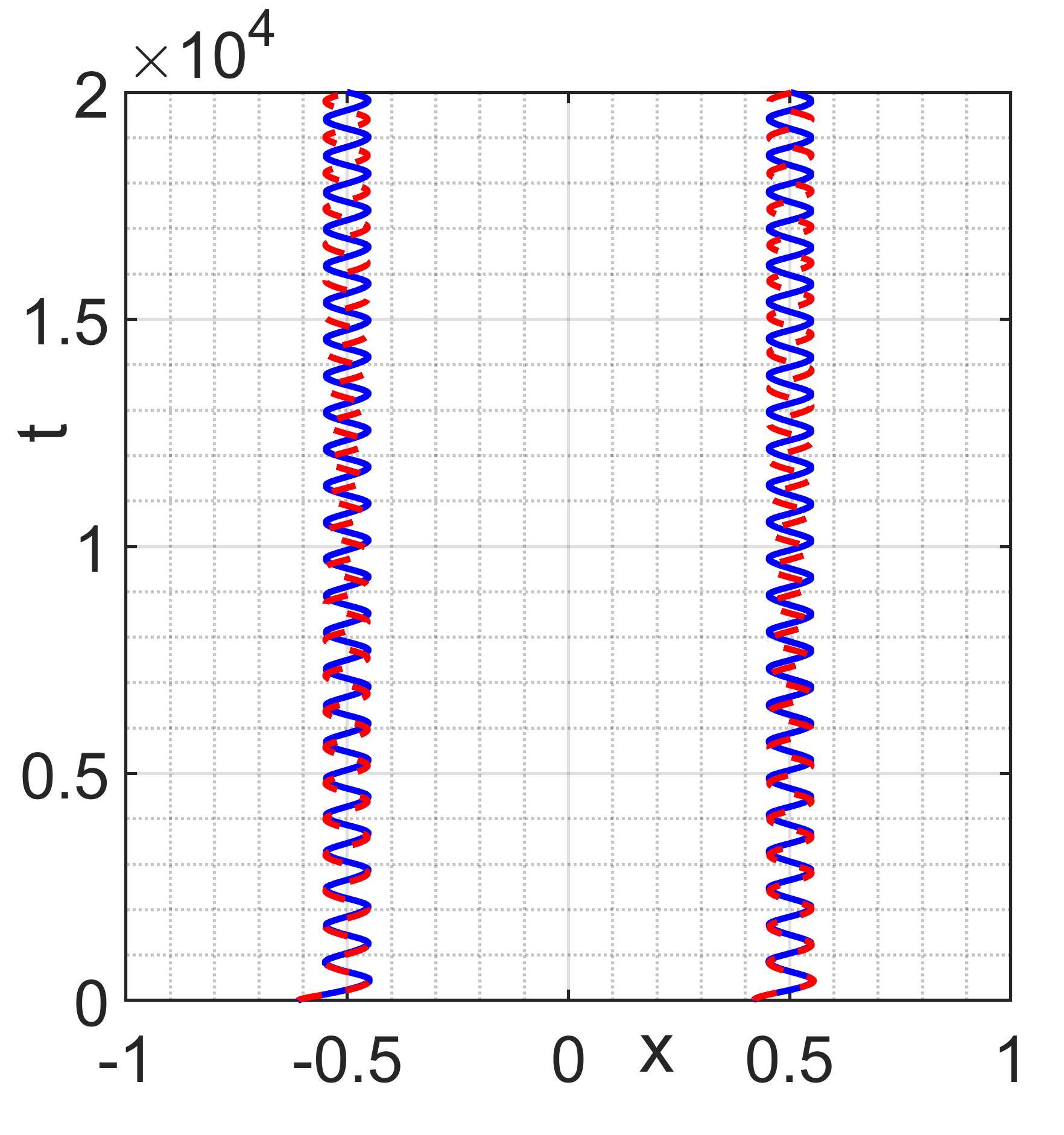}
\end{subfigure}
\centering
\begin{subfigure}[b]{0.23\textwidth}
\centering
\includegraphics[width=\textwidth]{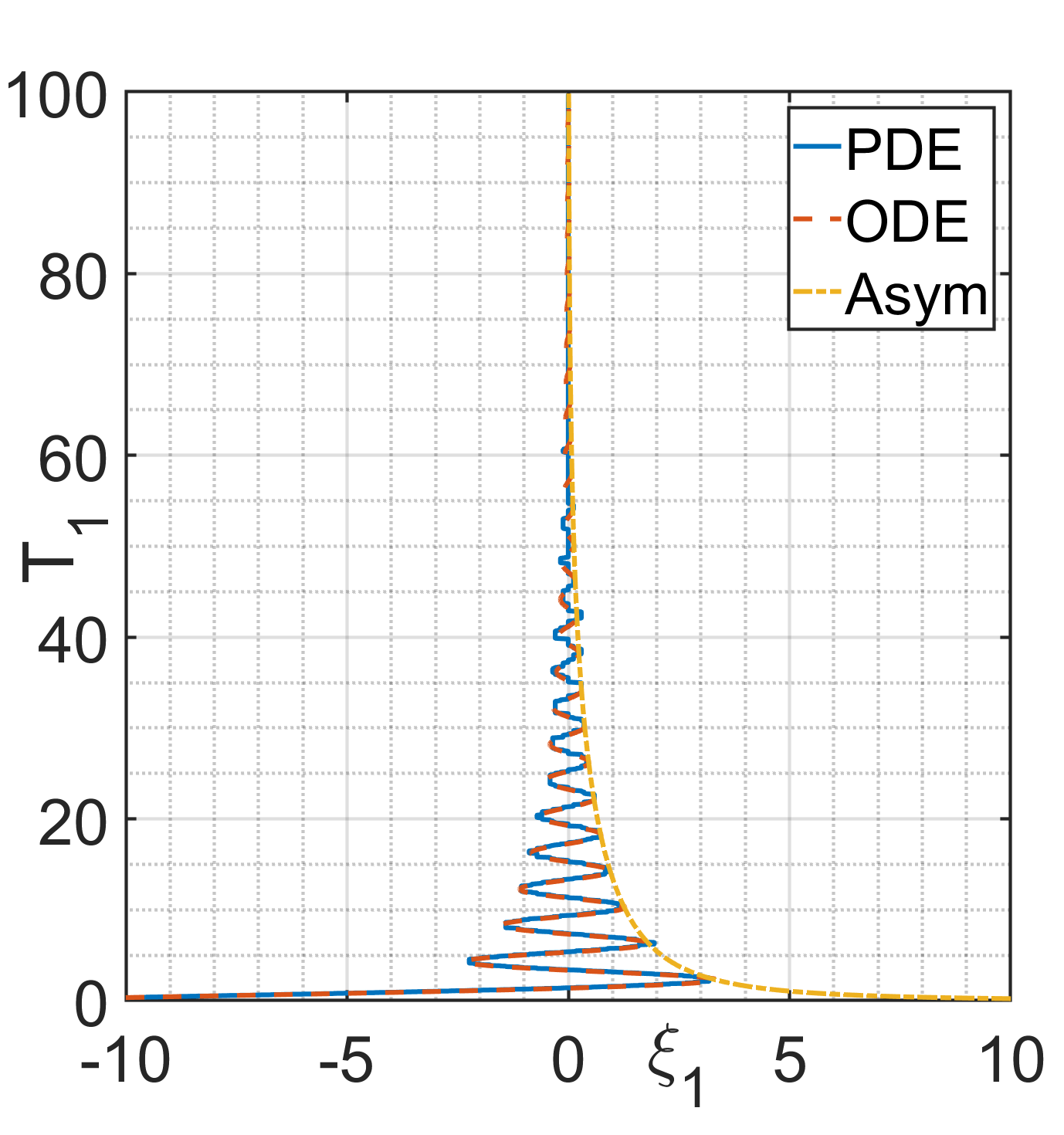}
\end{subfigure}
\hfill
\begin{subfigure}[b]{0.23\textwidth}
\centering
\includegraphics[width=\textwidth]{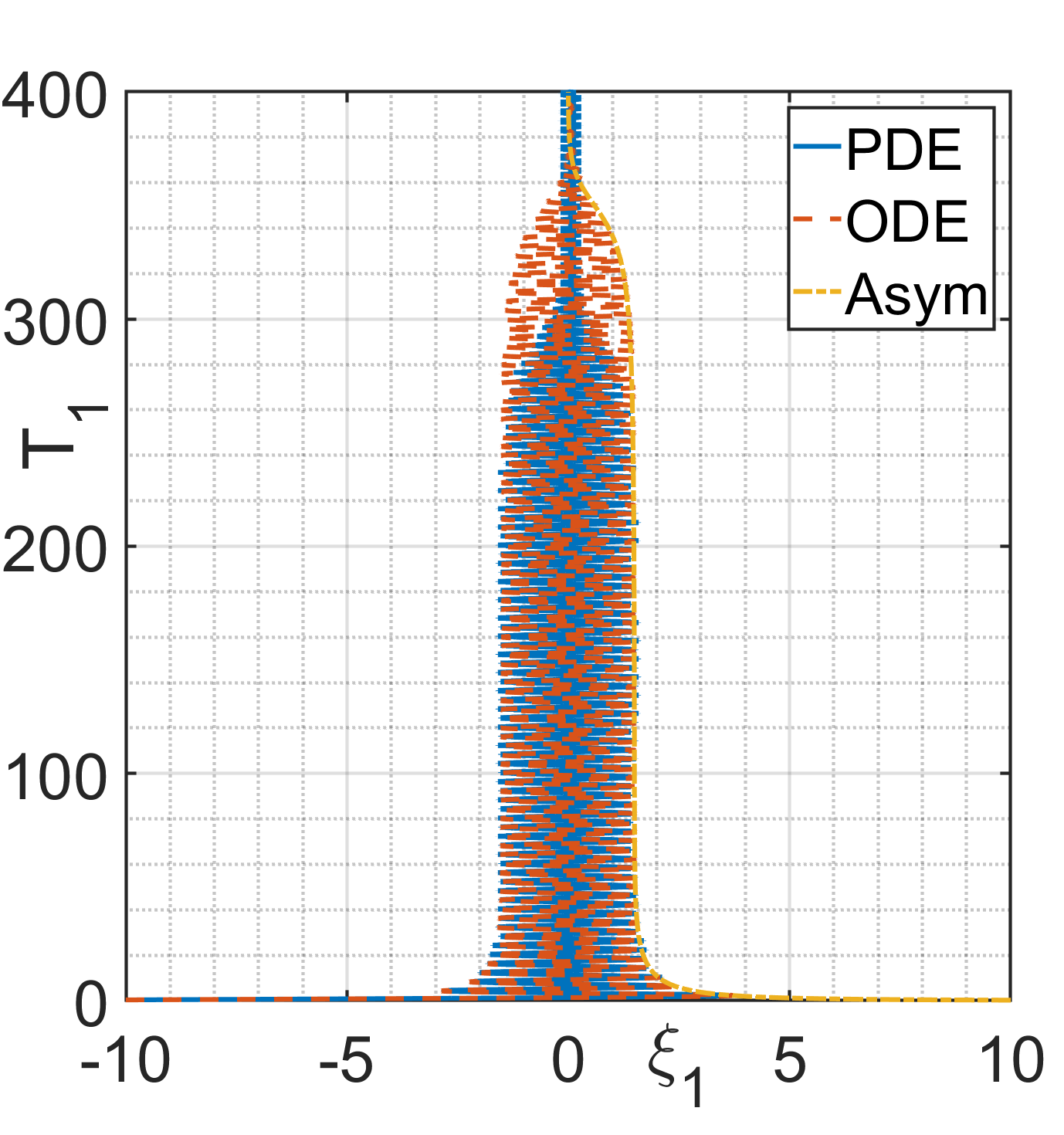}
\end{subfigure}
\hfill\begin{subfigure}[b]{0.23\textwidth}
\centering
\includegraphics[width=\textwidth]{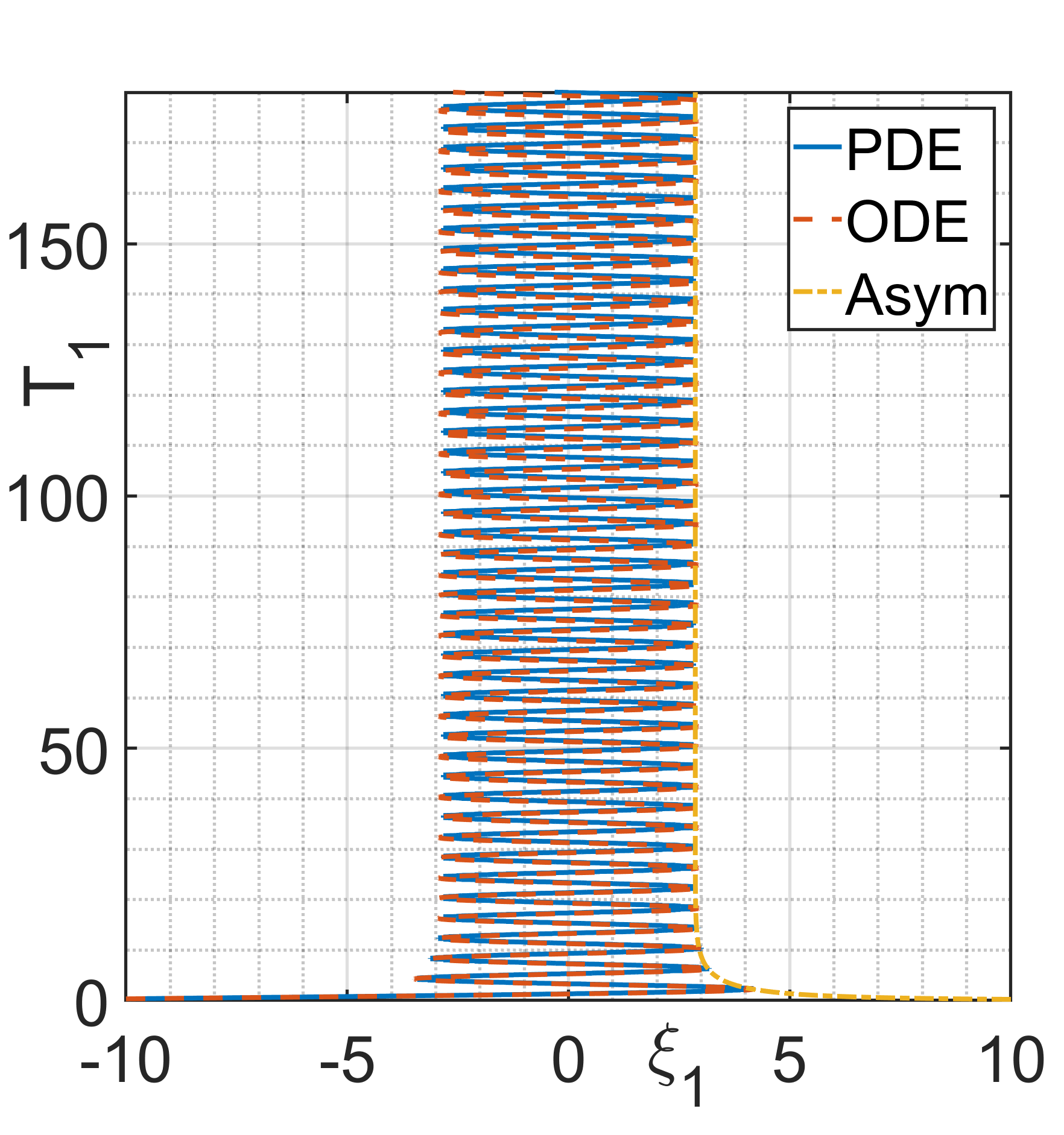}
\end{subfigure}
\hfill\begin{subfigure}[b]{0.23\textwidth}
\centering
\includegraphics[width=\textwidth]{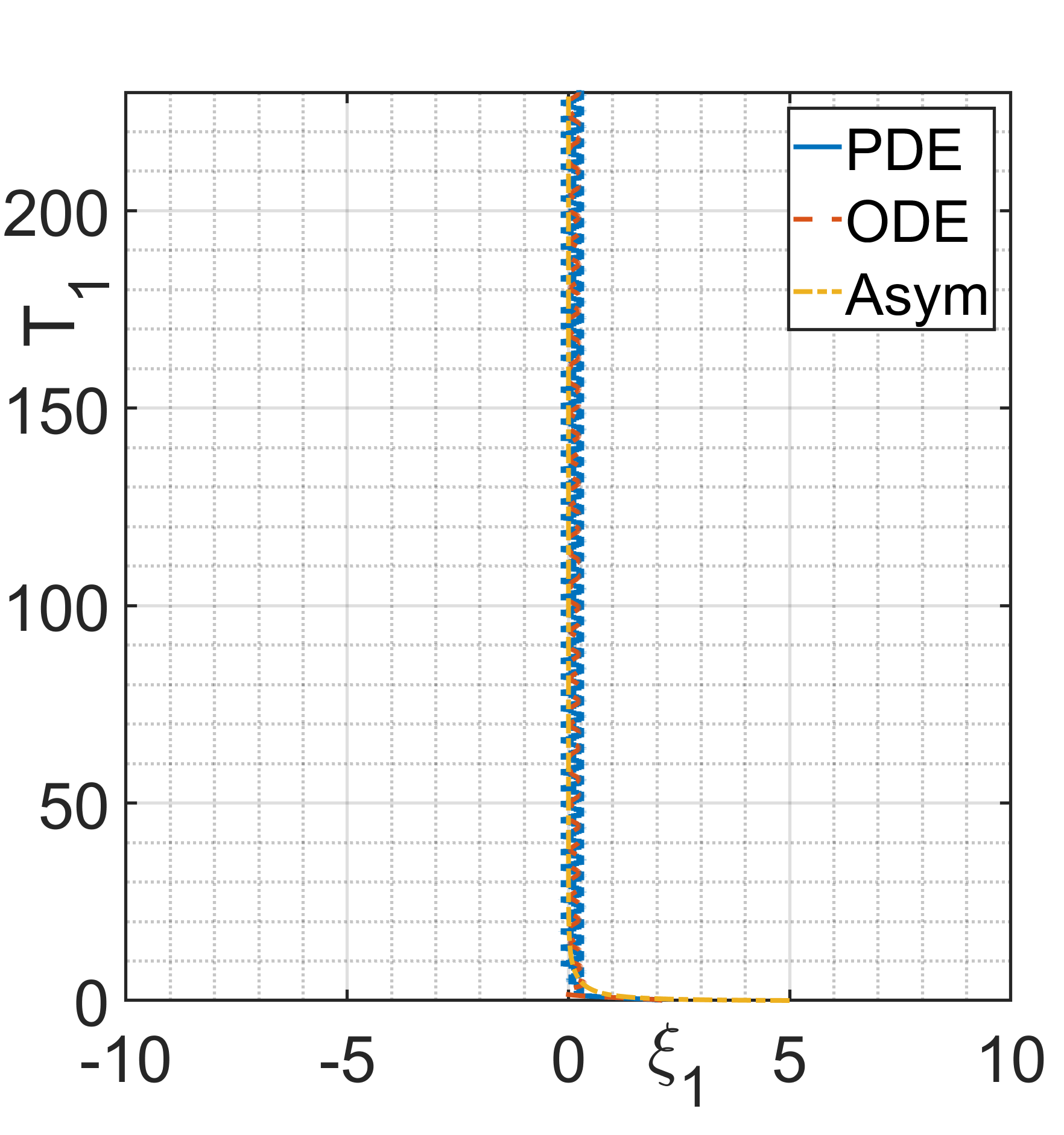}
\end{subfigure}
\hfill
\begin{subfigure}[b]{0.23\textwidth}
\centering
\includegraphics[width=\textwidth]{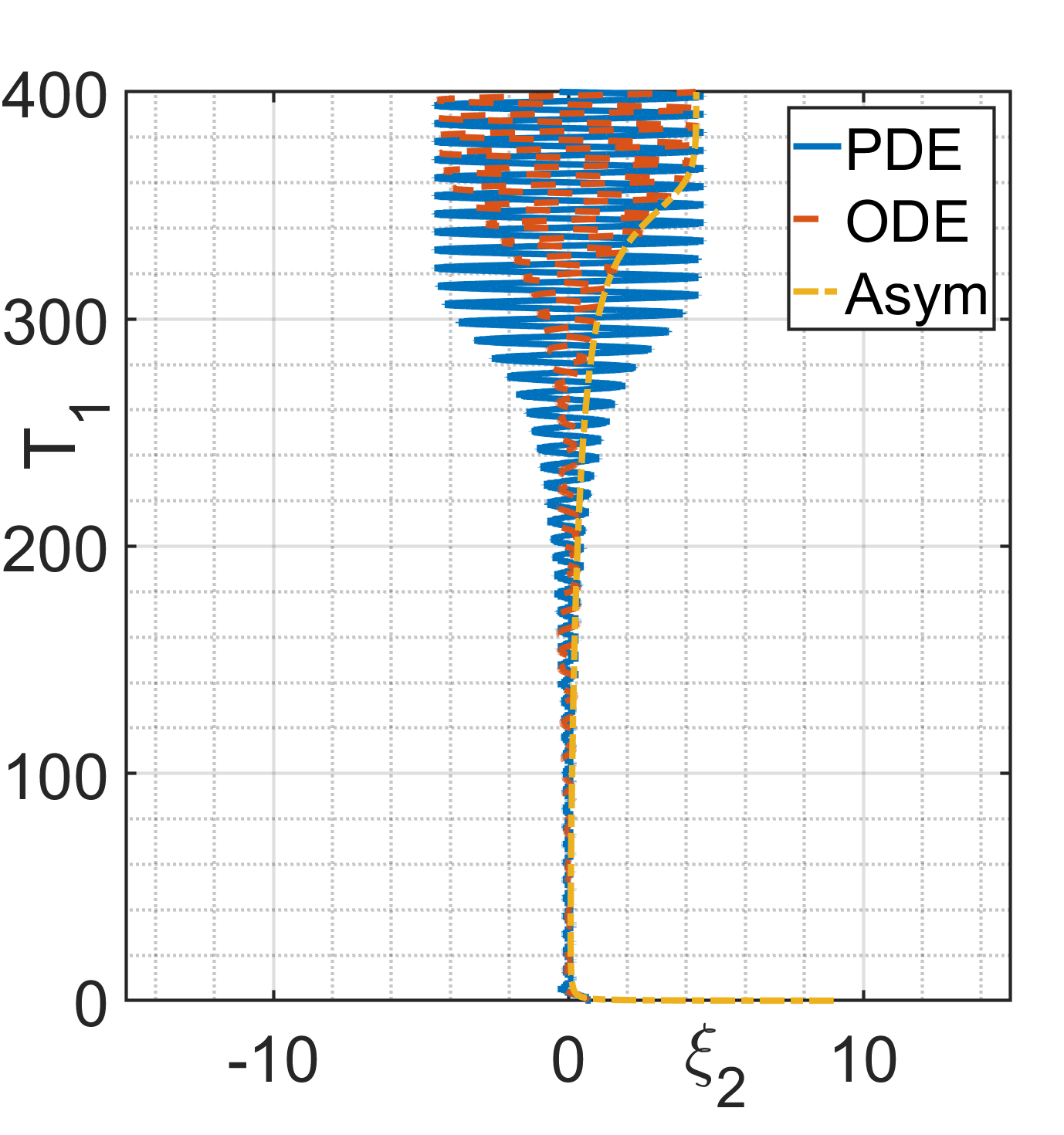}\caption{$\hat{\tau}=50$}\label{fig:502}
\end{subfigure}
\hfill
\begin{subfigure}[b]{0.23\textwidth}
\centering
\includegraphics[width=\textwidth]{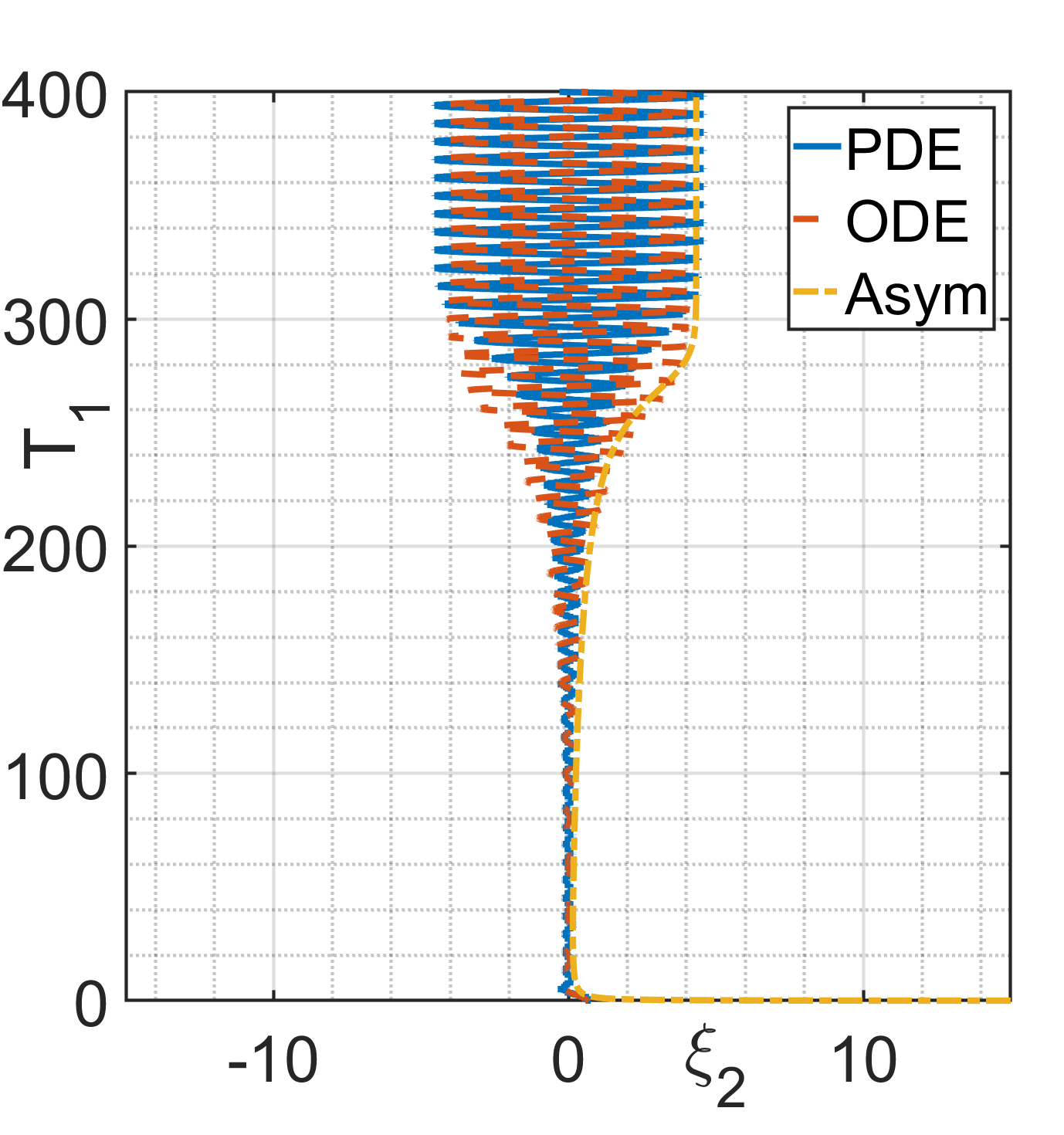}
\caption{$\hat{\tau}=500$}
\label{fig:5002}
\end{subfigure}
\hfill\begin{subfigure}[b]{0.23\textwidth}
\centering
\includegraphics[width=\textwidth]{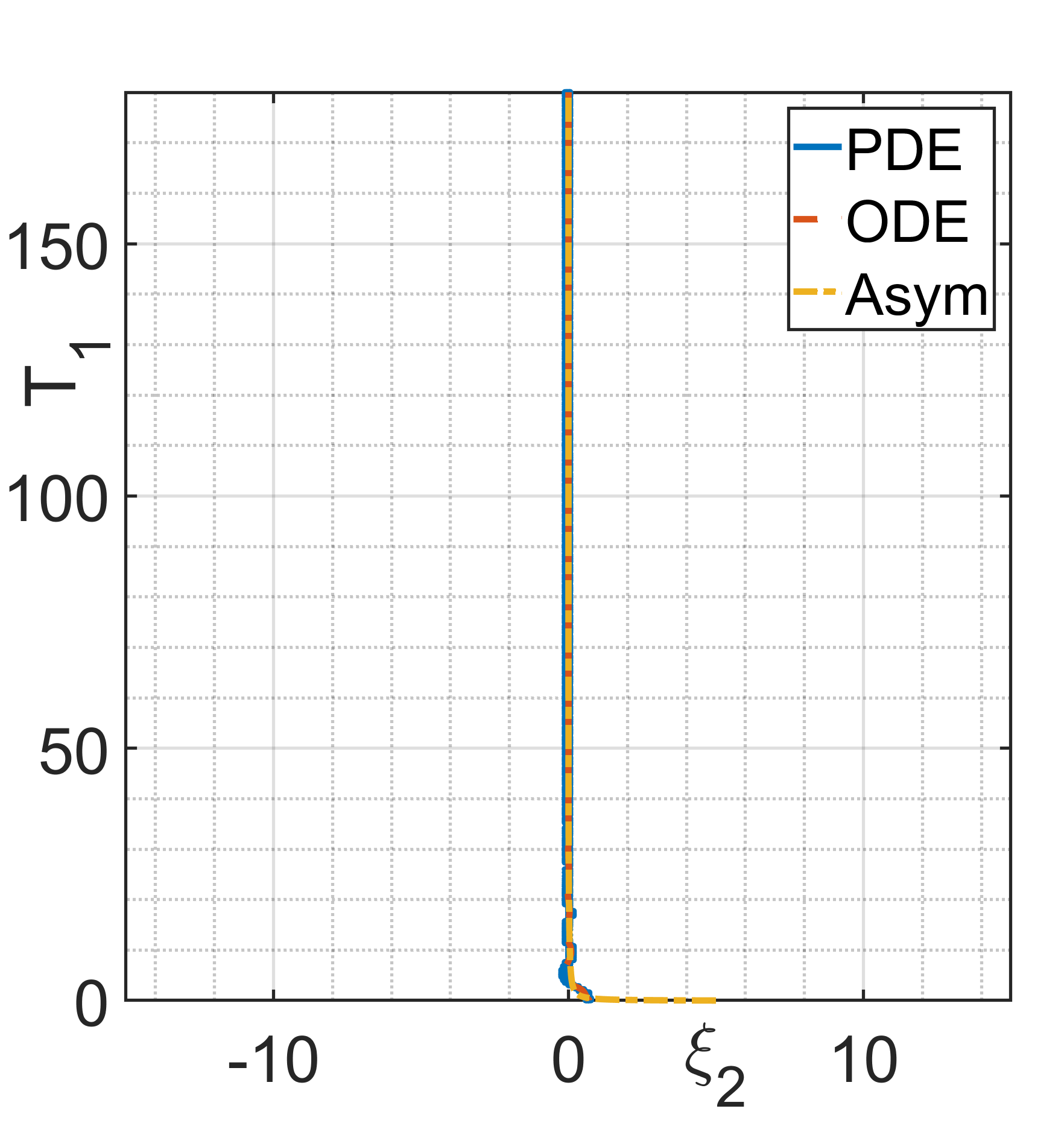}
\caption{$\hat{\tau}=1000$}
\label{fig:10002}
\end{subfigure}
\hfill\begin{subfigure}[b]{0.23\textwidth}
\centering
\includegraphics[width=\textwidth]{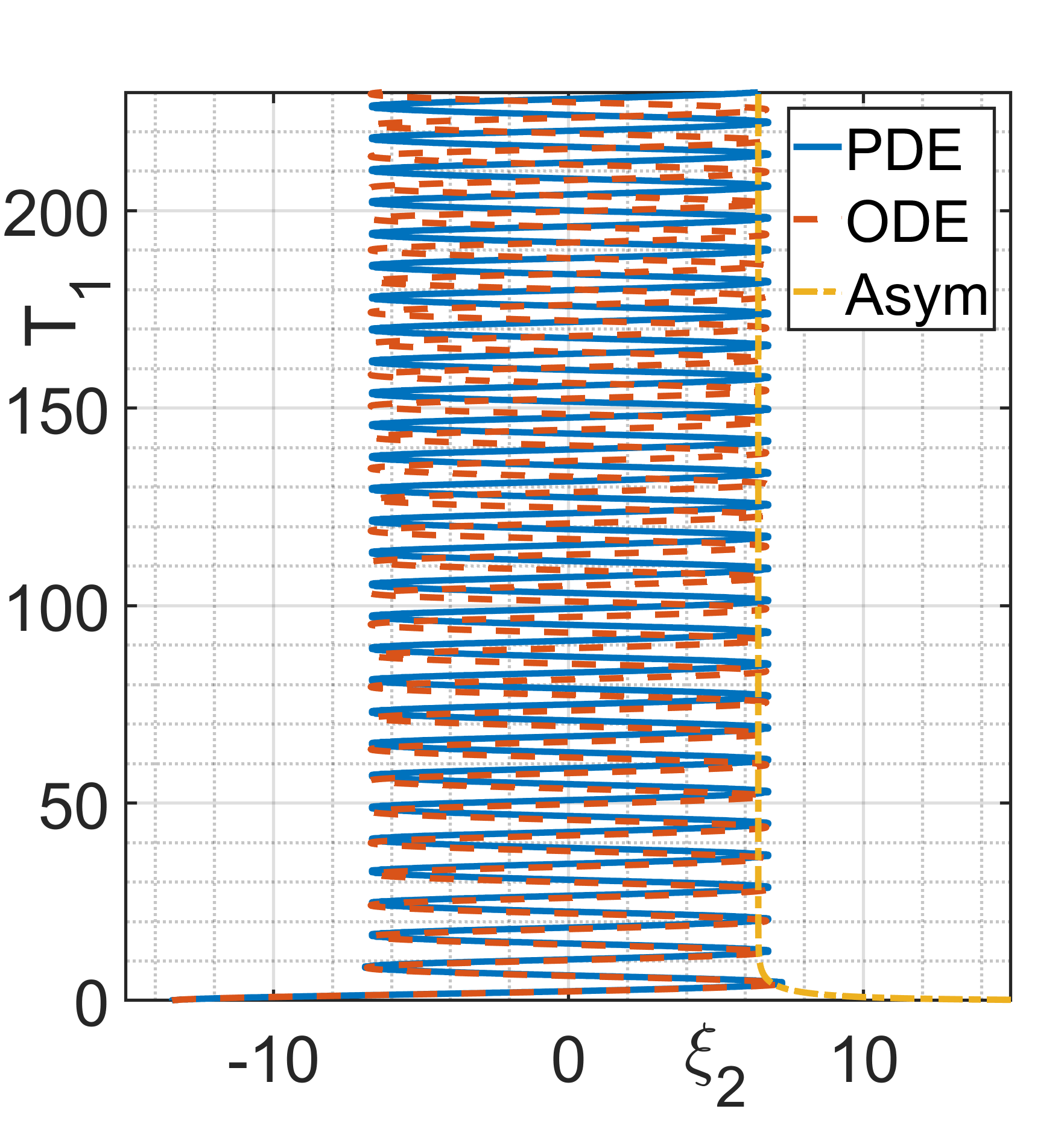}
\caption{$\hat{\tau}=1000$}
\label{fig:10003}
\end{subfigure}
\hfill\caption{(Color online) First row:  the locations of the spikes simulated
by the PDE  (solid lines) and the ODE (dashed lines) in the original variable. Other rows: slow time evolution of new variable $\xi_j,j=1,2$. First column (a): 
simulation results obtained at $\hat{\tau}=50$, when only $(\mathcal{B}_{1},\mathcal{B}%
_{2})=(0,0)$ is the stable equilibrium point; thus, both $|\mathcal{B}_{1}|$
and $|\mathcal{B}_{2}|$ decay. Second column (b): simulation results obtained at
$\hat{\tau}=500$. Although there exist three equilibrium points, only $(\mathcal{B}_{1},\mathcal{B}_{2})=(0,4.339)$ is stable. Thus, even though
the initial condition is close to the equilibrium point $(\mathcal{B}%
_{1},\mathcal{B}_{2})=(1.4967,0)$, which corresponds to the out-of-phase oscillations, the dynamics converge to $(\mathcal{B}_{1},\mathcal{B}_{2})=(0,4.339)$, which corresponds to the in-phase oscillations. Third and fourth columns (c) and (d): simulation results obtained at $\hat{\tau}=1000$, where both $(\mathcal{B}_{1},\mathcal{B}%
_{2})=(2.86,0)$ and $(0,6.435)$ are stable. Thus, with different initial conditions (out-of-phase oscillations or in-phase oscillations), we end up with different oscillatory states. As the images in the first row may look indiscernible, the reader is referred to the web version of this article for high image quality.}%
\label{fig:twospike}
\end{figure}
Conventional linear stability analysis yields the following results with respect to the equilibrium points and their stabilities,

\begin{enumerate}
\item When $\hat{\tau}<\frac{1-96D}{24D(1-48D){\kappa}^{2}}$,
Eq.~(\ref{Equil}) admits only one non-negative solution
\begin{equation}
(\mathcal{B}_{1},\mathcal{B}_{2})=(0,0),
\end{equation}
which is stable.
\item When $\frac{1-96D}{24D(1-48D){\kappa}^{2}}<\hat{\tau}<\frac
{1}{12D{\kappa}^{2}}$, Eq.~(\ref{Equil}) admits two non-negative solutions
\begin{equation}
(\mathcal{B}_{1},\mathcal{B}_{2})=(0,0)~\text{and}~\left(  0,\sqrt
{\frac{112(\hat{\tau}{\kappa}^{2}+\frac{\lambda_{2,0}}{6})}{-5\lambda_{2,0}}}\right)  .
\end{equation}
It is easy to check that $(0,0)$ is unstable and $\left(  0,\sqrt
{\frac{112(\hat{\tau}{\kappa}^{2}+\frac{\lambda_{2,0}}{6})}{-5\lambda_{2,0}}}\right)  $
is stable.

\item When $\frac{1}{12D{\kappa}^{2}}<\hat{\tau}<\frac{1-32D}{8D(1-48D){\kappa
}^{2}}$, Eq.~(\ref{Equil}) admits non-negative solutions
\begin{equation}
(\mathcal{B}_{1},\mathcal{B}_{2})=(0,0),~\left(  \sqrt{\frac{112(\hat{\tau
}{\kappa}^{2}+\frac{\lambda_{1,0}}{6})}{-5\lambda_{1,0}}},0\right)  ,~\left(0,
\sqrt{\frac{112(\hat{\tau}{\kappa}^{2}+\frac{\lambda_{2,0}}{6})}{-5\lambda_{2,0}}%
}\right)  .
\end{equation}
Again, $(0,0)$ is unstable,$\left(  0,\sqrt{\frac{112(\hat{\tau}{\kappa}%
^{2}+\frac{\lambda_{2,0}}{6})}{-5\lambda_{2,0}}}\right)  $ is stable, and $\left(
\sqrt{\frac{112(\hat{\tau}{\kappa}^{2}+\frac{\lambda_{1,0}}{6})}{-5\lambda_{1,0}}},0\right)  $is
unstable .
\item When $\hat{\tau}>\frac{1-32D}{8D(1-48D){\kappa}^{2}}$, Eq.~(\ref{Equil}%
) admits four non-negative solutions
\begin{equation}
(\mathcal{B}_{1},\mathcal{B}_{2})=(0,0),~\left(
\sqrt{\frac{112(\hat{\tau}{\kappa}^{2}+\frac{\lambda_{1,0}}{6})}{-5\lambda_{1,0}}},0\right)  ,~\left(
0,\sqrt{\frac{112(\hat{\tau}{\kappa}%
^{2}+\frac{\lambda_{2,0}}{6})}{-5\lambda_{2,0}}}\right)  , \left( \frac{1}{2}\sqrt{-\mathbf{b}^{-1}(\hat{\tau}{\kappa}^{2}%
+\frac{\lambda_{1,0}}{6},\hat{\tau}{\kappa}^{2}+\frac{\lambda_{2,0}}{6})} \right).
\end{equation}
Only $\left(
\sqrt{\frac{112(\hat{\tau}{\kappa}^{2}+\frac{\lambda_{1,0}}{6})}{-5\lambda_{1,0}}},0\right)~\text{and}~\left(
0,\sqrt{\frac{112(\hat{\tau}{\kappa}%
^{2}+\frac{\lambda_{2,0}}{6})}{-5\lambda_{2,0}}}\right)   $ are stable.
\end{enumerate}

 The system has three branch points:
$\hat{\tau}=\frac{1-96D}{24D(1-48D){\kappa}^{2}},~\frac{1}{12D{\kappa}^{2}%
},~\text{and}~\frac{1-32D}{8D(1-48D){\kappa}^{2}}$.  The first two branch points are the same as the Hopf bifurcation points obtained from the PDE stability analysis in Section \textsection\ref{sec2}. 
The third branch point is the critical one to determine whether both in-phase and out-of-phase oscillations can occur simultaneously. Below this point, only the in-phase oscillations are stable. Fig.~(\ref{bd}) is the bifurcation diagram for $\hat{\tau}$ obtained via Matcont $7.1$, \cite{govaerts2018matcont}. When $D=\frac{1}{150},~\kappa=0.2$, the three branch points are located at $\hat{\tau}=82.72,~312.50,~\text{and}~542.28$. It should be noted that there exist two equilibrium points when $312.5<\hat{\tau}<542.28$, which correspond to two types of oscillations, but only one is stable, indicating that only one type of oscillation is stable. Indeed, we can see that only the in-phase oscillations are stable in the PDE and ODE simulation results when $\hat{\tau}=500$ (see Fig.~\ref{fig:5002}). The coexistence of stable in-phase and out-of-phase oscillations occurs when $\tau>542.28$ (see Figs.~\ref{fig:10002} and \ref{fig:10003}).%

Fig.~\ref{fig:twospike} shows the numerical simulation of the ODEs
(\ref{NspikesB}) when $N=2$ for various values of $\hat{\tau}$. The results agree closely with the spike dynamics of the original PDE system with respect to the amplitudes and frequencies.

\begin{figure}[!htb]
\centering
\includegraphics[width=0.6\textwidth]{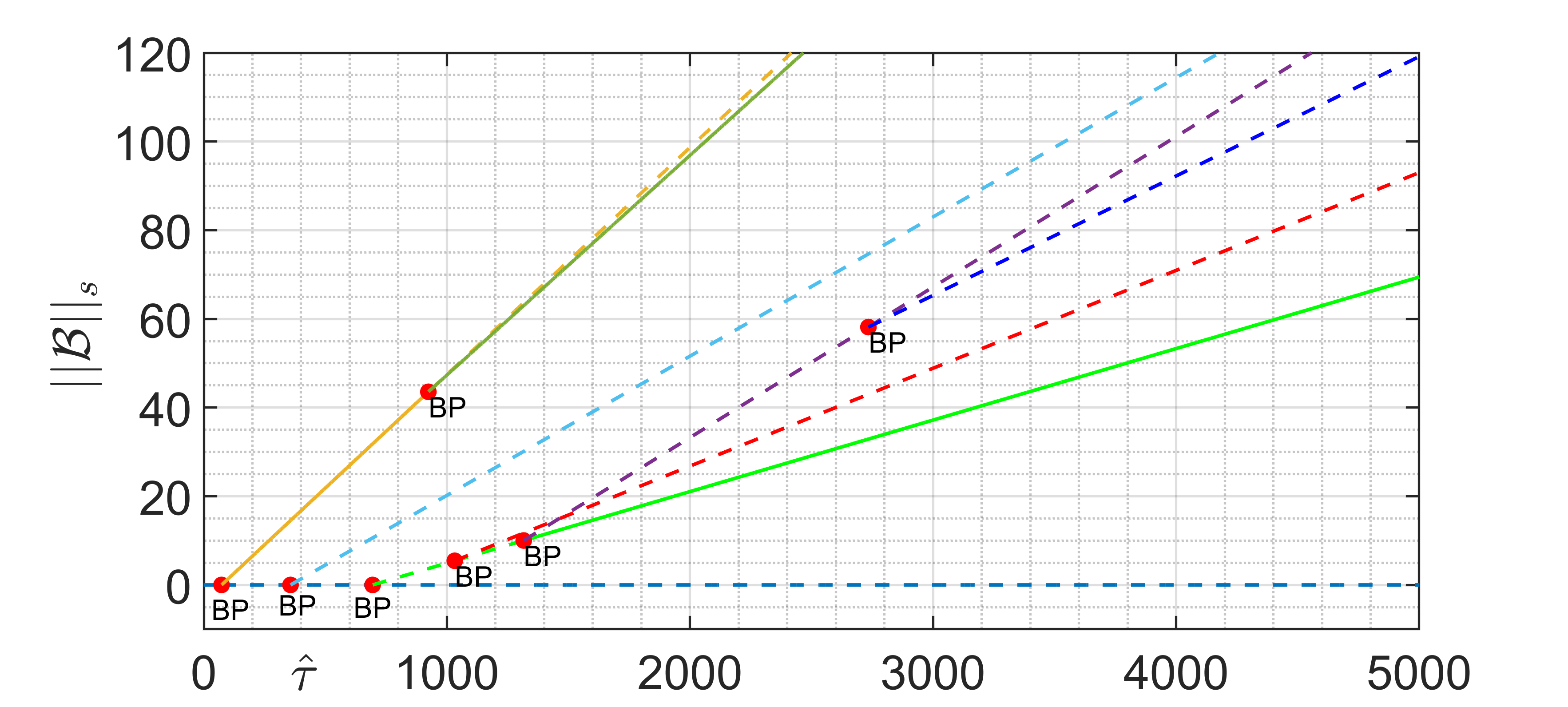}\caption{(Color online) Bifurcation
diagram for $\hat{\tau}$ of Eq.~(\ref{NspikesB}) in the three-spike case. The horizontal axis is $\hat{\tau}$, and the vertical axis is $||\mathcal{B}||_{s}%
^{2}=2\mathcal{B}_{1}^{2}+\mathcal{B}_{2}^{2}+3\mathcal{B}_{3}^{2}$. The solid lines are the stable parts, and the dashed lines are the unstable parts. The red dots marked $BP$ indicate the bifurcation branch points. The parameters are $D_{v}=\frac{1}{500},~{\kappa}=0.2$.
The branch point from left to right are at $\hat{\tau}%
=72.93,~356.68,~694.45,~923.75,~1032.22,~1315.99,~\text{and}~2733.90$. }%
\label{bd3}
\end{figure}

\subsection{Numerical investigation of the amplitude modulation equations for the three-spike dynamics.}
\begin{figure}[!htb]
\centering
\begin{subfigure}[b]{0.24\textwidth}
\centering
\includegraphics[width=\textwidth]{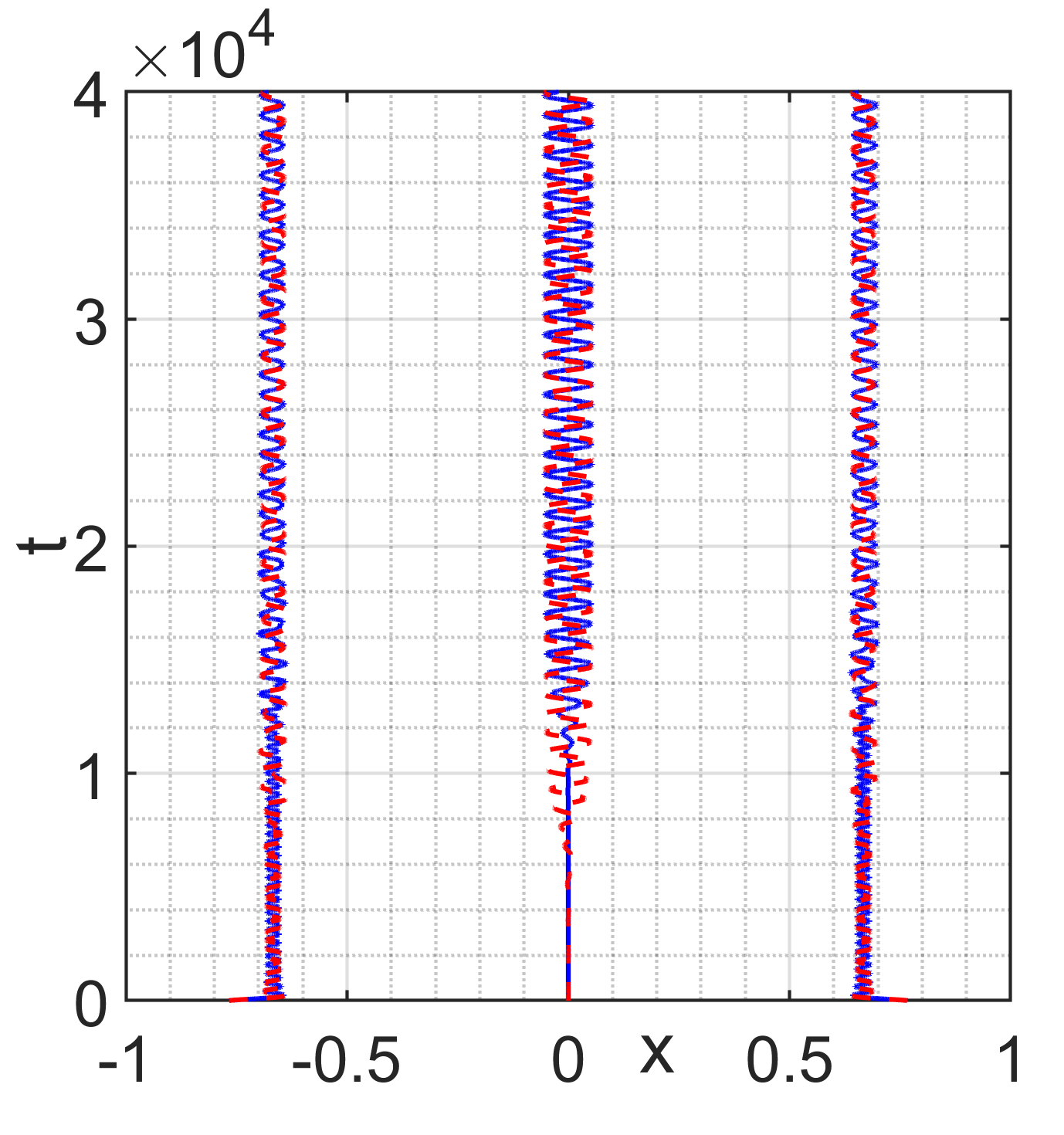}
\end{subfigure}
\hfill\begin{subfigure}[b]{0.24\textwidth}
\centering
\includegraphics[width=\textwidth]{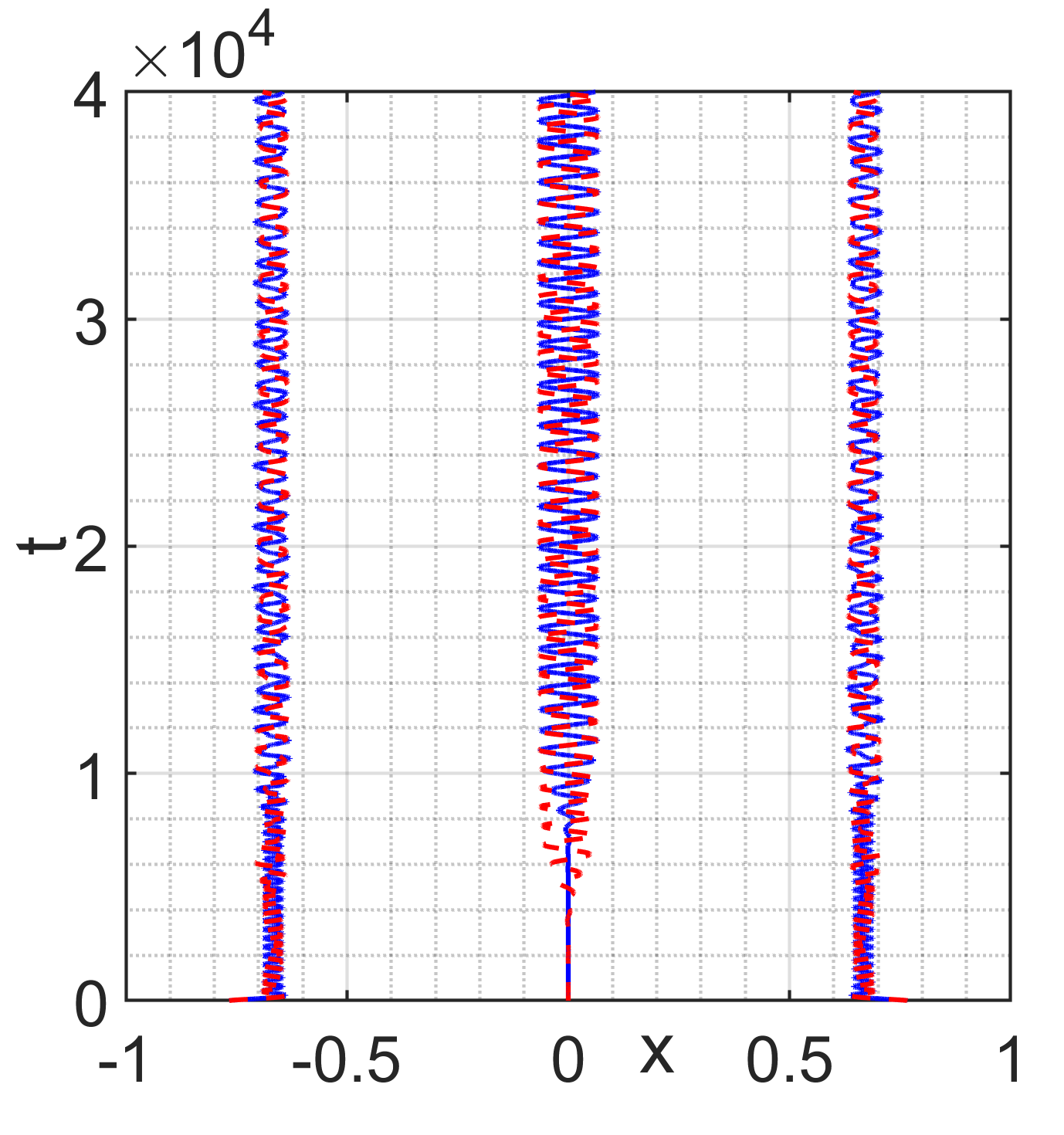}
\end{subfigure}
\hfill\begin{subfigure}[b]{0.24\textwidth}
\centering
\includegraphics[width=\textwidth]{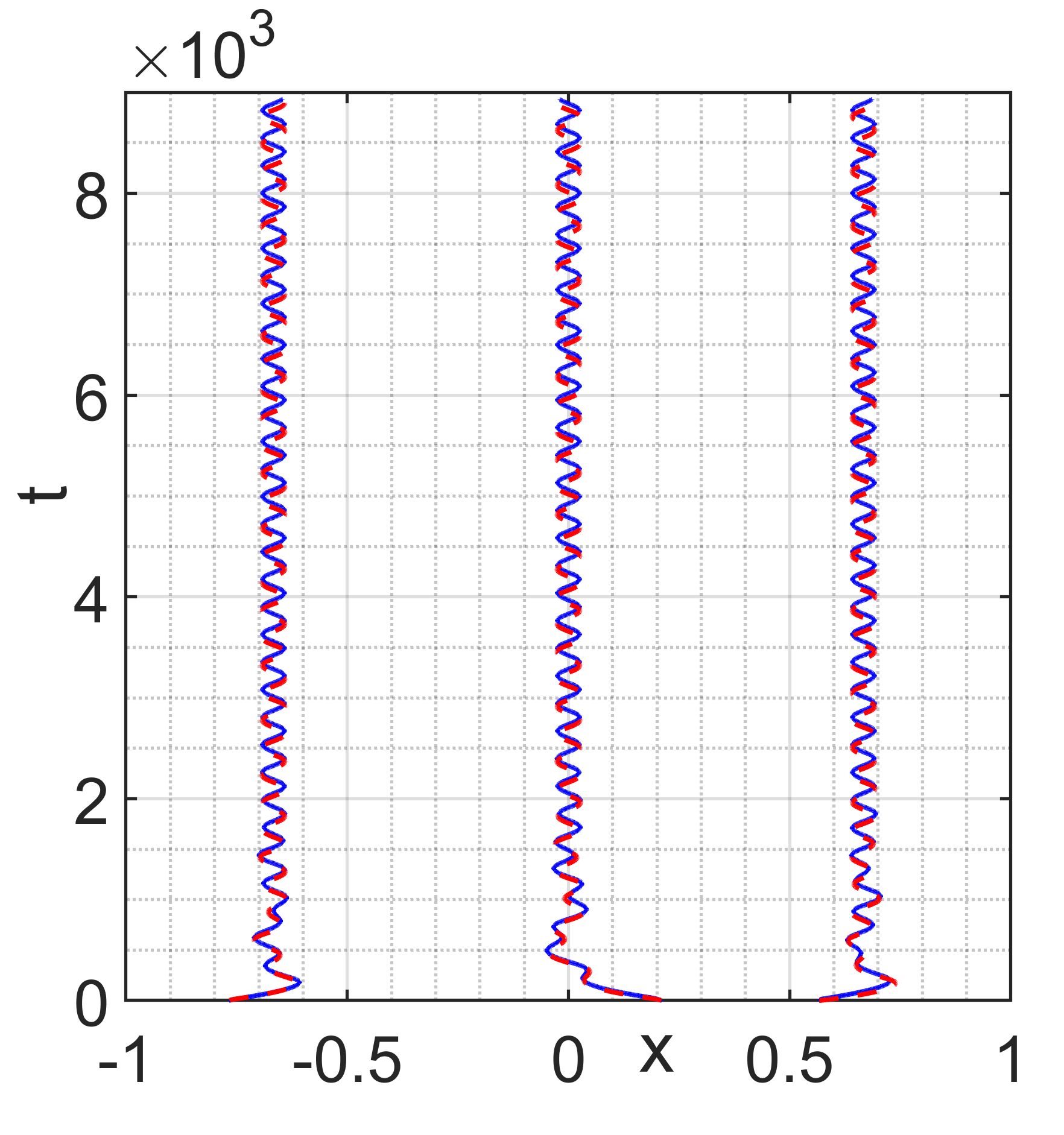}
\end{subfigure}
\hfill\begin{subfigure}[b]{0.24\textwidth}
\centering
\includegraphics[width=\textwidth]{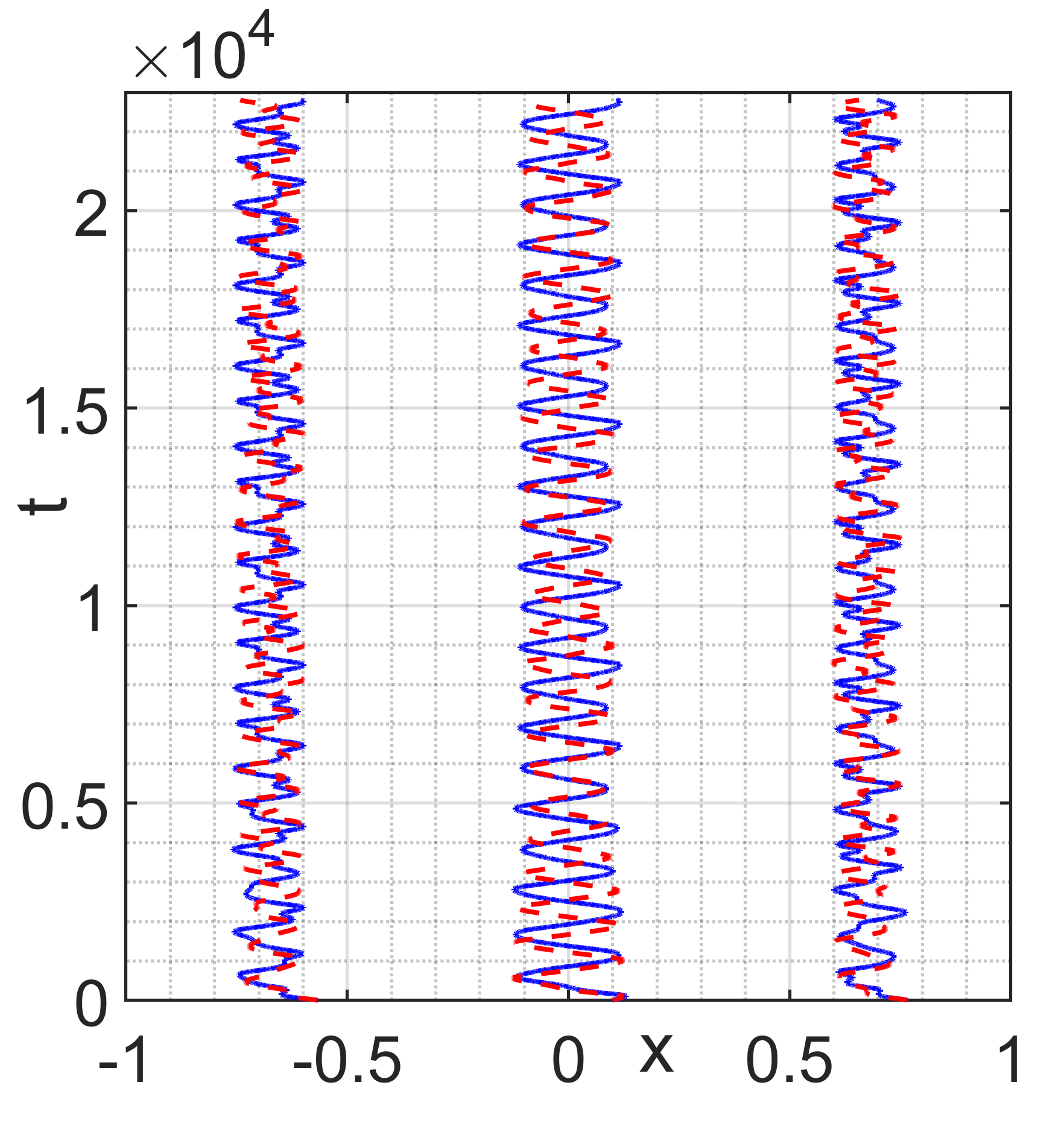}
\end{subfigure}
\hfill
\begin{subfigure}[b]{0.24\textwidth}
\centering
\includegraphics[width=\textwidth]{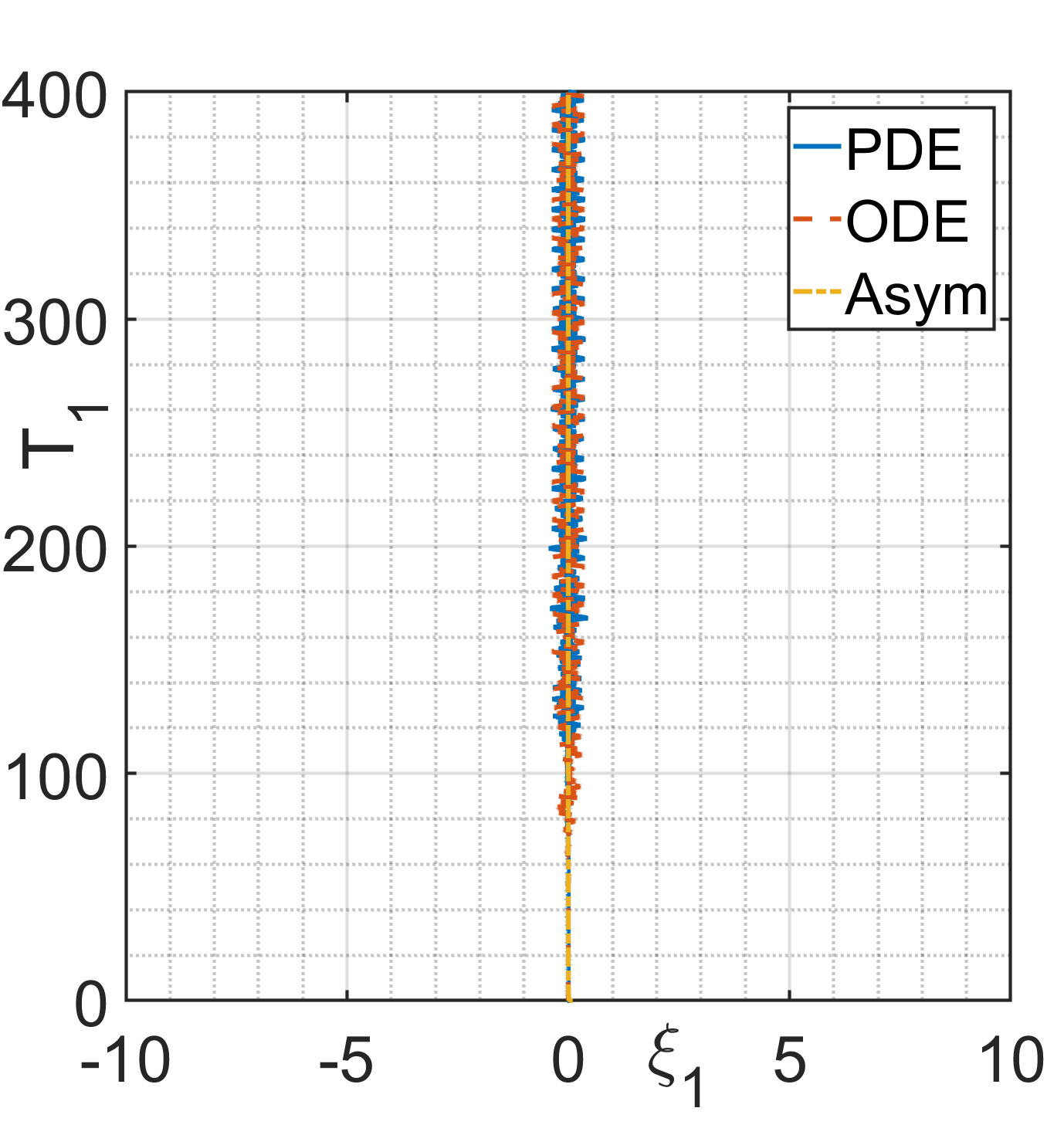}
\end{subfigure}
\hfill\begin{subfigure}[b]{0.24\textwidth}
\centering
\includegraphics[width=\textwidth]{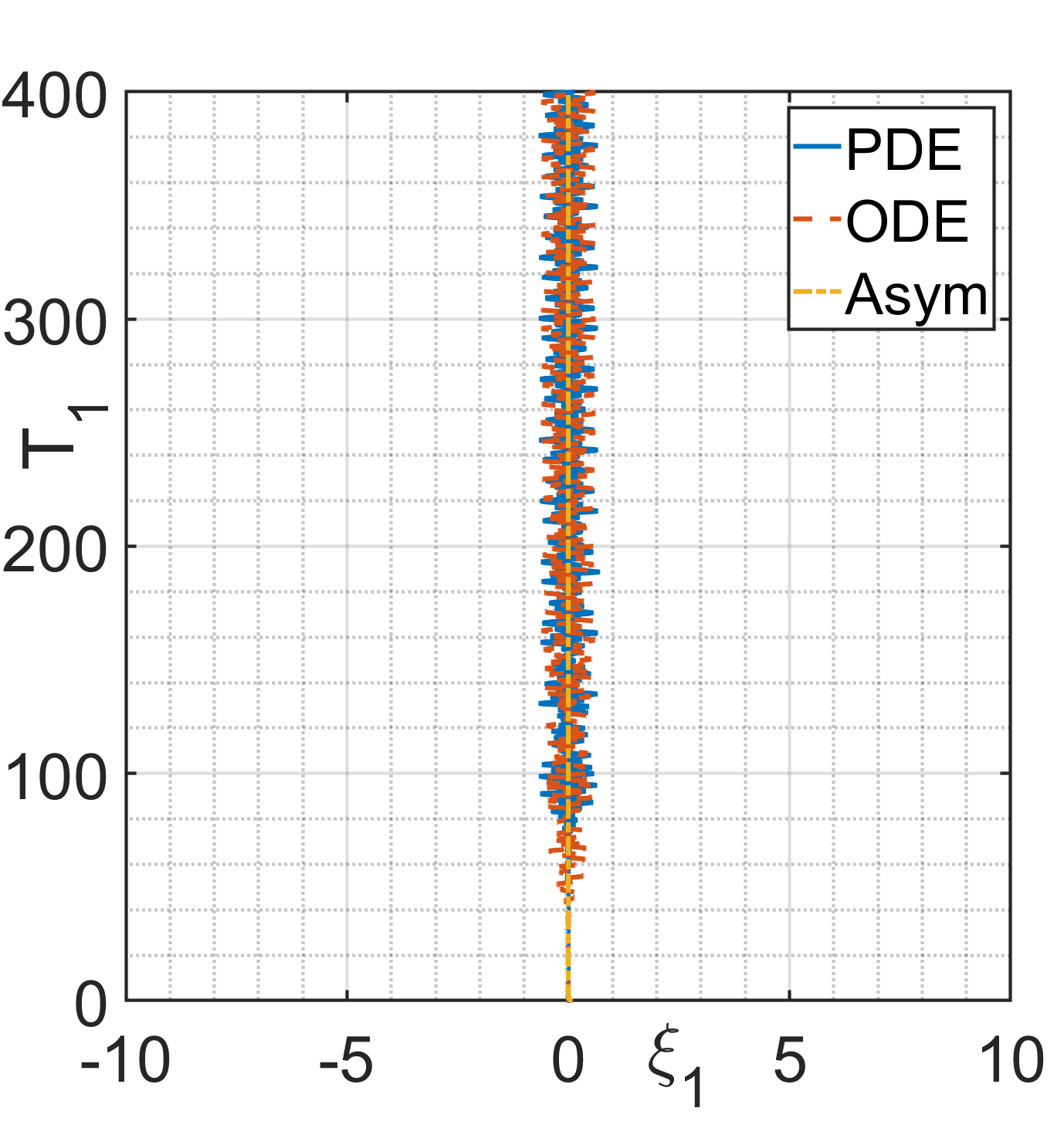}
\end{subfigure}
\hfill\begin{subfigure}[b]{0.24\textwidth}
\centering
\includegraphics[width=\textwidth]{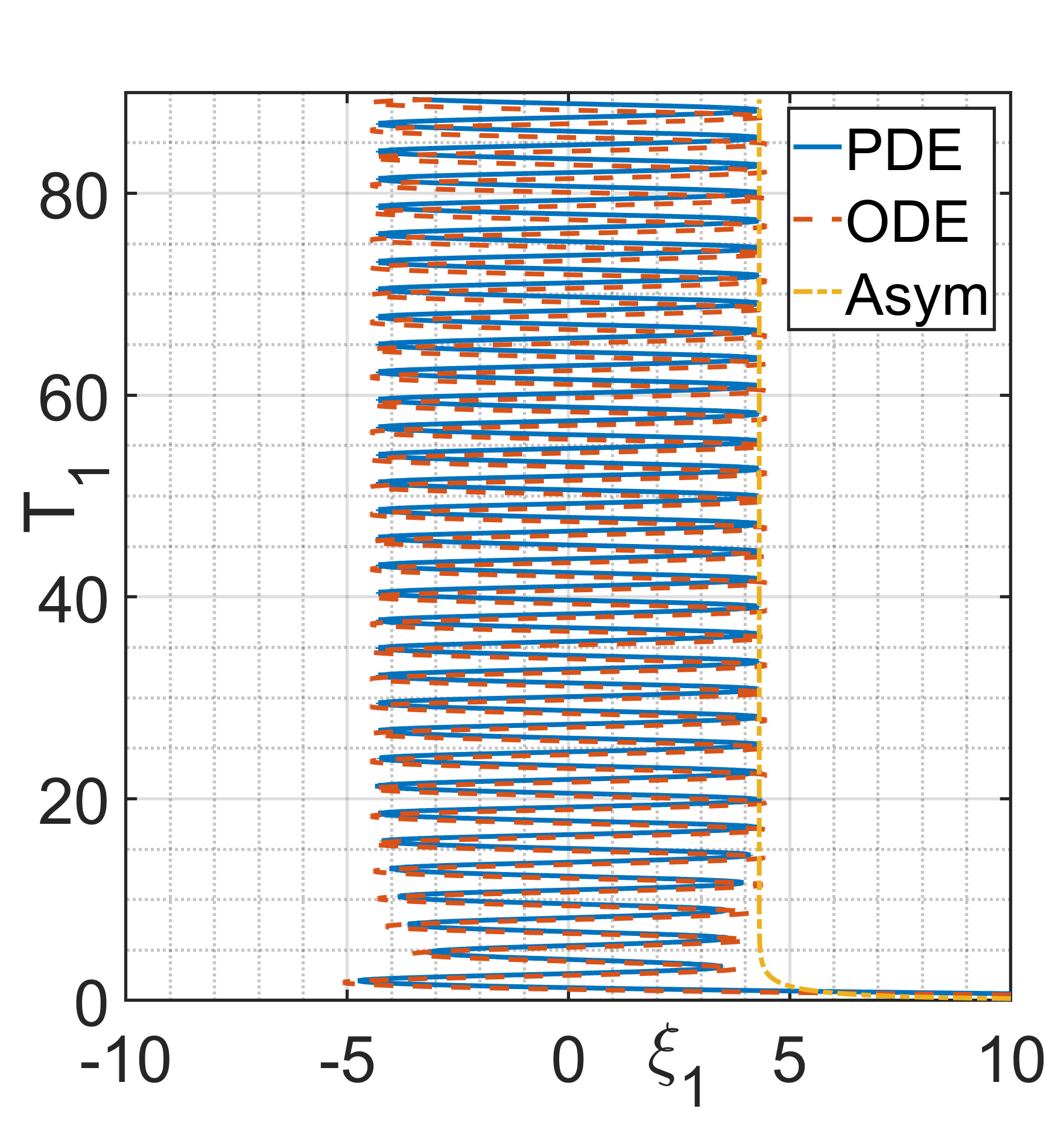}
\end{subfigure}
\hfill\begin{subfigure}[b]{0.24\textwidth}
\centering
\includegraphics[width=\textwidth]{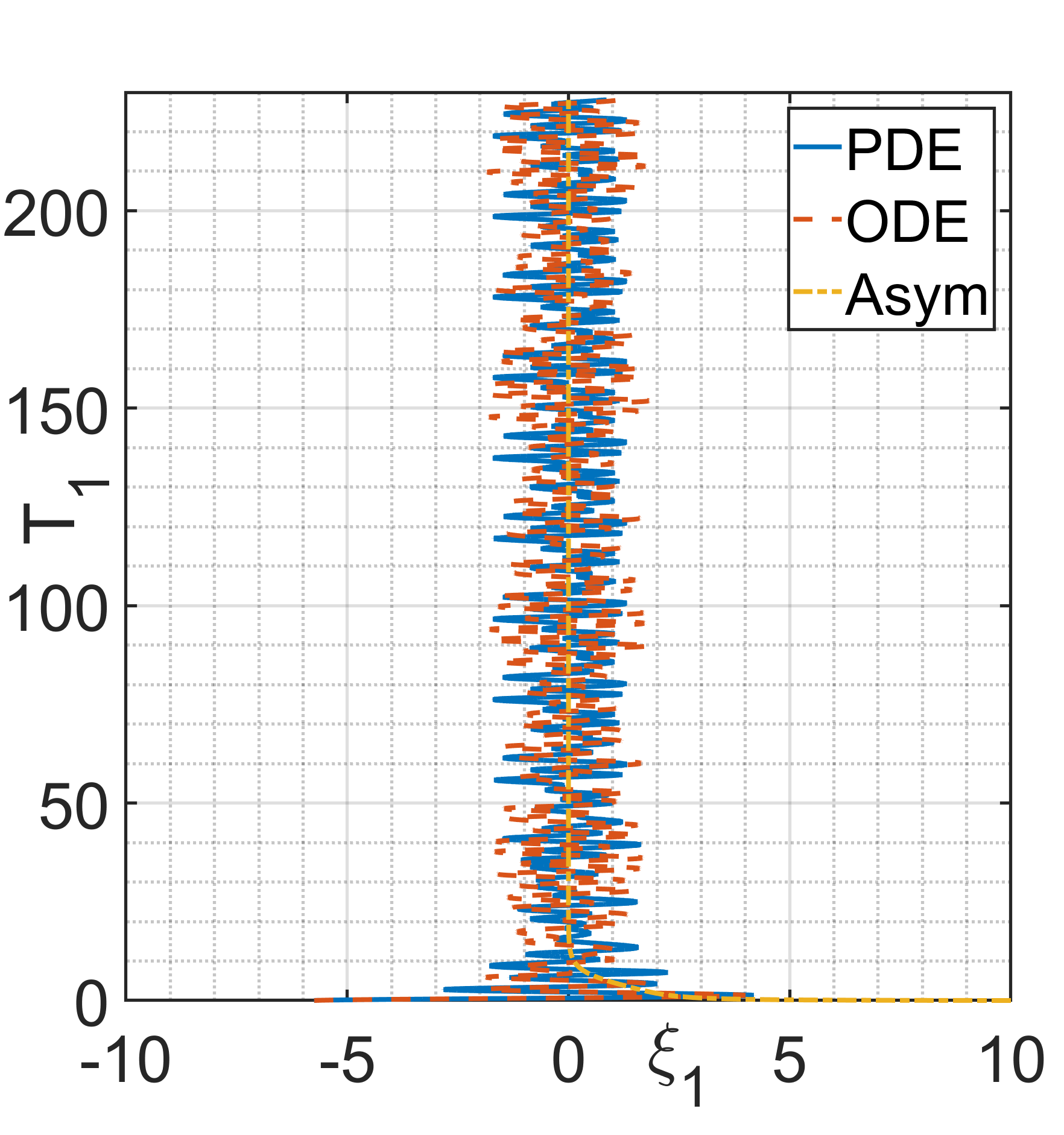}
\end{subfigure}
\hfill

\begin{subfigure}[b]{0.24\textwidth}
\centering
\includegraphics[width=\textwidth]{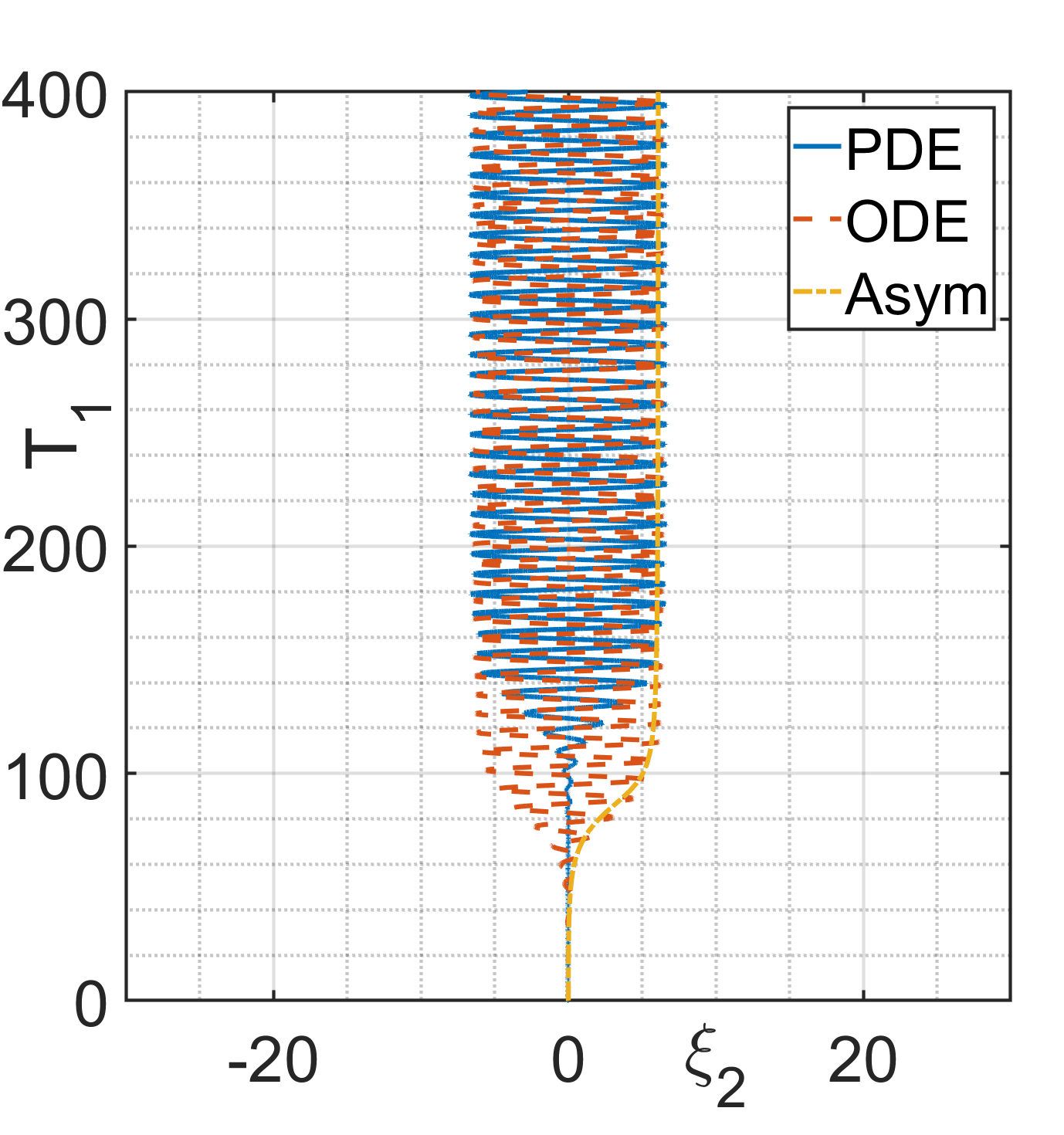}
\end{subfigure}
\hfill\begin{subfigure}[b]{0.24\textwidth}
\centering
\includegraphics[width=\textwidth]{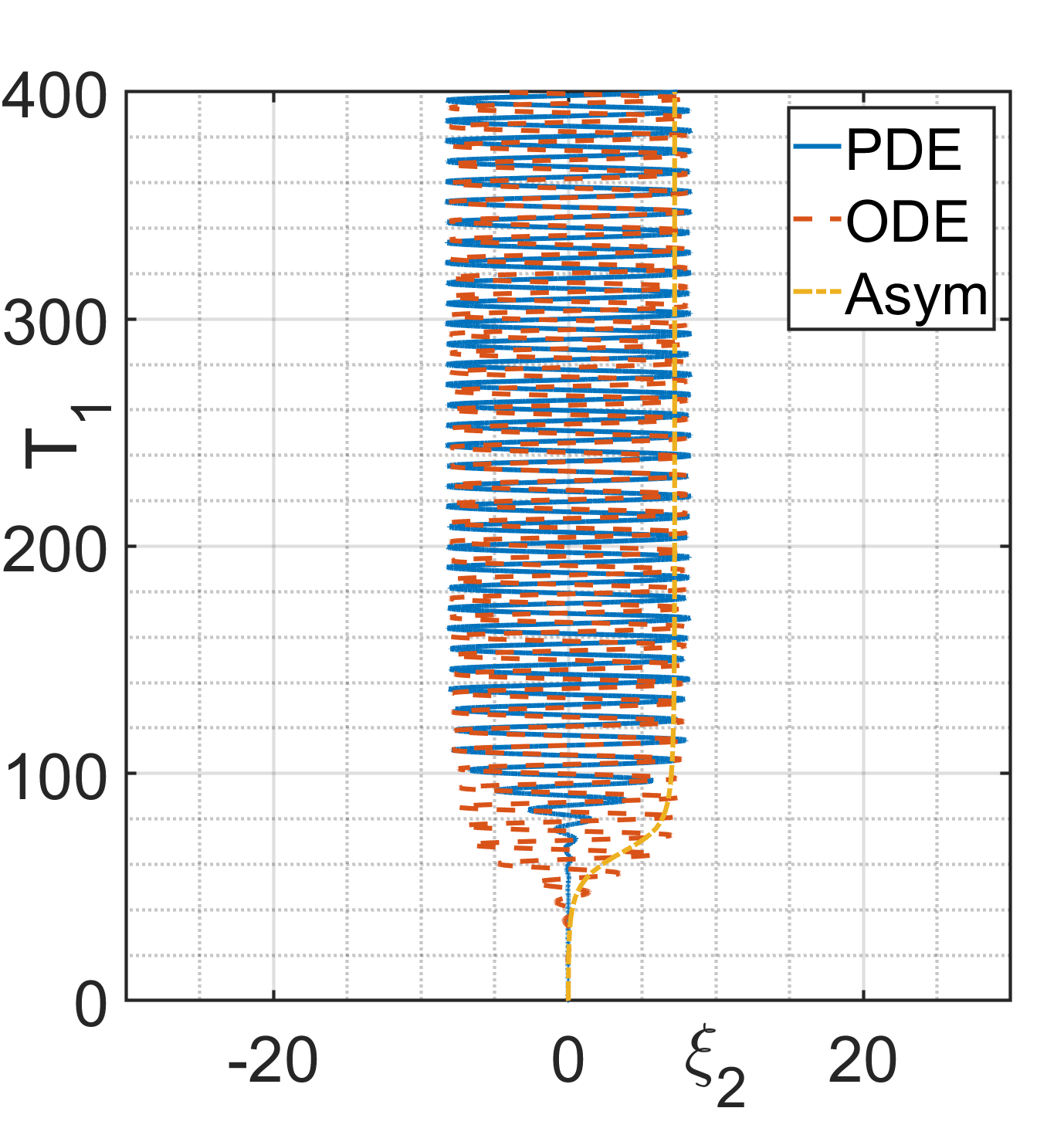}
\end{subfigure}
\hfill\begin{subfigure}[b]{0.24\textwidth}
\centering
\includegraphics[width=\textwidth]{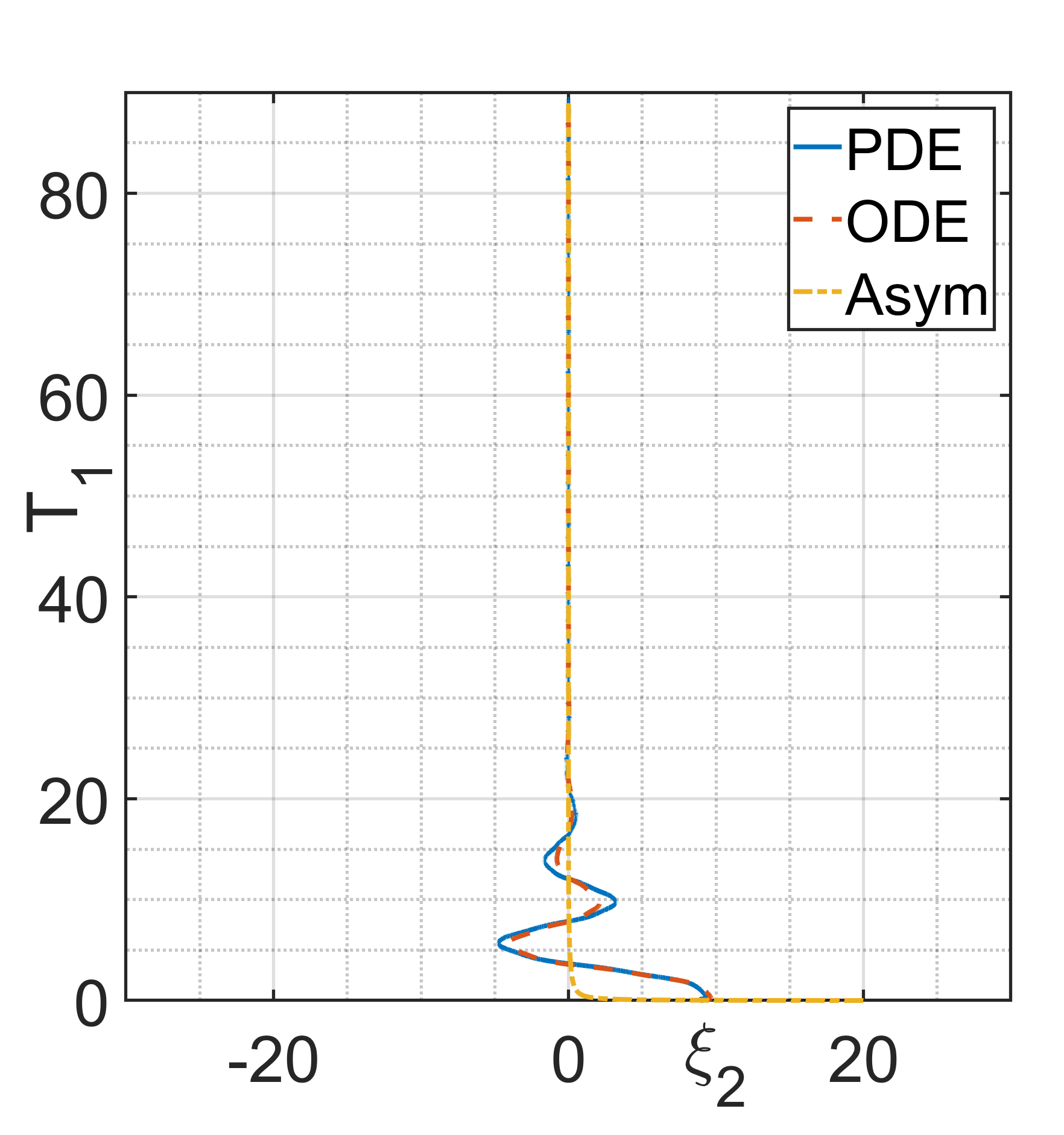}
\end{subfigure}
\hfill\begin{subfigure}[b]{0.24\textwidth}
\centering
\includegraphics[width=\textwidth]{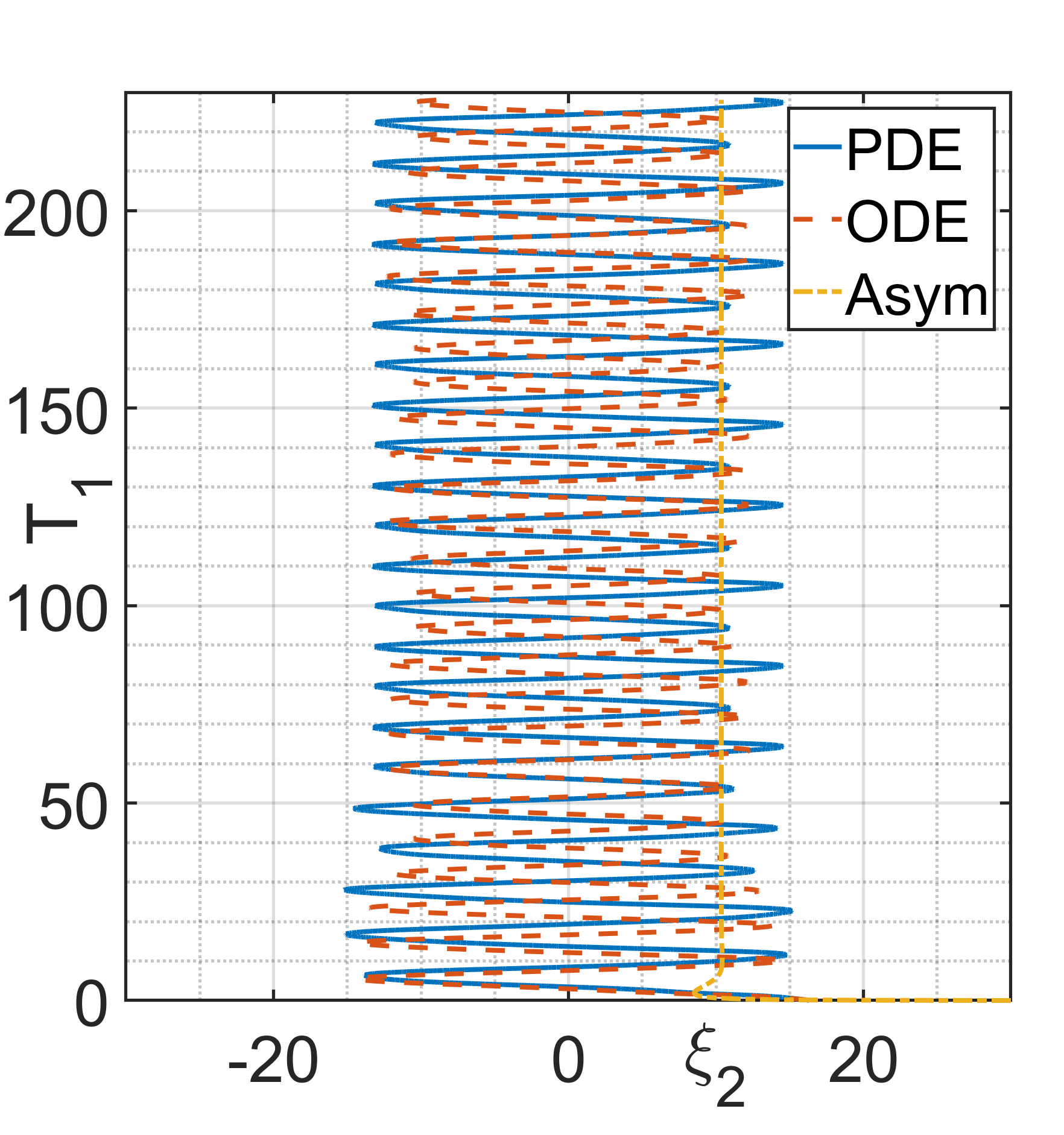}
\end{subfigure}
\hfill
\begin{subfigure}[b]{0.24\textwidth}
\centering
\includegraphics[width=\textwidth]{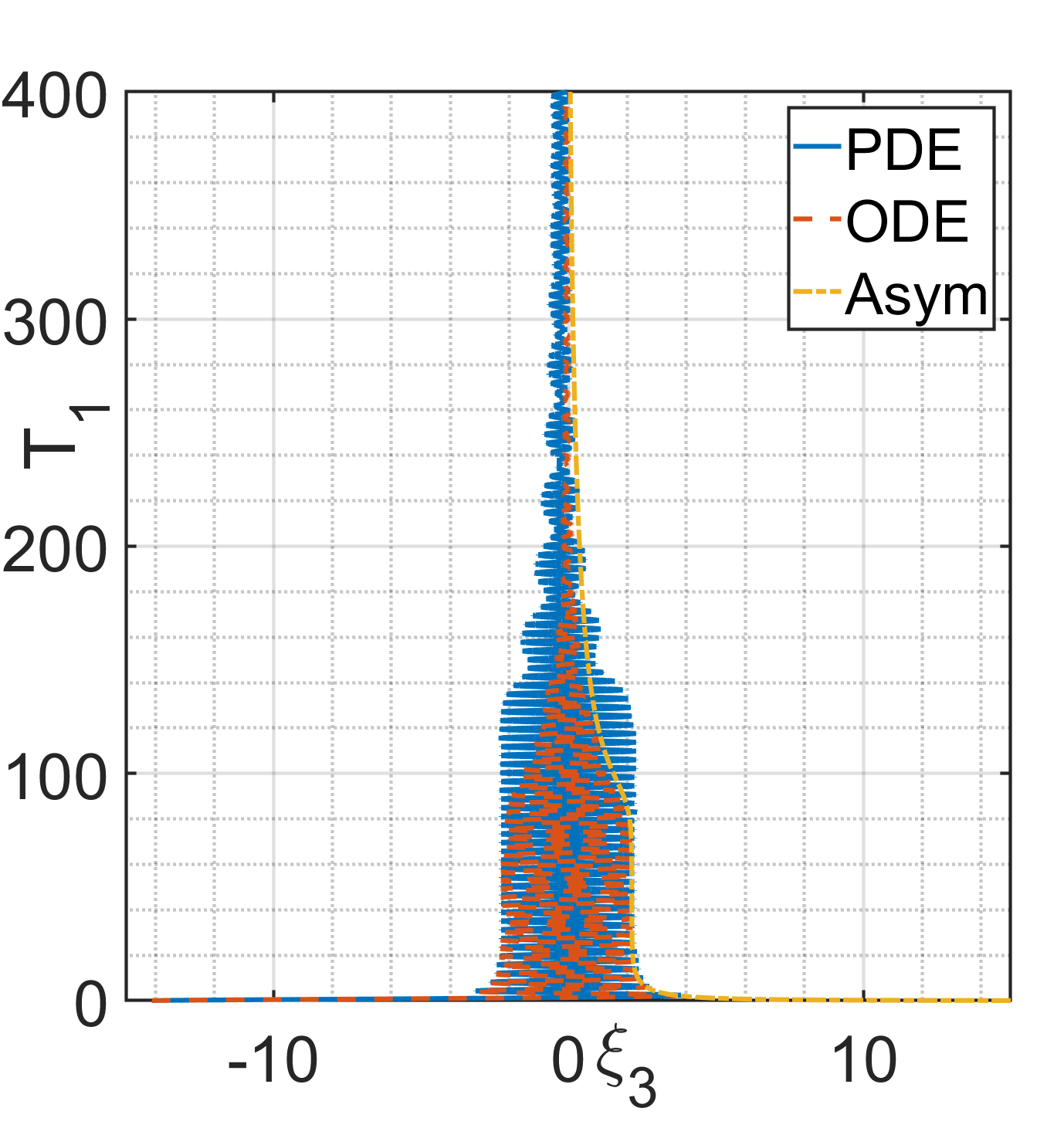}
\caption{$\hat{\tau}=800$}
\label{fig:threespike800}
\end{subfigure}
\hfill\begin{subfigure}[b]{0.24\textwidth}
\centering
\includegraphics[width=\textwidth]{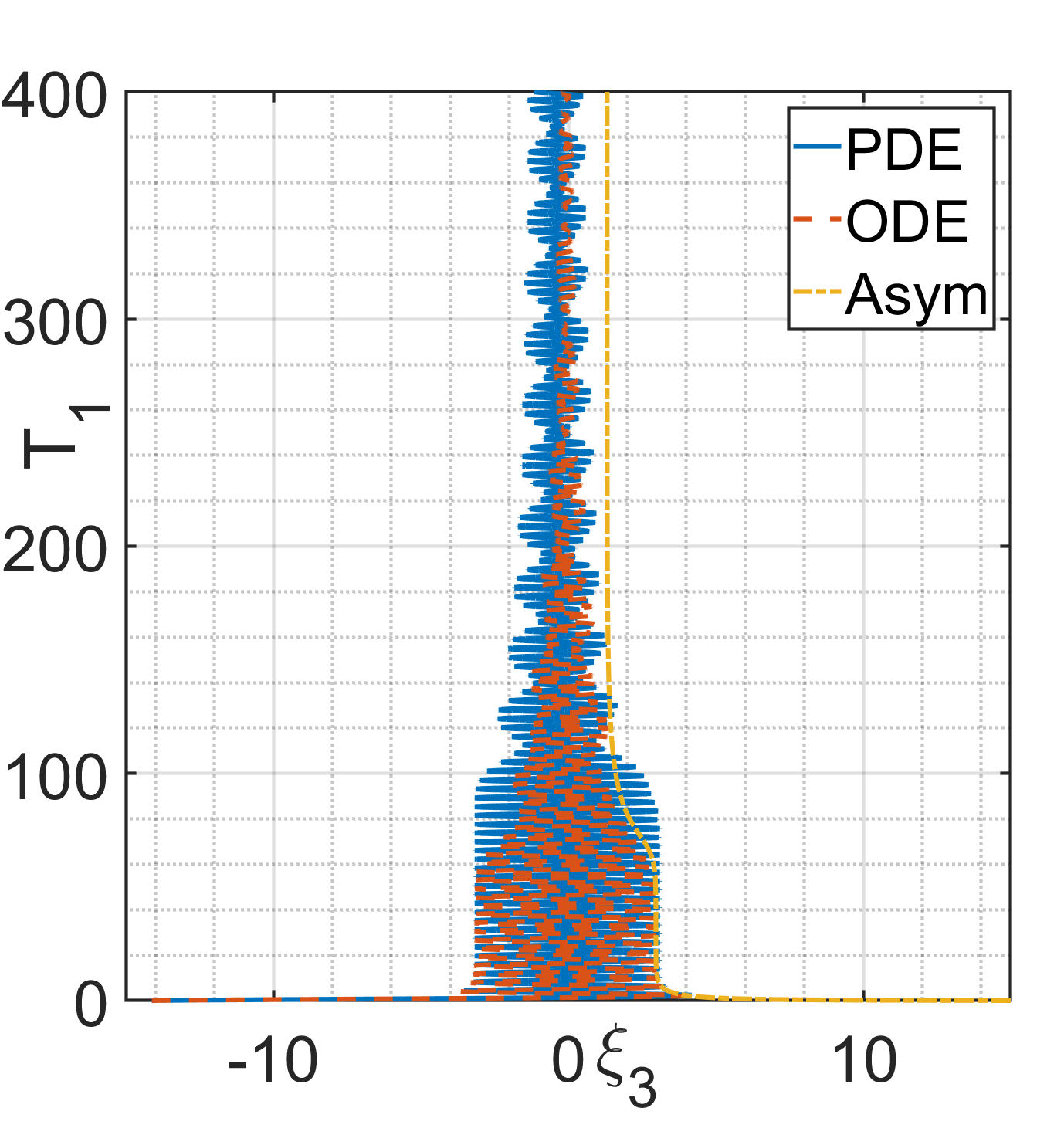}
\caption{$\hat{\tau}=1200$}
\label{fig:threespike1200}
\end{subfigure}
\hfill\begin{subfigure}[b]{0.24\textwidth}
\centering
\includegraphics[width=\textwidth]{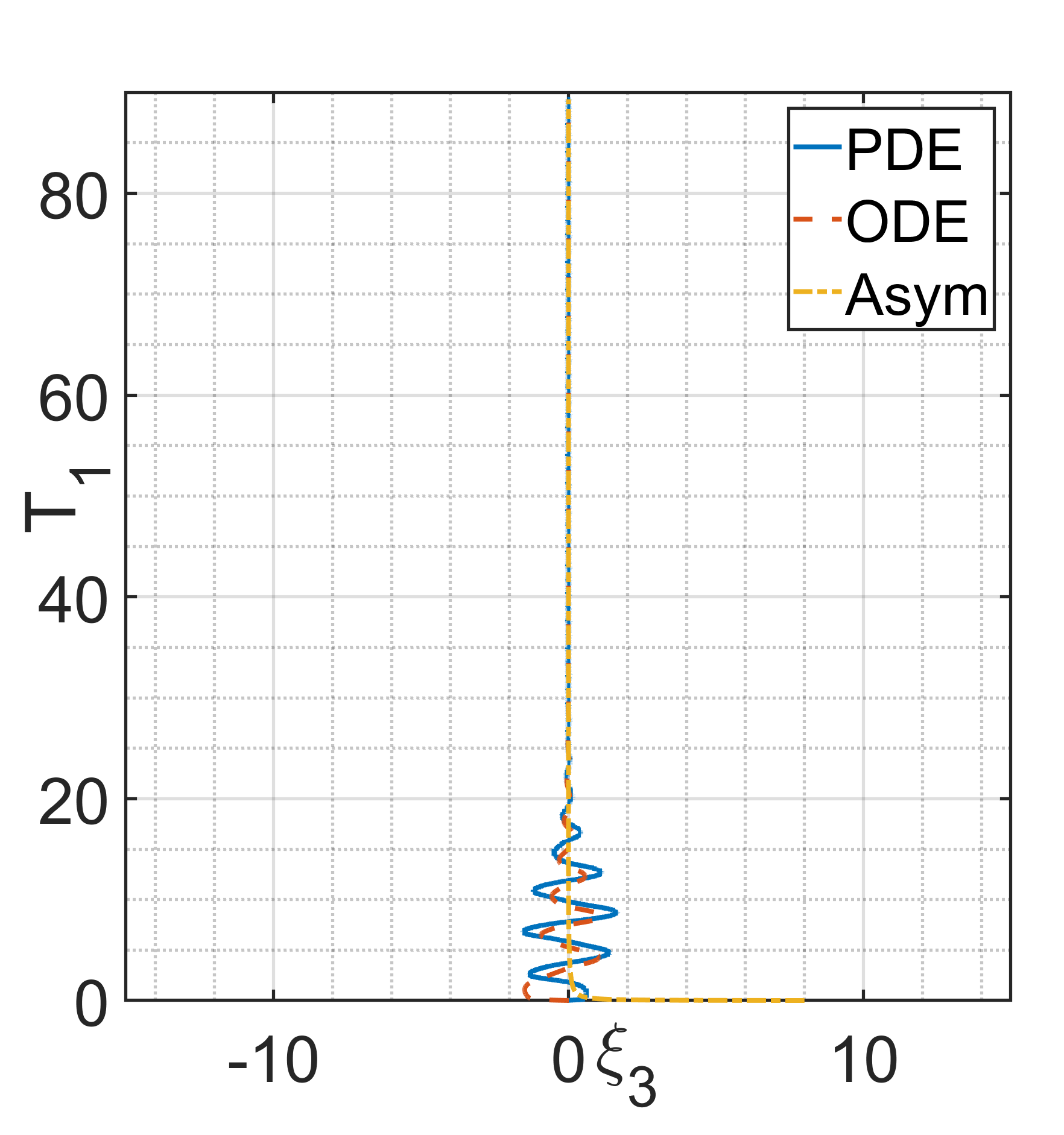}
\caption{$\hat{\tau}=3000$}
\label{fig:threespike30001}
\end{subfigure}
\begin{subfigure}[b]{0.24\textwidth}
\centering
\includegraphics[width=\textwidth]{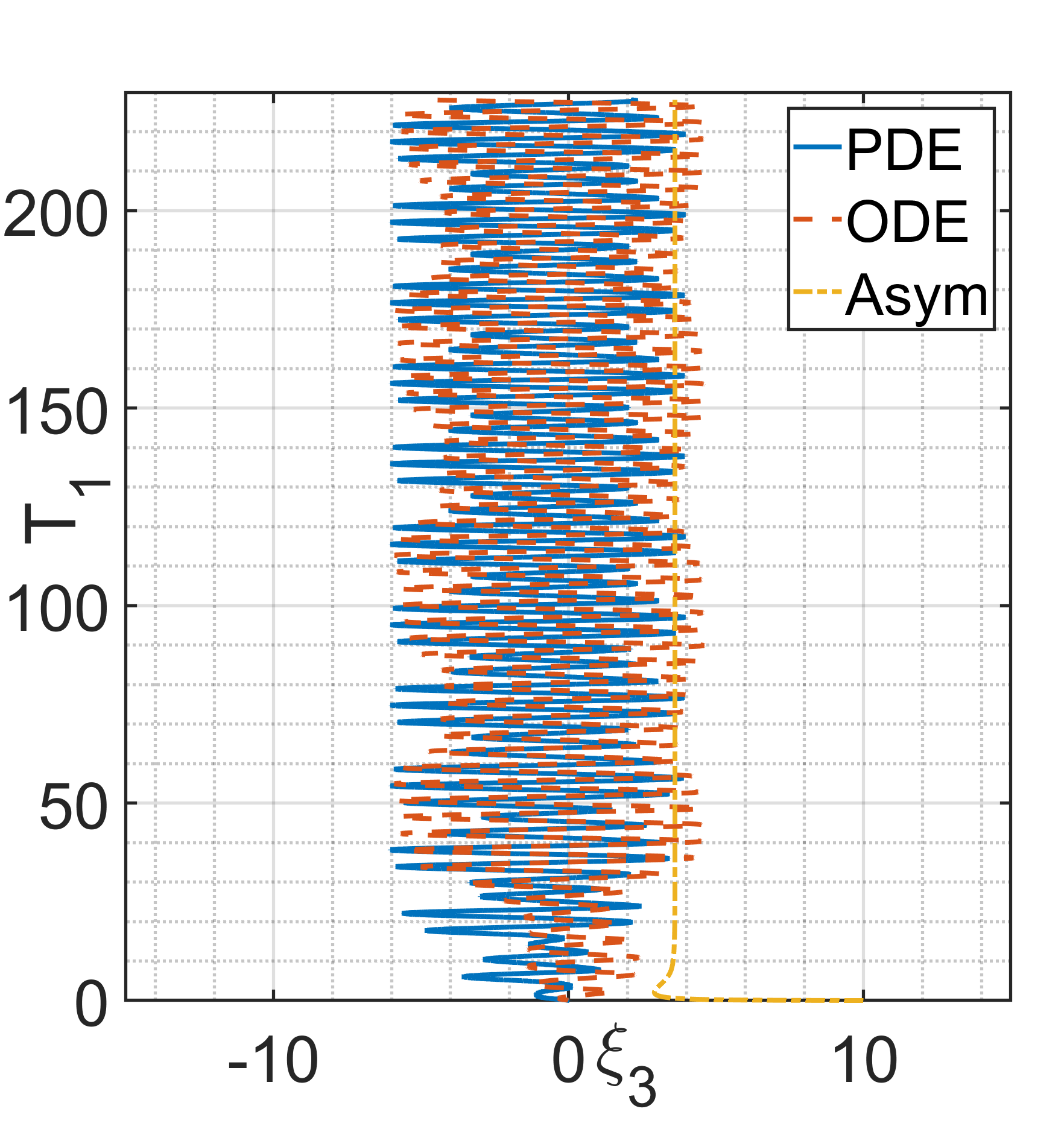}
\caption{$\hat{\tau}=3000$}
\label{fig:threespike30002}
\end{subfigure}
\hfill\caption{(Color online)  First row: locations of the spikes simulated by the PDE (solid lines)  and  ODE  (dashed  lines) in the original variable. The other rows: slow time evolution of new variable $\xi_j,j=1,\ldots,3$. Columns: simulation results obtained at $\hat{\tau}=800,~1200,\text{and}~3000$. Note that columns $1$ and $2$ look the same, but they correspond to different branches of the bifurcation diagram in Fig~(\ref{bd3}), the yellow branch and sushi green branch respectively, which differ by $\xi_{3}$.}%
\label{fig:threespike}
\end{figure}

In this subsection, we discuss the numerical investigation of the ODEs and amplitude modulation equations (\ref{NspikesB}) of the spike locations for the three-spike dynamics. There are at most two stable equilibrium points for the amplitude modulation equations, indicating that the three-spike pattern can at most have two stable oscillatory states. 

When $N=3$, the constants in Eqs.~(\ref{NspikesB}) are:
\begin{equation}
q_{1}=\frac{1}{\sqrt{3}}(1,-1,1)^{\prime},~~q_{2}=\sqrt{\frac{2}{3}}(\frac
{1}{2},1,\frac{1}{2})^{\prime},~~q_{3}=\sqrt{\frac{2}{3}}(\frac{\sqrt{3}}%
{2},0,-\frac{\sqrt{3}}{2})^{\prime},%
\end{equation}%
\begin{equation}
  \lambda_{1,0}=-\frac{1}{2D},~~
  \lambda_{2,0}=-\frac{1}{2D}-\frac{1}{72D^{2}}\left(  1-\frac{1}{81D}\right)
^{-1},~~
  \lambda_{3,0}=-\frac{1}{2D}-\frac{1}{648D^{2}}\left(  1-\frac{1}%
{243D}\right)^{-1}.%
\end{equation}
We choose $D=\frac{1}{500}$ and $\kappa=0.2$ to obtain the bifurcation diagram in Fig.~\ref{bd3}. The first three branch points are located at $\hat{\tau}=72.93,~356.68,~\text{and}~694.45$, corresponding to the three Hopf bifurcation points obtained in the stability analysis of the PDE system. Again, even though there are three Hopf modes becoming unstable, one or two  oscillation states are stable. Then, we choose several special values of $\hat{\tau}=800,1200,~\text{and}~3000$ to validate our asymptotic results. In Fig.~\ref{bd3}, we can see one stable equilibrium point at $\hat{\tau}=800,~1200$ and two stable equilibrium points at $\hat{\tau}=3000$. Fig.~\ref{fig:threespike} shows the stable oscillations of the spike locations corresponding to these values.  The long term behaviors of the amplitude modulations of the spike locations from the PDE simulation and reduced ODE simulation are in good agreement.  We remark that higher-order approximations are required to capture the difference between the amplitude modulations of the spike locations from the PDE, ODE and modulation equations more fully when the corresponding modulation approaches $0$.

\section{Discussion}
\label{sec:discuss}

\begin{figure}[!htb]
\includegraphics[width=1\textwidth]{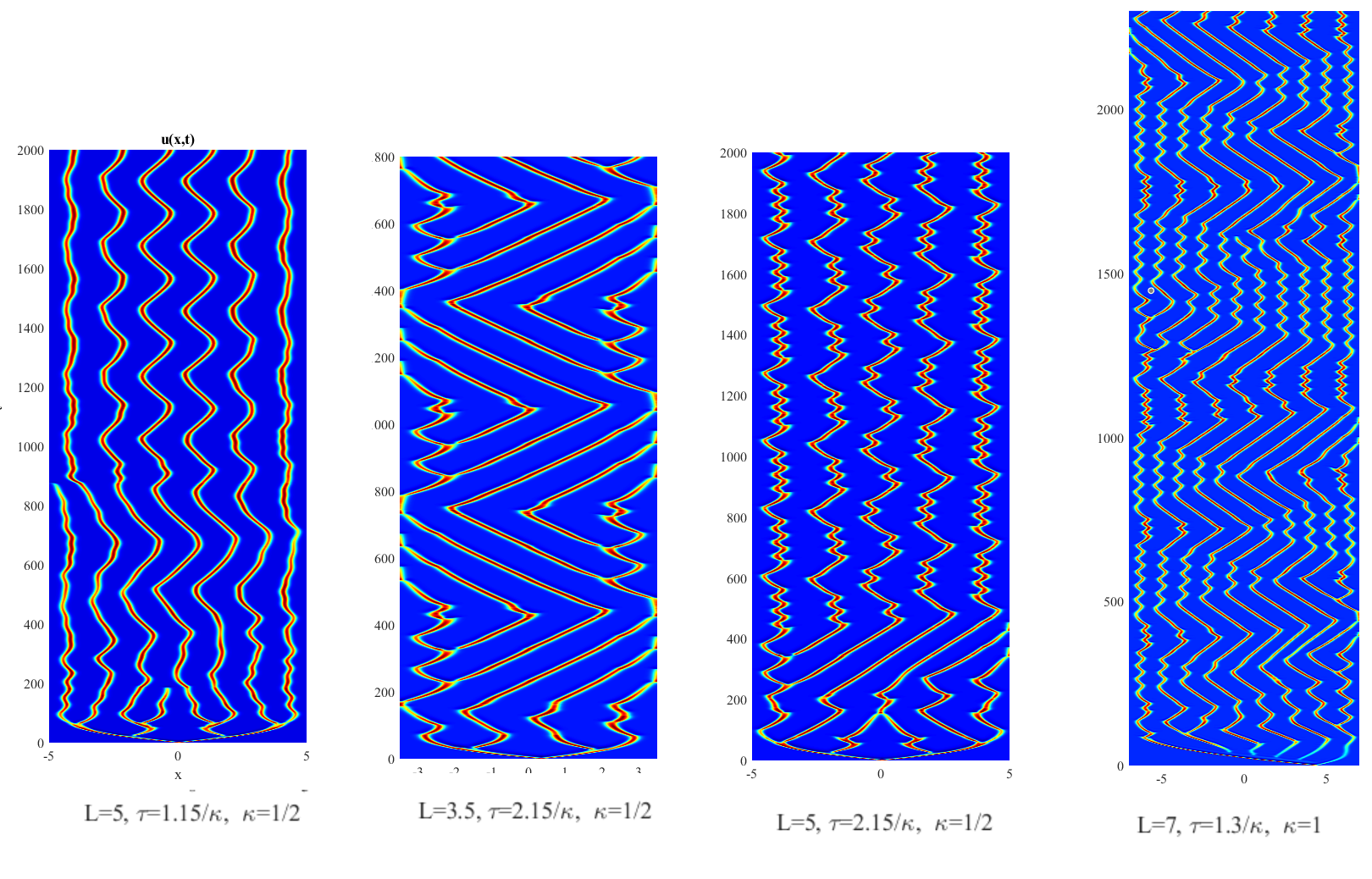}\caption{(Color online) Motion, spike
creation-destruction loops, and coexistence of multi-mode.}%
\label{fig:zoo1d}%
\end{figure}
\begin{figure}[ht!]
\includegraphics[width=1\textwidth]{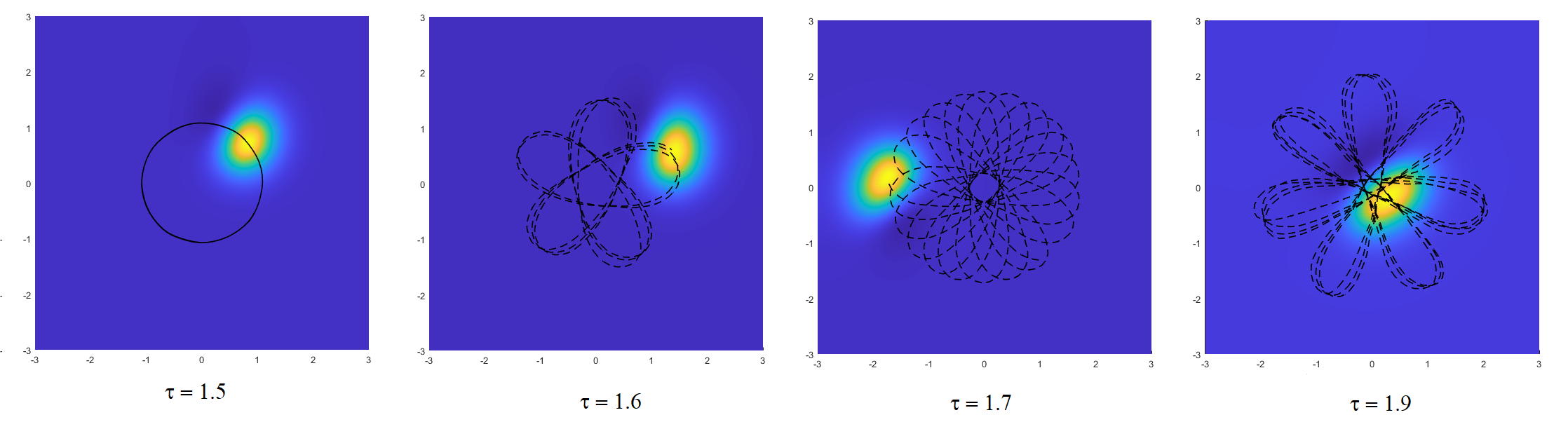}\caption{ (Color online) Exploration of a
single-spot motion in the two-dimensional analogue of system~(\ref{eq0}), obtained by
replacing $\partial_{xx}$ with $\partial_{xx}+\partial_{yy},$ on a square domain
$\left(  x,y\right)  \in\left(  -3,3\right)  ^{2}.$ Here, $\varepsilon
=0.2,\ D=4,\ \kappa=1$ and $\tau$ is as given in the caption. The spike is
rotating clockwise. Higher values of $\tau$ lead to complex procession
orbits.}%
\label{fig:zoo2d}%
\end{figure}
We  presented an extension of the Schnakenberg model, similar to the three-component gas discharge system \cite{schenk1997interacting,bode2002interaction, or1998spot}, in which a second inhibitor is added. For a solution consisting of $N$ spikes, this extension exhibits $N$ distinct and nearly simultaneous Hopf bifurcations in the spike positions, leading to very complex oscillatory dynamics, which cannot be observed in the usual two-component RD models. We analyzed the spike motion near the onset of oscillatory dynamics by first deriving a system of $2N$ ODEs for the spike positions and their velocities, and second by using multiple-scale techniques to further elucidate the dominant dynamics near the N-fold bifurcations.

The reduction techniques (PDE $\rightarrow$ ODEs) are related to those used in a series of papers on the gas discharge model \cite{schenk1997interacting,
bode2002interaction,gurevich2013moving, or1998spot, teramoto2004phase}.   Similarly, our analysis is only valid near the multi-Hopf bifurcation points. While this is a rather limited parameter regime, it allows for a more complete description of the dynamics, including the untangling of the complex interaction between the simultaneous oscillatory modes using multiple scales analysis.   See \cite{veerman2015breathing} for related methods of obtaining leading order expressions for the center manifold expansion  of localised structures near a Hopf bifurcation in a general two-component RD system. Note that the translational mode is not excited near the Hopf bifurcation in  \cite{veerman2015breathing}, thus, the breathing pulses are reported with reference to the pulse tip oscillations.

In a broader context, oscillatory localized patterns with respect to location due to Hopf instabilities in RD systems have been intensively studied. Spike oscillations have been observed previously in two-component RD systems such as the Gray-Scott model (of which Schnakenberg model is a limiting case) \cite{doelman2000slowly, muratov2001traveling, chen2009oscillatory,
kolokolnikov2005existence}. In these works, it was found that even oscillations are the dominant behavior when two spikes oscillate (as well as the so-called breather oscillations).  The oscillatory instability of a single front solution for a two-component activator-inhibitor model was first studied analytically by \cite{nishiura1989layer}. In the two-layer case,  a competition occurs between the in-phase and out-of-phase oscillations . The selection mechanism between them on a finite interval was discussed in \cite{ikeda1994pattern}. Note that the coexistence of those two phases was not observed in \cite{ikeda1994pattern}. In \cite{mckay2012interface,xie2016oscillations}, it was shown that for a certain large class of RD systems, even oscillations dominate the dynamics. In contrast, for the system (\ref{eq0})\ we found multiple coexisting periodic motions, supporting both odd and even oscillations.

Many open questions remain. Among them is to study the doubly-reduced ODE system (\ref{NspikesB}) for $N\geq3$ spikes. A preliminary study with $N=2$ spikes shows a rich bifurcation structure as well as the coexistence of multiple frequency oscillations. The question that how many stable oscillations can coexist for multiple spikes ($N\geq3$) remains open. Further away from the Hopf bifurcations, there is a zoo of interesting dynamics. Some of these are shown in Fig.~\ref{fig:zoo1d}. In particular, for larger domain sizes we observe \textquotedblleft chaotic\textquotedblright%
\ dynamics and spike creation-destruction cycles. 
A similar creation-destruction process and chaotic motion for the two-component Gray-Scott model were analyzed numerically in \cite{nishiura2001spatio}. 
See also a brief survey \cite{nishiura2009dynamics} in this direction. For moderate $\tau$ values, multiple oscillatory modes are seen to coexist leading to rich dynamics. For larger values of $\tau,$ \textquotedblleft zigzag\textquotedblright\ spike motion dominates.  It would be very interesting to derive the reduced equations of motion in this regime, far from the bifurcation points.

Very rich dynamics are observed in two or higher dimensions, even for a single spike. Fig.~\ref{fig:zoo2d} shows complex \textquotedblleft
flower\textquotedblright\ orbits for a single spot in a square domain. Some of these are reminiscent of the trace of a meandering tip of a spiral wave
\cite{golubitsky1997meandering}. It is a completely open question to analyze these; however, see \cite{xie2017moving} for analysis of simple (circular)\ orbit in the two-dimensional Schnakenberg model inside a disk. In addition, it was shown in \cite{teramoto2009rotational} that rotational motion of spot can emerge due to a combination of drift and peanut instabilities.

In conclusion, three-component RD systems exhibit very rich oscillatory spike motions. Combining PDE $\rightarrow$ ODE reduction and multiple scales techniques sheds light on the long-time behavior of spikes in these systems.

\section*{Acknowledgement}
S.X. and Y.N.  acknowledge partial support by the Council for Science, Technology and Innovation (CSTI), Japan, Cross-Ministerial Strategic Innovation Promotion Program (SIP), Japan, ‘Materials Integration’ for Revolutionary Design System of Structural Materials. Also, Y.N. gratefully acknowledges the support of KAKENHI, Japan Grants-in-Aid no.20K20341. T.K is supported by NSERC discovery grant, Canada.

\begin{appendices}
\section{Integral evaluations.}\label{appendix}
In this appendix, we provide the solution for $U_1$ and evaluations of the integrals used in \S \ref{sec3}. The constants in the integrals are evaluated using Maple software, \cite{maple}.

We solve for $U_1$ first. Solving for $W_1$ from (\ref{order2_3}) and substituting it into (\ref{order2_1}) yields
\begin{equation}\label{U_1even}
    U_{1yy}- U_1+2U_0V_0U_1=-U_0^2V_1.
\end{equation}
Substituting $U_0$ and $V_1$ into (\ref{U_1even}) gives
\begin{equation}\label{U_1even2}
    U_{1yy}- U_1+2\rho U_1=-\frac{\rho^2}{C_{0,k}^2}\left( \frac{1}{D} \int_0^y \int_0^z \frac{1}{C_{0,k}} \rho^2 d\hat{y} dz +C_{1,k} \right).
\end{equation}
Assuming that  $U_1$  has the form
\begin{equation}
    U_1=-\frac{ C_{1k} \rho}{C_{0,k}^2} -\frac{ \rho \int_0^y \int_0^z  \rho^2 d\hat{y} dz-f }{DC_{0,k}^3} ,
\end{equation}
then $f$ satisfies
\begin{equation}\label{F}
    f_{yy}- f+2\rho f=\rho^3+2\rho'\int_0^y \rho^2 dz.
\end{equation}
Defining the operator $H(f)=(\frac{\partial^2}{\partial y^2}-1+2\rho)f$,  we have the following identity by direct computation:
\begin{equation}\label{operatorH}
\begin{split}
        H(\rho)=\rho^2,\\
        H(y\rho)=2\rho-2\rho^2, \\
        H(\rho^2)=3\rho^2-\frac{4}{3}\rho^3.
\end{split}
\end{equation}
Since $\rho=\frac{3}{2}\text{sech}(\frac{y}{2})$, the integral $\rho'\int_0^y \rho^2 dz=\frac{2}{3}\rho^3+\rho^2-3\rho$ can be computed explicitly.
Thus, with a linear combination of Eq.~(\ref{operatorH}), $f$ can be solved as
\begin{equation}\label{Af}
    f=-\frac{7}{4}\rho^2+\frac{5}{4}\rho-3y\rho'.
\end{equation}

Now we presented the evaluation of the integrals used in \S \ref{sec3}.

$\bullet$   Firstly, we address the integrals 
\begin{equation}
\int_{-\infty}^{\infty}(U_{1}^{2}V_{0}+2U_{0}U_{1}V_{1})U_{0y}dy, \quad \text{and}\quad     \int_{-\infty}^{\infty} U^{3}_{0}(y) \int_{0}^y (2U_{0}V_{0}%
U_{1}+U_{0}^{2}V_{1})d\hat{y} dy.
\end{equation}
Since $U_0,~V_0,~U_1,~\text{and}~V_1$ are even functions,  $(U_{1}^{2}V_{0}+2U_{0}U_{1}V_{1})U_{0y}$ is an odd function and
\begin{equation}\label{A8}
\int_{-\infty}^{\infty}(U_{1}^{2}V_{0}+2U_{0}U_{1}V_{1})U_{0y}dy=0.
\end{equation}
Similarly, $2U_{0}V_{0}U_{1}+U_{0}^{2}V_{1}$ is even; thus, $\int_{0}^y (2U_{0}V_{0}U_{1}+U_{0}^{2}V_{1})d\hat{y} $ is odd, and we can conclude that $U^{3}_{0}(y) \int_{0}^y (2U_{0}V_{0}
U_{1}+U_{0}^{2}V_{1})d\hat{y}$ is odd. Consequently,
\begin{equation}\label{A9}
    \int_{-\infty}^{\infty} U^{3}_{0}(y) \int_{0}^y (2U_{0}V_{0}%
U_{1}+U_{0}^{2}V_{1})d\hat{y} dy=0.
\end{equation}

$\bullet$ Next, we evaluate the integral
\begin{equation}
I_1=\int_{-\infty}^{\infty}U_{0y}(U_{1}^{2}V_{1}+2U_{0}V_{1}U_2+2U_{1}V_{0}%
U_2+2U_{0}U_{1}V_2)dy.
\end{equation}
Solving for $W_3$ from (\ref{order3-3}) and substituting it into (\ref{order3-1}) produces
\begin{equation}\label{ru}
\frac{\partial^2 U_2}{\partial y^2}-U_2+2U_0V_0U_2=-U_0^2V_3-U_1^2V_0-2U_0U_1V_1-\frac{\partial \alpha_k}{\partial t}U_{0y}+\alpha_{k}\frac{\partial p_{k}}{\partial t}
U_{0yy}.
\end{equation}
$U_2$ can be decomposed as
\begin{equation}
    U_2=U_{2,e}+U_{2,o},
\end{equation}
where $U_{2,e}$ is an even function, satisfying
\begin{equation}\label{rueven}
\frac{\partial^2 U_{2,e}}{\partial y^2}-U_{2,e}+2U_0V_0U_{2,e}=-U_1^2V_0-2U_0U_1V_1+\alpha_{k}\frac{\partial p_{k}}{\partial t}
U_{0yy},
\end{equation}
and $U_{2,o}$ is an odd function, satisfying
\begin{equation}\label{ruodd}
\frac{\partial^2 U_{2,o}}{\partial y^2}-U_{2,o}+2U_0V_0U_{2,o}=-B_{2,k}y U_0^2-\frac{B_{2,k} \int_{-\infty}^\infty U_0^3dy}{3\int_{-\infty}^{\infty} U^{2}_{0y} dy}U_{0y}=-B_{2,k}y U_0^2-\frac{2B_{2,k}}{C_{0,k}} U_{0y}.
\end{equation}
With the orthogonal condition, $\int_0^\infty U_{0y} U_{2,o} dy =0$, $U_{2,o}$ can be solved as
\begin{equation}
    U_{2,o}=\frac{-B_{2k}  }{C_{0,k}^2}\left(y\rho(y) -\frac{\int_{-\infty}^{\infty} y\rho(y) \rho'(y) dy }{\int_{-\infty}^{\infty}  (\rho'(y))^2 dy}\rho'(y) \right)=\frac{-B_{2k}  }{C_{0,k}^2}\left(y\rho(y) +\frac{5}{2}\rho'(y) \right).
\end{equation}
 We can also rewrite $V_2$ as
\begin{equation}
    V_2=V_{2,e}+V_{2,o},
\end{equation}
where
\begin{equation}
    V_{2,e}=\frac{1}{D}\int_{0}^y \int_{0}^{z}(2U_{0}V_{0}%
U_{1}+U_{0}^{2}V_{1})d\hat{y}dz+C_{2,k},\quad V_{2,o}=B_{2,k}y.
\end{equation}
Therefore
\begin{equation}\label{I1A}
\begin{split}
&I_1=\int_{-\infty}^{\infty}U_{0y}(U_{1}^{2}V_{1}+2U_{0}V_{1}U_2+2U_{1}V_{0}%
U_2+2U_{0}U_{1}V_2)dy\\
&=\int_{-\infty}^{\infty}U_{0y}(U_{1}^{2}V_{1}+2U_{0}V_{1}r_{u,e}+2U_{1}V_{0}%
U_{2,e}+2U_{0}U_{1}r_{v,e})dy+\int_{-\infty}^{\infty}U_{0y}(2U_{0}V_{1}U_{2,o}+2U_{1}V_{0}%
U_{2,e}+2U_{0}U_{1}V_{2,o})dy\\
&=\int_{-\infty}^{\infty}U_{0y}(2U_{0}V_{1}U_{2,o}+2U_{1}V_{0}%
U_{2,o}+2U_{0}U_{1}V_{2,o})dy\\
&=\int_{-\infty}^{\infty} 2U_{0}V_{1}U_{2,o}U_{0y}dy-\frac{5B_{2,k}}{C^2_{0,k}}\int_{-\infty}^{\infty}  (\rho')^2 U_{1}   dy\\
&=\int_{-\infty}^{\infty}  \frac{-2B_{2k}  }{C_{0,k}^4} \rho \rho' \left(y\rho+\frac{5}{2}\rho'(y) \right) \left( \frac{1}{D C_{0,k}} \int_0^y \int_0^z \rho^2 d\hat{y} dz +C_{1k} \right)  dy+\frac{5B_{2,k}}{C^2_{0,k}}\int_{-\infty}^{\infty}  (\rho')^2 \left(\frac{C_{1,k}\rho}{C_{0,k}^2}+\frac{\rho \int_0^y \int_0^z  \rho^2 d\hat{y} dz-f}{DC_{0,k}^3}  \right)dy\\
&=\frac{-2B_{2k}  }{C_{0,k}^4}\int_{-\infty}^{\infty}   \rho^2 \rho' y \left( \frac{1}{D C_{0,k}} \int_0^y \int_0^z \rho^2 d\hat{y} dz +C_{1k} \right)  dy-\frac{5B_{2,k}}{C^2_{0,k}}\int_{-\infty}^{\infty}    \frac{(\rho')^2 f}{DC_{0,k}^3}  dy\\
&=-\frac{B_{2,k}}{DC_{0,k}^5} \left(5\int_{-\infty}^{\infty}    (\rho')^2 f dy+2\int_{-\infty}^{\infty}   \rho^2 \rho' y  \int_0^y \int_0^z \rho^2 d\hat{y} dz\right)+\frac{2B_{2,k}C_{1k}}{3DC_{0,k}^5} \int_{-\infty}^{\infty} \rho^3dy\\
&= -(\frac{144}{5}\ln{2}-\frac{3438}{175})\frac{1}{DC_{0,k}^5}+\frac{24B_{2,k}C_{1k}}{5C_{0,k}^4}.
\end{split}
\end{equation}

$\bullet$ Next, we address the integral $I_3=\int_{-\infty}^{\infty} U_0^2R_{v,o} U_{0y} dy$.
\begin{equation}\label{I2A}
\begin{split}
   I_3= \int_{-\infty}^{\infty} U_0^2R_{v,o} U_{0y} dy&=-\frac{1}{3} \int_{-\infty}^{\infty} U_0^3 \frac{\partial R_{v,o}}{\partial y} dy\\
    &=-\frac{1}{3} \int_{-\infty}^{\infty} U_0^3 \left(\frac{1}{D
    } \int_0^y U_0^2r_{v,o}+2U_0V_0r_{u,o} dz +B_{3,k} \right) dy\\
    &=\frac{1}{3} \int_{-\infty}^{\infty} U_0^3 \left(\frac{1}{D
    } \int_0^y \frac{B_{2,k}}{C_{0,k}^2}\left( \rho^2 y +5 \rho \rho' \right) dz -B_{3,k} \right) dy \\
    &=(\frac{72}{5}\ln{2}-\frac{4143}{350})  \frac{B_{2,k}}{DC_{0,k}^5}-\frac{12B_{3,k}}{5C_{0,k}^3}.
\end{split}
\end{equation}

$\bullet$  Finally, we evaluate the integral $\frac{1}{2}\left(\int_0^\infty U_0^2V_2+2U_0V_0U_2dy+\int_{0}^{-\infty} U_0^2V_2+2U_0V_0U_2dy\right)$.
\begin{equation}\label{A20}
\begin{split}
&\frac{1}{2}\left(\int_{0}^\infty U_0^2V_2+2U_0V_0U_2dy+\int_{0}^{-\infty} U_0^2V_2+2U_0V_0U_2dy\right)\\
&=\int_{0}^\infty U_0^2V_{2,o}+2U_0V_0U_{2,o}dy\\
&=\frac{B_{2,k}}{C_{0,k}^2 }\int_{0}^\infty \left( \rho^2 y +5 \rho \rho' \right) dy\\ 
&= (6\ln{2}-\frac{57}{8}) \frac{B_{2,k}}{C_{0,k}^2 }.
\end{split}
\end{equation}

\end{appendices}

\bibliographystyle{unsrt}
\bibliography{main}

\end{document}